\documentclass{CSML}
\pdfoutput=1

\usepackage{lastpage}
\newcommand{\dOi}{14(2:18)2018}
\lmcsheading{}{1--\pageref{LastPage}}{}{}%
{Jan.~31, 2017}{Jun.~25, 2018}{}

\usepackage{hyperref}
\hypersetup{hidelinks}

\usepackage{amsfonts}
\usepackage{empheq}
\usepackage{multirow}
\usepackage{tikz}
\usepackage{array}
\usetikzlibrary{calc,arrows,positioning}
\allowdisplaybreaks

\usepackage{comment}
\usepackage{color}

\newif\ifdraft\drafttrue

\DeclareFontFamily{U}{mathb}{\hyphenchar\font45}
\DeclareFontShape{U}{mathb}{m}{n}{
      <5> <6> <7> <8> <9> <10> gen * mathb
      <10.95> mathb10 <12> <14.4> <17.28> <20.74> <24.88> mathb12
}{}
\DeclareSymbolFont{mathb}{U}{mathb}{m}{n}
\DeclareMathSymbol{\sqdoublecup} {2}{mathb}{"5F} 
\DeclareMathSymbol{\boxplus} {2}{mathb}{"60} 

\DeclareMathOperator{\logic}{\mathcal L}

\DeclareMathOperator{\logicready}{\logic_{\mathrm{r}}}
\DeclareMathOperator{\logicstateready}{\logic_{\mathrm{r}}^{\mathrm{s}}}
\DeclareMathOperator{\logicdistready}{\logic_{\mathrm{r}}^{\mathrm{d}}}
\DeclareMathOperator{\logicplus}{\logic_{+}}
\DeclareMathOperator{\logicstateplus}{\logic_{+}^{\mathrm{s}}}
\DeclareMathOperator{\logicdistplus}{\logic_{\mathrm{+}}^{\mathrm{d}}}

\newcommand{\ready}{\sqsubseteq_{\mathrm r}}

\newcommand{\HH}{\mathcal H}

\newcommand{\w}{\mathfrak{w}}
\newcommand{\W}{\mathfrak{W}}
\newcommand{\f}{\mathfrak{f}}
\newcommand{\K}{\mathcal K}
\newcommand{\n}{\mathfrak{n}}

\newcommand{\dgsos}{$\Sigma$-distribution }

\newcommand{\diam}[1]{\langle #1 \rangle}

\newcommand{\init}[1]{\mathrm{init}(#1)}
\newcommand{\der}[1]{\mathrm{der}(#1)}

\newcommand{\rel}{\,{\mathcal R}\,}
\newcommand{\reldist}{\,{\mathcal R}^{\dagger}\,}

\newcommand{\ProbDist}[1]{\Delta(#1)}

\newcommand{\var}[1]{\mathrm{var}(#1)}
\newcommand{\Act}{\mathcal A}

\newcommand{\powset}[1]{{\mathcal P}(#1)}

\newcommand{\rhs}{\mathrm{rhs}}
\newcommand{\lhs}{\mathrm{lhs}}

\newcommand{\src}[1]{\mathrm{src}(#1)}
\newcommand{\trg}[1]{\mathrm{trg}(#1)}
\newcommand{\lit}[1]{\mathrm{Lit}(#1)}
\newcommand{\pick}{\mathbf{pick}}
\newcommand{\opp}{\mathbf{opp}}

\newcommand{\ReT}{\Re^{P}}
\newcommand{\ReD}{\Re^{\Sigma}}

\newcommand{\trans}[1][]{\xrightarrow{\, {#1} \, }}
\newcommand{\ntrans}[1][]{\mathrel{{\trans[#1]}\makebox[0em][r]{$\not$\hspace{2ex}}}{\!}}

\DeclareMathOperator{\logicstate}{\logic^{\mathrm{s}}}
\DeclareMathOperator{\logicdist}{\logic^{\mathrm{d}}}

\newcommand{\leaveout}[1]{}

\newcommand{\SOSrule}[2]{\frac{\displaystyle #1}{\displaystyle #2}}

\newcommand{\D}{\textsf{D}}
\newcommand{\T}{\textsf{T}}

\newcommand{\ddedrule}[2]{\frac{\displaystyle #1}{\displaystyle #2}}
\newcommand{\dedrule}[2]{\frac{#1}{#2}}

\newcommand{\openDTerms}{\mathbb{DT}(\Sigma)}
\newcommand{\closedDTerms}{{\mathcal DT}(\Sigma)}

\newcommand{\openT}{\mathbb{T}(\Sigma)}
\newcommand{\openST}{\mathbb{T}}
\newcommand{\openTerms}{\openSTerms \cup \openDTerms}
\newcommand{\openSTerms}{\openST(\Sigma)}

\newcommand{\closedTerms}{\mathcal{T}(\Sigma)}

\newcommand{\pprem}[1]{\mathrm{pprem}(#1)}
\newcommand{\nprem}[1]{\mathrm{nprem}(#1)}

\newcommand{\prem}[1]{\mathrm{prem}(#1)}
\newcommand{\conc}[1]{\mathrm{conc}(#1)}

\newcommand{\Red}[2]{\textrm{Red}}

\newcommand{\SVar}{\mathcal{V}\!_s}
\newcommand{\DVar}{\mathcal{V}\!_d}
\newcommand{\Var}{\mathcal{V}}

\newcommand{\support}{\mathsf{supp}}

\newcommand{\M}{\mathcal{M}}

\usepackage{color}

\newenvironment{apx-proof}[1] 
        {\noindent \rm \textbf{Proof of #1.}} 
        {\qed}

\definecolor{lightblue}{RGB}{224,224,255}
\definecolor{lightred}{RGB}{255,224,224}
\definecolor{lightgreen}{RGB}{224,255,224}
\definecolor{lightyellow}{RGB}{255,255,224}
\definecolor{lightpurple}{RGB}{255,224,255}
\definecolor{darkerred}{RGB}{64,0,0}
\definecolor{darkred}{RGB}{128,0,0}
\definecolor{darkblue}{RGB}{0,0,128}
\definecolor{darkgreen}{RGB}{0,128,0}
\definecolor{darkpurple}{RGB}{128,0,128}
\definecolor{grey}{rgb}{0.745098,0.745098,0.745098}
\definecolor{lightgrey}{rgb}{0.9,0.9,0.9}
\definecolor{darkgrey}{rgb}{0.6,0.6,0.6}

\newcommand{\colorpar}[3]{\colorbox{#1}{\parbox{#2}{#3}}}

\newcommand{\marginremark}[3]{\marginpar{\colorpar{#2}{\linewidth}{\color{#1}#3}}}

\makeatletter
\def\THICKhrulefill{\leavevmode \leaders \hrule height 5pt\hfill \kern \z@}
\makeatother

\newcommand{\remarkST}[1]{\marginremark{darkgreen}{lightgreen}{\tiny{[ST]~ #1}}}
\newcommand{\remarkVC}[1]{\marginremark{darkblue}{lightyellow}{\tiny{[VC]~ #1}}}

\ifdraft
\else
\newcommand{\remarkST}[1]{}
\newcommand{\remarkVC}[1]{}

\fi

\begin{document}

\title{SOS-based Modal Decomposition on Nondeterministic Probabilistic Processes}

\author{Valentina Castiglioni\rsuper{a}}
\address{\lsuper{a}University of Insubria, Como, Italy}
\email{v.castiglioni2@uninsubria.it}
\email{simone.tini@uninsubria.it}

\author{Daniel Gebler\rsuper{b}}
\address{\lsuper{b}VU University, Amsterdam, The Netherlands}
\email{e.d.gebler@vu.nl}

\author{Simone Tini\rsuper{a}}


\titlecomment{A preliminary version of this paper appeared as \cite{CGT16b}}

\begin{abstract}
We propose a method for the decomposition of modal formulae on processes with nondeterminism and probability with respect to Structural Operational Semantics.
The purpose is to reduce the satisfaction problem of a formula for a process to verifying whether its subprocesses satisfy certain formulae obtained from the decomposition. 
To deal with the probabilistic behavior of processes, and thus with the decomposition of formulae characterizing it, we introduce a SOS-like machinery allowing for the specification of the behavior of open distribution terms.
By our decomposition, we obtain (pre)congruence formats for probabilistic bisimilarity, ready similarity and similarity.
\end{abstract}

\maketitle


\section{Introduction}
\label{sec:intro}

Behavioral equivalences and modal logics have been successfully employed for the specification and verification of communicating concurrent systems, henceforth processes.
The former ones, in particular the family of bisimulations, provide a simple and elegant tool for the comparison of the observable behavior of processes.
The latter ones allow for an immediate expression of the desired properties of processes.
Since the work of \cite{HM85} on the Hennessy-Milner logic (HML), these two approaches are connected by means of \emph{logical characterizations} of behavioral equivalences: two processes are behaviorally equivalent if and only if they satisfy the same formulae in the logic.
Hence, the characterization of an equivalence subsumes both the fact that the logic is as expressive as the equivalence and the fact that the equivalence preserves the logical properties of processes.

However, the connection between behavioral equivalences and modal logics goes even further: \emph{modal decomposition} of formulae exploits the characterization of an equivalence to derive its compositional properties. 
Roughly speaking, the definition of the semantic behavior of processes by means of the \emph{Structural Operational Semantics} (SOS)  framework \cite{Plo04} allowed for decomposing the satisfaction problem of a formula for a process into the verification of the satisfaction problem of certain formulae for its subprocesses (see \cite{BFvG04,FvG16,FvGL17,FvGW06,FvGW12,LX91}) by means of the notion of \emph{ruloid} \cite{BIM95}, namely inference transition rules that are derived from the SOS specification and define the behavior of open processes in terms of the behavior of their variables.
Then, in \cite{BFvG04,FvG16,FvGL17,FvGW12}, the decomposition of modal formulae is used to systematically derive expressive congruence (precongruence) formats for several behavioral equivalences (preorders) from their modal characterizations.
Further, in \cite{GF12} the semantic model of reactive probabilistic labeled transition systems \cite{vGSS95} is considered and a method for decomposing formulae from a probabilistic version of HML \cite{PS07} characterizing probabilistic bisimilarity wrt.\ a probabilistic transition system specification in the format of \cite{LT09} is proposed.

Our purpose is to extend the SOS-driven decomposition approach to processes in which the nondeterministic behavior coexists with probability.
To this aim we take the very general semantic model of nondeterministic probabilistic transition systems (PTSs) of \cite{DL12,S95}.
In the PTS model, processes perform actions and evolve to probability distributions over processes, i.e.\ an $a$-labeled transition is of the form $t \trans[a] \pi$, with $t$ a process and $\pi$ a distribution holding all information on the probabilistic behavior arising from this transition.
All modal logics developed for the PTS model are equipped with modalities allowing for the specification of the quantitative properties of processes.
In essence, this means that some modal formulae are (possibly indirectly) evaluated on distributions.
In order to decompose this kind of formulae, we introduce a SOS-like machinery, called \emph{distribution specification}, in which we syntactically represent open distribution terms as probability distributions over open terms.
More precisely, our distribution specification, consisting in a set of \emph{distribution rules} defined on a signature, will allow us to infer the expression $\Theta \trans[q] t$ whenever a closed distribution term $\Theta$ assigns probability weight $q$ to a closed term $t$.
Then, from these distribution rules we derive the \emph{distribution ruloids}, which will play a fundamental r\^ole in the decomposition method.
In fact, as happens for ruloids on terms, our distribution ruloids will allow us to derive expressions of the form $\Theta \trans[q] t$, for an arbitrary open distribution term $\Theta$ and open term $t$, by considering only the behavior of the variables occurring in $\Theta$.
Hence, they will allow us to decompose the formulae capturing the quantitative behavior of processes since through them we can relate the satisfaction problem of a formula of this kind for a closed distribution term to the satisfaction problem of certain derived formulae for its subterms.
We stress that our distribution ruloids can support the decomposition of formulae in any modal logic for PTSs and moreover the distribution specification we have developed can be easily generalized to cover the case of models using sub-distributions in place of probability distributions (see for instance \cite{LdV15,LdV16}).

We present the decomposition of formulae from the two-sorted boolean-valued modal logic $\logic$ of \cite{DD11}.
This is an expressive logic, which characterizes probabilistic bisimilarity \cite{DD11} and bisimilarity metric \cite{CGT16a}.
We apply our decomposition method also to two subclasses of formulae in $\logic$, denoted by $\logicready$ and $\logicplus$, which we prove to characterize, respectively, probabilistic ready similarity and similarity.
Finally, to show the robustness of our approach we apply it to derive the congruence theorem for probabilistic bisimilarity wrt.\ the PGSOS format \cite{DGL14} and the precongruence theorem for probabilistic ready similarity and similarity wrt.\ the PGSOS format and the positive PGSOS format, respectively.
Summarizing:
\begin{enumerate}
\item We present new logical characterizations of probabilistic ready similarity and similarity obtained by means of two sublogics of $\logic$, resp.\ $\logicready$ and $\logicplus$ (Theorem~\ref{thm:ready_sim_adequate}).
\item We define a SOS machinery for the specification of the probabilistic behavior of processes, which can support the decomposition of any modal logic for PTSs. 
\item We develop a method of decomposing formulae in $\logic$ and in its sublogics $\logicready$ and $\logicplus$ (Theorem~\ref{thm:decomposition} and Theorem~\ref{cor:decomposition}).
\item We derive (pre)congruence formats for probabilistic bisimilarity, ready similarity and similarity by exploiting our decomposition method on the logics characterizing them (Theorem~\ref{thm:congruence}). 
\end{enumerate}

The paper is organized as follows: in Section~\ref{sec:back} we recall some basic notions on the PTS model, the PGSOS specification, probabilistic (bi)simulations and their logical characterizations. 
In particular we provide the characterization results for probabilistic ready similarity and similarity. 
In Section~\ref{sec:new_rules} we introduce the SOS-like machinery for the specification of the behavior of distribution terms and in Section~\ref{sec:ruloids} we define the two classes of ruloids: the $P$-\emph{ruloids}, built on a PGSOS specification $P$, and the \emph{distribution ruloids}, derived from a distribution specification.
Section~\ref{sec:decompose_decomposition} is the core of our paper and provides our decomposition method and the derivation of the (pre)congruence formats for probabilistic bisimilarity, ready similarity and similarity.
Finally we end with some conclusions and discussion about future work in Section~\ref{sec:conclusions}.


\section{Probabilistic Transition Systems}
\label{sec:back}

\subsection{The PTS model}

A \emph{signature} is given by a countable set $\Sigma$ of \emph{operators}.
We let $f$ range over $\Sigma$ and $\n$ range over the rank of $f$.
We assume a countable set of (state, or term) \emph{variables} $\SVar$ disjoint from $\Sigma$. 
The set $\T(\Sigma, V)$ of \emph{terms} over $\Sigma$ and a set of variables
$V \subseteq \SVar$ is the least set such that:
\begin{enumerate*}[label=(\emph{\roman*})]
\item $x \in \T(\Sigma, V)$ for all $x \in V$, and 
\item $f(t_1,\ldots,t_{\n}) \in \T(\Sigma, V)$ whenever $f \in \Sigma$ and $t_1,\ldots,t_{\n} \in \T(\Sigma, V)$.
\end{enumerate*}
The set of the \emph{closed terms} $\T(\Sigma, \emptyset)$ will be denoted also with $\closedTerms$.
The set of all \emph{open terms} $\T(\Sigma, \SVar)$ wil be denoted also with $\openT$.

Nondeterministic probabilistic labelled transition systems (PTSs) \cite{S95} extend LTSs by allowing for probabilistic choices in the transitions.
The state space is the set of the closed terms $\closedTerms$. 
The \emph{transitions} are of the form $t\trans[a]\pi$, with $t$  a term in $\closedTerms$, $a$ an action label, and $\pi$ a probability distribution over $\closedTerms$, i.e.\ a mapping $\pi \colon \closedTerms \to [0,1]$ with $\sum_{t \in \closedTerms} \pi(t) = 1$.
By $\ProbDist{\closedTerms}$ we denote the set of all probability distributions over $\closedTerms$.

\begin{defi}
[PTS, \cite{S95}]
A \emph{nondeterministic probabilistic labeled transition system (PTS)} is a triple $P=(\closedTerms,\Act,\trans[])$, where: 
\begin{enumerate*}[label=(\emph{\roman*})]
\item $\Sigma$ is a signature,
\item $\Act$ is a countable set of \emph{actions}, and 
\item ${\trans[]} \subseteq {\closedTerms \times \Act \times \ProbDist{\closedTerms}}$ is a \emph{transition relation}. 
\end{enumerate*}
\end{defi}
We write $t \trans[a]$ if there is a distribution $\pi \in \ProbDist{\closedTerms}$ with $t\trans[a] \pi$, and $t \ntrans[a]$ otherwise.

We define the \emph{initials} of term $t$ as the set $\init{t} =\{ a \in \Act \mid  t \trans[a]\}$ of the actions that can be performed by $t$.
For each action $a \in \Act$, the set of $a$-\emph{derivatives} of term $t$ is defined as the set $\der{t,a} =\{\pi\in\ProbDist{\closedTerms} \mid t\trans[a]\pi\}$ of the distributions reachable from $t$ through action $a$.

We need some notation on distributions.
For a distribution $\pi \in \ProbDist{\closedTerms}$, we denote by $\support(\pi)$ the \emph{support} of $\pi$, namely $\support(\pi) = \{ t \in \closedTerms\mid \pi(t) >0\}$.
For a term $t \in \closedTerms$, we denote by $\delta_t$ the \emph{Dirac distribution} such that $\delta_t(t)= 1$ and $\delta_t(s)=0$ for all $s \neq t$.
For $f \in \Sigma$ and $\pi_i \in \ProbDist{\closedTerms}$, $f(\pi_1,\ldots,\pi_{\n})$ is the distribution defined by 
\[
f(\pi_1,\ldots,\pi_{\n})(t) =
\begin{cases}
\prod_{i = 1}^{\n}\pi_i(t_i) & \text{ if } t = f(t_1,\ldots,t_{\n}) \\
0 & \text{ otherwise.}
\end{cases}
\]
Finally, the convex combination $\sum_{i \in I} p_i \pi_i$ of a family of distributions $\{\pi_i\}_{i \in I} \subseteq \ProbDist{\closedTerms}$ with $p_i \in (0,1]$ and $\sum_{i \in I} p_i = 1$ is defined by $(\sum_{i \in I} p_i \pi_i)(t) = \sum_{i \in I} (p_i \pi_i(t))$ for all $t \in \closedTerms$.


\subsection{Probabilistic bisimulation}

A probabilistic bisimulation is an equivalence relation over $\closedTerms$ equating two terms if they can mimic each other's transitions and evolve to distributions related, in turn, by the same bisimulation.
To formalize this intuition, we need to lift relations over terms to relations over distributions on terms.

\begin{defi}
[Relation lifting, \cite{DvG10}]
\label{def:relation_lifting}
The \emph{lifting} of a relation $\rel \subseteq \closedTerms \times \closedTerms$ is the relation $\reldist \subseteq \ProbDist{\closedTerms} \times \ProbDist{\closedTerms}$ with $\pi \reldist \pi'$ whenever there is a countable set of indexes $I$ such that:
\begin{enumerate*}[label=(\emph{\roman*})]
\item $\pi = \sum_{i \in I} p_i \delta_{s_i}$,
\item $\pi' = \sum_{i \in I} p_i \delta_{t_i}$, and
\item $s_i \; \rel \; t_i$ for all $i \in I$.
\end{enumerate*}
\end{defi}

We recall a definition equivalent to Definition~\ref{def:relation_lifting} which will be useful in our proofs.

\begin{prop}
[\protect{\cite[Proposition\ $2.3$]{DD11}}]
\label{prop:equiv_rel_lifting}
Consider a relation $\rel \subseteq \closedTerms \times \closedTerms$. 
Then $\reldist \subseteq \ProbDist{\closedTerms} \times \ProbDist{\closedTerms}$ is the smallest relation satisfying:
\begin{enumerate}
\item $s \rel t$ implies $\delta_s \reldist \delta_t$;
\item  $\pi_i \reldist \pi'_i$ implies $(\sum_{i \in I}p_i \pi_i) \reldist (\sum_{i \in I} p_i \pi'_i)$, where $I$ is  an arbitrary set of indexes and $\sum_{i \in I} p_i =1$.
\end{enumerate}
\end{prop}

\begin{defi}
[Probabilistic (bi)simulations, \cite{LS91,S95}]
\label{def:prob_(bi)sim}
Assume a PTS $(\closedTerms,\Act,\trans[])$.
Then:
\begin{enumerate}
\item
A binary relation $\rel \,\subseteq \closedTerms \times \closedTerms$ is a \emph{probabilistic simulation} if, whenever $s \rel  t$,
if $s \trans[a] \pi_s$ then there is a transition $t \trans[a] \pi_t$ such that $\pi_s \rel^{\dagger} \pi_t$.
\item
A probabilistic simulation $\rel$ is a \emph{probabilistic ready simulation} if, whenever $s \rel t$,
if $s \ntrans[a]$ then $t \ntrans[a]$.
\item
A \emph{probabilistic bisimulation} is a symmetric probabilistic simulation.
\end{enumerate}
\end{defi}

The union of all probabilistic simulations (resp.: ready simulations, bisimulations) is the greatest probabilistic simulation (resp.: ready simulation, bisimulation) \cite{LS91,S95},  denoted by $\sqsubseteq$ (resp.: $\ready$, $\sim$), called \emph{probabilistic similarity} (resp.: \emph{ready similarity}, \emph{bisimilarity}), and is a preorder (resp.: preorder, equivalence).


\subsection{Logical characterization}

As a logic expressing behavioral properties over terms, we consider the modal logic $\logic$ of \cite{DD11}, which extends the Hennessy-Milner Logic \cite{HM85} with a probabilistic choice modality.

\begin{defi}
[Modal logic $\logic$, \cite{DD11}]
\label{def:logic}
The classes of \emph{state formulae} $\logicstate$ and \emph{distribution formulae} $\logicdist$ over $\Act$ are defined by the following BNF-like grammar: 
\[
\begin{array}{rcl} 
 \logicstate\colon & \varphi ::= & 
                 \top  \ |\  
                 \neg \varphi   \ |\  
                \bigwedge_{j \in J}\varphi_j
                 \ | \ \diam{a}\psi \\[1.0 ex]

\logicdist\colon & \psi ::= & 
\bigoplus_{i\in I}r_i\varphi_i  
\end{array}
\]
where: 
\begin{enumerate*}[label=(\emph{\roman*})]
\item $\varphi, \varphi_i, \varphi_j$ range over $\logicstate$, 
\item $\psi$ ranges over $\logicdist$,
\item $a \in \Act$,
\item $J$ is an at most countable set of indexes with $J \neq \emptyset$, and
\item $I$ is an at most countable set of indexes with $I \neq \emptyset$, $r_i \in (0,1]$ for each $i \in I$ and $\sum_{i\in I} r_i=1$.
\end{enumerate*}
\end{defi}
We shall write $\diam{a}\varphi$ for $\diam{a} \bigoplus_{i\in I}r_i\varphi_i$ with $I = \{i\}$, $r_i=1$ and $\varphi_i = \varphi$.


Formulae are interpreted over a PTS.
In particular, formulae $\bigoplus_{i \in I} r_i \varphi_i$ are evaluated on distributions.
Intuitively, a probability distribution $\pi$ satisfies the distribution formula $\bigoplus_{i \in I} r_i \varphi_i$ if, for each $i \in I$, $\pi$ assigns probability (at least) $r_i$ to processes satisfying the state formula $\varphi_i$.
This is formalized by requiring that $\pi$ can be rewritten as a convex combination of probability distributions $\pi_i$, using the $r_i$ as weights of the combination, such that all the processes in the support of $\pi_i$ are guaranteed to satisfy the state formula $\varphi_i$.

\begin{defi}
[Semantics of $\logic$, \cite{DD11}]
\label{def:satisfiability}
Assume a PTS $(\closedTerms, \Act, \trans[])$.
The \emph{satisfaction relation} $\models \, \subseteq (\closedTerms \times \logicstate) \cup (\ProbDist{\closedTerms} \times \logicdist)$ is defined by structural induction on formulae by
\begin{itemize}
\item $t \models \top$ always;
\item $t \models \neg\varphi$ if{f} $t \models \varphi$ does not hold;
\item $t \models \bigwedge_{j \in J} \varphi_j$ if{f} $t \models \varphi_j$ for all $j \in J$;
\item $t \models \diam{a}\psi$ if{f} $t \trans[a] \pi$ for a distribution $\pi \in \ProbDist{\closedTerms}$ with $\pi\models \psi$;
\item $ \pi \models \bigoplus_{i\in I}r_i\varphi_i$ if{f} $\pi = \sum_{i \in I}r_i \pi_i$ for distributions $\pi_i$ with $t \models \varphi_i$ for all  $t \in \support(\pi_i)$.
\end{itemize}
\end{defi}

Dealing with $\logic$ is motivated by its characterization of bisimilarity, proved in \cite{DD11} (see Theorem~\ref{thm:adequacy_DD11} below), bisimilarity metric, proved in \cite{CGT16a}, and similarity and ready similarity, proved here (see Theorem~\ref{thm:ready_sim_adequate} below).

\begin{thm}[\protect{\cite{DD11}}]
\label{thm:adequacy_DD11}
Assume a PTS $(\closedTerms,\Act,\trans[])$ and terms $s,t \in \closedTerms$.
Then, $s \sim t$ if and only if they satisfy the same formulae in $\logicstate$.
\end{thm}

The characterization of ready similarity and similarity requires two subclasses of $\logic$.

\begin{defi}[Ready and positive formulae]
The class of \emph{ready formulae} $\logicready$ is defined as
\[ 
\begin{array}{rcl} 
\logicstateready \colon & \varphi ::= & 
                 \top  \ |\  
                 \bar{a}   \ |\ 
                  \bigwedge_{j \in J}\varphi_j 
                 \ | \ \diam{a}\psi 
\\
\logicdistready \colon & \psi ::= & 
                 \bigoplus_{i \in I} r_i \varphi_i 
\end{array}
\]
where $\bar{a}$ stays for $\neg\diam{a}\top$, and the class of \emph{positive formulae} $\logicplus$ is defined as
\[
\begin{array}{rcl} 
\logicstateplus \colon & \varphi ::= & 
                 \top  \ |\  
\bigwedge_{j \in J}\varphi_j 
                 \ | \ \diam{a}\psi 
\\
\logicdistplus \colon & \psi ::= & 
                 \bigoplus_{i \in I} r_i \varphi_i .
\end{array}
\]
\end{defi}

The classes $\logicready$ and $\logicplus$ are strict sublogics of the one proposed in \cite{DvGHM08} for the characterization of failure similarity and forward similarity \cite{S95}.
In particular, the logic used in \cite{DvGHM08} allows for arbitrary formulae to occur after the diamond modality. 

We can show that these sublogics are powerful enough for the characterization of ready similarity and similarity.

\begin{thm}
\label{thm:ready_sim_adequate}
Assume a PTS $(\closedTerms,\Act,\trans[])$ and terms $s,t \in \closedTerms$. Then:
\begin{enumerate}
\item $s \ready t$ if{f} for any formula $\varphi \in \logicstateready$, $s \models \varphi$ implies $t \models \varphi$.
\item $s \sqsubseteq t$ if{f} for any formula $\varphi \in \logicstateplus$, $s \models \varphi$ implies $t \models \varphi$.
\end{enumerate}
\end{thm}


\begin{proof}
We prove only the first item, namely the characterization of the ready simulation preorder.
The proof for simulation is analogous.
The proof that $s \ready t$ implies that for all formulae $\varphi \in \logicstateready$ we have that $s \models \varphi$ implies $t \models \varphi$ is by structural induction over $\varphi$.
The other implication is proved by showing that the relation 
$ \{(s,t) \mid s\models \varphi \text{ implies } t \models \varphi \text{ for all } \varphi \in \logicstateready\}$ is a ready simulation.
\begin{description}
\item[($\Rightarrow$)]
Let $\varphi \in \logicstateready$.
We aim to prove that
\begin{equation}
\label{eq:thm_ready_sim_adequate_1}
\text{whenever } s \ready t \text{ and } s \models \varphi \text{, then } t \models \varphi.
\end{equation}
We proceed by structural induction over $\varphi$.
\begin{itemize}
\item Base case $\varphi = \top$. 
The proof obligation Equation~\eqref{eq:thm_ready_sim_adequate_1} immediately follows.

\item Base case $\varphi = \bar{a}$.
By Definition~\ref{def:satisfiability}, $s \models \bar{a}$ gives $s \ntrans[a]$.
Since $s \ready t$, this implies that $t \ntrans[a]$ from which we draw $t \models \bar{a}$.
Therefore, the proof obligation Equation~\eqref{eq:thm_ready_sim_adequate_1} follows also in this case.

\item Inductive step $\varphi = \bigwedge_{j \in J} \varphi_j$.
By Definition~\ref{def:satisfiability}, $s \models \bigwedge_{j \in J} \varphi_j$ gives that $s \models \varphi_j$ for each $j \in J$.
Hence, by structural induction we obtain that $t \models \varphi_j$ for each $j \in J$, thus implying $t \models \bigwedge_{j \in J} \varphi_j$.
Therefore, the proof obligation Equation~\eqref{eq:thm_ready_sim_adequate_1} follows also in this case.

\item Inductive step $\varphi = \diam{a}\bigoplus_{i \in I} r_i \varphi_i$.  
By Definition~\ref{def:satisfiability}, $s \models \diam{a}\bigoplus_{i \in I} r_i \varphi_i$ gives that there exists a distribution $\pi_s$ such that $s \trans[a] \pi_s$ and $\pi_s \models \bigoplus_{i \in I} r_i \varphi_i$.
Since $s \ready t$, $s \trans[a] \pi_s$ implies the existence of a distribution $\pi_t$ such that $t \trans[a] \pi_t$ and $\pi_s \ready^{\dagger} \pi_t$.
Hence, to derive the proof obligation Equation~\eqref{eq:thm_ready_sim_adequate_1} we need to prove that 
\begin{equation}
\label{eq:thm_ready_sim_adequate_2}
\pi_t \models \bigoplus_{i \in I} r_i \varphi_i\, .
\end{equation}
From $\pi_s \models \bigoplus_{i \in I} r_i \varphi_i$ we gather that $\pi_s = \sum_{i \in I} r_i \pi_i$ for some distributions $\pi_i$ such that whenever $s' \in \support(\pi_i)$ then $s' \models \varphi_i$ (Definition~\ref{def:satisfiability}).
Moreover, by Definition~\ref{def:relation_lifting} and Proposition~\ref{prop:equiv_rel_lifting}, $\pi_s \ready^{\dagger} \pi_t$ and $\pi_s = \sum_{i \in I} r_i \pi_i$ together imply the existence of distributions $\pi_i'$ such that $\pi_t = \sum_{i \in I} r_i \pi_i'$ and for each $s' \in \support(\pi_i)$ there is a $t' \in \support(\pi_i')$ such that $s' \ready t'$.
Thus, from $s' \ready t'$ and $s' \models \varphi_i$, structural induction over $\varphi_i$ gives $t' \models \varphi_i$.
Hence, for each $t' \in \support(\pi_i')$ it holds that $t' \models \varphi_i$ thus giving Equation~\eqref{eq:thm_ready_sim_adequate_2}.
Therefore, we can conclude that $t \models \diam{a}\bigoplus_{i \in I} r_i \varphi_i$ and the proof obligation Equation~\eqref{eq:thm_ready_sim_adequate_1} follows also in this case.
\end{itemize}

\item[($\Leftarrow$)] Assume now that, for any $\varphi \in \logicstateready$, $s \models \varphi$ implies $t \models \varphi$.
We define the relation 
\[
\rel = \{(s,t) \mid s\models \varphi \text{ implies } t \models \varphi \text{ for all } \varphi \in \logicstateready\}.
\]
We aim to show that $\rel$ is a probabilistic ready simulation.

Let $s \rel t$.
We aim to prove that
\begin{flalign}
& \label{eq:thm_ready_sim_adequate_3}
\text{whenever } s \ntrans[b] \text{ then } t \ntrans[b] \\
& \label{eq:thm_ready_sim_adequate_4}
\text{whenever } s \trans[a] \pi_s \text{ then there is a transition } t \trans[a] \pi_t \text{ with } \pi_s \rel^{\dagger} \pi_t.
\end{flalign}
Assume first that $s \ntrans[b]$.
Then, by Definition~\ref{def:satisfiability}, we derive $s \models \bar{b}$.
From $s \rel t$ we gather $t \models \bar{b}$ thus giving $t \ntrans[b]$ and the proof obligation Equation~\eqref{eq:thm_ready_sim_adequate_3} follows.

Next, consider any transition $s \trans[a] \pi_s$.
To prove the proof obligation Equation~\eqref{eq:thm_ready_sim_adequate_4} we need to show that there exists a probability distribution $\pi_t$ such that $t \trans[a] \pi_t$ and $\pi_s \reldist \pi_t$.
We recall that by definition of lifting of a relation (Definition~\ref{def:relation_lifting}) we have $\pi_s \reldist \pi_t$ if{f} whenever $\pi_s = \sum_{i \in I} p_i \delta_{s_i}$, for some set of indexes $I$, then $\pi_t = \sum_{i \in I} p_i \delta_{t_i}$ for some processes $t_i$ such that $s_i \rel t_i$ for each $i \in I$.
Since it is immediate to see that $\pi_s = \sum_{s' \in \support(\pi_s)} \pi_s(s')\delta_{s'}$, by Proposition~\ref{prop:equiv_rel_lifting} to prove the proof obligation Equation~\eqref{eq:thm_ready_sim_adequate_4} we need to show that there exists a probability distribution $\pi_t$ such that $\pi_t = \sum_{s' \in \support(\pi_s)} \pi_s(s') \pi_{s'}$ for a family of probability distributions $\{\pi_{s'}\}_{s' \in \support(\pi_s)}$ s.t. whenever $t' \in \support(\pi_{s'})$ then $s' \rel t'$.
Thus, let us consider the set
\[
\Pi_{t,a} = \{\pi \mid t \trans[a] \pi \,\wedge\, \pi = \sum_{s' \in \support(\pi_s)} \pi_s(s') \pi_{s'} \,\wedge\, \exists\, s' \in \support(\pi_s),\, t' \in \support(\pi_{s'})\colon s' \not\!\!\!\rel\, t'\}.
\]
Our aim is to prove that there is at least one probability distribution $\pi_t \in \der{t,a}$ which does not belong to the set $\Pi_{t,a}$.

By construction, for each $\pi \in \Pi_{t,a}$ there are some processes $s'_{\pi} \in \support(\pi_s)$ and $t'_{\pi} \in \support(\pi_{s'_{\pi}})$ and a ready state formula $\varphi_{\pi}$ for which $s'_{\pi} \models \varphi_{\pi}$ but $t'_{\pi} \not\models \varphi_{\pi}$.
Thus, for each $s' \in \support(\pi_s)$ we have $s' \models \displaystyle\bigwedge_{\{\pi \in \Pi_{t,a} \mid s'_{\pi} = s'\}} \varphi_{\pi}$.
Moreover, for each $\pi \in \Pi_{t,a}$ with $s'_{\pi} = s'$ there is some $t'_{\pi} \in \support(\pi_{s'})$ such that $t'_{\pi} \not\models \displaystyle\bigwedge_{\{\pi \in \Pi_{t,a} \mid s'_{\pi} = s'\}} \varphi_{\pi}$.

Consider now that ready state formula 
\[
\varphi = \diam{a} \bigoplus_{s' \in \support(\pi_s)} \pi_s(s') \bigwedge_{\{\pi \in \Pi_{t,a} \mid s'_{\pi} = s'\}} \varphi_{\pi}.
\]
Then, it is clear that $s \models \varphi$ thus implying $t \models \varphi$, as by hypothesis $s \rel t$. 
From $t \models \varphi$ it follows that there exists a distribution $\pi_t$ such that $t \trans[a] \pi_t$ and 
\[
\pi_t \models \bigoplus_{s' \in \support(\pi_s)} \pi_s(s') \bigwedge_{\{\pi \in \Pi_{t,a} \mid s'_{\pi} = s'\}} \varphi_{\pi}
\]
 namely $\pi_t = \sum_{s' \in \support(\pi_s)} \pi_s(s') \pi'_{s'}$ for some distributions $\pi'_{s'}$ such that whenever $t' \in \support(\pi'_{s'})$ then $t' \models \displaystyle \bigwedge_{\{\pi \in \Pi_{t,a} \mid s'_{\pi} = s'\}} \varphi_{\pi}$.
Consequently, $\pi_t \not \in \Pi_{t,a}$ and hence for all $s' \in \support(\pi_s)$ each $t' \in \support(\pi'_{s'})$ is such that $s' \rel t'$.
Therefore, from Proposition~\ref{prop:equiv_rel_lifting} we obtain $\delta_{s'} \rel^{\dagger} \pi'_{s'}$ and consequently (from the same Proposition~\ref{prop:equiv_rel_lifting}) $\pi_s \rel^{\dagger} \pi_t$, thus proving the proof obligation Equation~\eqref{eq:thm_ready_sim_adequate_4}.\qedhere
\end{description}
\end{proof}



\subsection{Probabilistic transition system specifications}
\label{sec:ptss}

PTSs are usually defined by means of SOS rules, which are syntax-driven inference rules allowing us to infer the behavior of terms inductively with respect to their structure.
Here we consider rules in the probabilistic GSOS format \cite{DGL14} (see examples in Example~\ref{tab:ex_PGSOS} below), which allow for the specification of the semantics of most of probabilistic process algebras operators \cite{GLT15,GT18}.

In these rules we need syntactic expressions that denote probability distributions. 
We assume a countable set of \emph{distribution variables} $\DVar$. 
We denote by $\Var$ the set of state and distribution variables $\Var = \SVar \cup \DVar$.
We let $\mu,\nu,\ldots$ range over $\DVar$ and $\zeta$ range over $\Var$.

\begin{defi}
[Distribution terms, \cite{DGL14}]
Let $V_s \subseteq \SVar$ and 
$V_d \subseteq \DVar$.
The set of \emph{distribution terms} $\D\T(\Sigma, V_s, V_d)$, over $\Sigma$, $V_s$ and $V_d$
is the least set satisfying: 
\begin{enumerate*}[label=(\emph{\roman*})]
\item \label{def:DT:var}
$V_d \subseteq \D\T(\Sigma, V_s, V_d)$,
\item \label{def:DT:dirac}
$\{\delta_t \mid t \in \T(\Sigma, V_s)\} \subseteq \D\T(\Sigma, V_s, V_d)$, 
\item \label{def:DT:prod} 
$f(\Theta_1, \ldots, \Theta_n) \in \D\T(\Sigma, V_s, V_d)$ whenever $f \in \Sigma$ and $\Theta_i \in \D\T(\Sigma, V_s, V_d)$, and
\item \label{def:DT:sum} 
${\textstyle \sum_{i\in I} p_i \Theta_i \in \D\T(\Sigma, V_s, V_d)}$ whenever $\Theta_i \in \D\T(\Sigma, V_s, V_d)$ and $p_i \in (0,1]$ with $\sum_{i\in I} p_i = 1$.
\end{enumerate*}
We write $\closedDTerms$ for the set of the \emph{closed distribution terms} $\D\T(\Sigma, \emptyset,\emptyset)$, and
$\openDTerms$ for the set of all \emph{open distribution terms} $\D\T(\Sigma, \SVar,\DVar)$.
\end{defi}

Notice that closed distribution terms denote distributions.
Open distribution terms instantiate to distributions through closed substitutions. 
We recall that a \emph{substitution} is a mapping $\sigma \colon \SVar \cup \DVar \to \openTerms$ with $\sigma(x) \in \openT$, if $x\in \SVar$, and $\sigma(\mu) \in \openDTerms$, if $\mu\in \DVar$.
Then, $\sigma$ is \emph{closed} if it maps term variables to closed terms and distribution variables to closed distribution terms.

By $\var{t}$ (resp.\ $\var{\Theta}$) we denote the set of the variables occurring in term $t$ (resp.\ distribution term $\Theta$).

A \emph{positive (resp.\ negative) literal} is an expression of the form $t \trans[a] \Theta$ (resp.\ $t \ntrans[a]$) with $t \in \openT$, $a \in \Act$ and $\Theta \in \openDTerms$.
The literals $t\trans[a] \Theta$ and $t \ntrans[a]$ are said to \emph{deny} each other.

\begin{defi}
[PGSOS rules, \cite{DGL14}]
\label{def:pgsos_rule}
A \emph{PGSOS rule} $r$ has the form:
\[
\SOSrule{\{x_i\trans[a_{i,m}]\mu_{i,m} \mid i\in I, m\in M_i\} \qquad \{x_i\ntrans[a_{i,n}] \mid i\in I, n\in N_i\}}{f(x_1, \ldots, x_{\n})\trans[a] \Theta}
\]
where $f\in \Sigma$, $I = \{1,\ldots,\n\}$, $M_i,N_i$ are finite indexes sets, $a_{i,m},a_{i,n},a\in\Act$ are actions, $x_i\in\SVar$ and $\mu_{i,m}\in\DVar$ are variables and $\Theta \in \openDTerms$ is a distribution term.
Furthermore, it is required that
\begin{enumerate*}[label=(\emph{\roman*})]
\item
\label{item:pgsos:cond_muik_different}
all $\mu_{i,m}$ for $i\in I$ and $m\in M_i$ are distinct, 
\item
all $x_1, \ldots , x_{\n}$ are distinct, and 
\item
\label{item:target_no_free_vars} 
$\var{\Theta}\subseteq\{\mu_{i,m} \mid i\in I, m\in M_i\} \cup \{x_1,\ldots, x_{\n}\}$.
\end{enumerate*}
\noindent

A \emph{PGSOS probabilistic transition system specification (PGSOS-PTSS)} is a tuple $P=(\Sigma,\Act,R)$, with $\Sigma$ a signature, $\Act$ a countable set of actions and $R$ a finite set of PGSOS rules.
\end{defi}

The constraints~\eqref{item:pgsos:cond_muik_different}--\eqref{item:target_no_free_vars} in Definition~\ref{def:pgsos_rule} above, are exactly the constraints of the nondeterministic GSOS format \cite{BIM95} with the difference that we have distribution variables as right hand sides of positive literals.

\begin{exa}
\label{tab:ex_PGSOS}
The operators of synchronous parallel composition $|$ and probabilistic alternative composition $+_p$, with $p \in (0,1]$, are specified by the following PGSOS rules:
\[
\SOSrule{x\trans[a] \mu \quad y \trans[a] \nu}{x \mid y \trans[a] \mu \mid \nu} \qquad 
\SOSrule{x \trans[a] \mu \quad y \ntrans[a]}{x +_p y \trans[a] \mu} \qquad
\SOSrule{x \ntrans[a] \quad y \trans[a] \nu}{x +_p y \trans[a] \nu} \qquad
\SOSrule{x \trans[a] \mu \quad y \trans[a] \nu}{x +_p y \trans[a] p \mu + (1-p) \nu} \; \text{.}
\]
\end{exa}

For a PGSOS rule $r$, the positive (resp.~negative) literals above the line are the \emph{positive premises}, notation $\pprem{r}$ (resp.~\emph{negative premises}, notation $\nprem{r}$).
The literal $f(x_1, \ldots, x_{\n})\trans[a] \Theta$ is called the \emph{conclusion}, notation $\conc{r}$, the term $f(x_1, \ldots, x_{\n})$ is called the \emph{source}, notation $\src{r}$, and the distribution term $\Theta$ is called the \emph{target}, notation $\trg{r}$. 

A PGSOS rule $r$ is said to be \emph{positive} if $\nprem{r} = \emptyset$.
Then we say that a PGSOS-PTSS $P = (\Sigma, \Act, R)$ is \emph{positive} if all the PGSOS rules in $R$ are positive.

A PTS is derived from a PTSS through the notions of substitution and proof.

A substitution $\sigma$
extends to terms, literals and rules by element-wise application.
A closed substitution instance of a literal (resp.~PGSOS rule) is called a \emph{closed literal} (resp.~\emph{closed PGSOS rule)}. 

\begin{defi}
[Proof]
\label{def:proof}
A \emph{proof} from a PTSS $P=(\Sigma,\Act,R)$ of a closed literal $\alpha$ is a well-founded, upwardly branching tree, with nodes labeled by closed literals, such that the root is labeled $\alpha$ and, if $\beta$ is the label of a node $\mathfrak{q}$ and $\K$ is the set of labels of the nodes directly above $\mathfrak{q}$, then:
\begin{itemize}
\item 
either $\beta$ is positive and $\K/\beta$ is a closed substitution instance of a rule in $R$,
\item 
or $\beta$ is negative and for each closed substitution instance of a rule in $R$ whose conclusion denies $\beta$, a literal in $\K$ denies one of its premises.
\end{itemize}
A literal $\alpha$ is \emph{provable} from $P$, notation $P \vdash \alpha$, if there exists a proof from $P$ of $\alpha$. 
\end{defi}

Each PGSOS-PTSS $P$ is \emph{strictly stratifiable} \cite{vG96} which implies that $P$ \emph{induces a unique model} corresponding to the PTS $(\closedTerms, \Act, {\trans})$ whose transition relation $\trans$ contains exactly the closed positive literals provable from $P$.
Moreover, the stratification implies that $P$ is also \emph{complete} \cite{vG96}, thus giving that for any term $t \in \closedTerms$ and action $a \in \Act$, either $P \vdash t \trans[a] \pi$ for some $\pi \in \ProbDist{\closedTerms}$ or $P \vdash t \ntrans[a]$.
Finally, the notion of provability in Definition~\ref{def:proof} (which is called \emph{supported} in \cite{vG96}) subsumes the \emph{negation as failure} principle of \cite{C77} for the derivation of negative literals: for each closed term $t$ we have that $P \vdash t \ntrans[a]$ if and only if $P \not\vdash t \trans[a] \pi$ for any distribution $\pi \in \ProbDist{\closedTerms}$.
Therefore, the PTS induced by $P$ contains literals that do not deny each other.


\section{Distribution specifications}
\label{sec:new_rules}

The idea behind the decomposition of state (resp.\ distribution) formulae is to establish which properties the closed substitution instances of the variables occurring in a term (resp.\ distribution term) must satisfy to guarantee that the closed substitution instance of that term (resp.\ distribution term) satisfies the chosen state (resp.\ distribution) formula.
To support the decomposition it is therefore necessary to relate the behavior of any open term (resp.\ distribution term) with the behavior of its variables.
In the case of a term $t$, such a relation will be given by ruloids, which are inference rules derived from the PGSOS rules having transitions from $t$ as conclusion and positive and negative transitions from the variables of $t$ as premises.
To deal in the same way with a distribution term $\Theta$, in this section we provide a set of inference rules, called $\Sigma$-\emph{distribution rules}, which will be generalized in the next section to distribution ruloids, which
will be inference rules relating the behavior of any open distribution term with the behavior of its variables.

An example of $\Sigma$-distribution rule is given by the inference rule
\[
\SOSrule{\{\mu \trans[q_i] x_i \mid i \in I\} \quad \{\nu \trans[q_j] x_j \mid j \in J \}}
{\{ \mu \mid \nu \trans[q_i \cdot q_j] x_i \mid x_j \;\big|\; i \in I, j \in J\}}
\]
which, intuitively, states that whenever the distribution variable $\mu$ is characterized as the distribution $\{\mu \trans[q_i] x_i \mid i \in I\}$ over the state variables $x_i$, and the distribution variable $\nu$ as the distribution $\{\nu \trans[q_j] x_j \mid j \in J\}$ over the state variables $x_j$, then the behavior of the distribution term $\mu \mid \nu$ can be inferred as the distribution $\{ \mu \mid \nu \trans[q_i \cdot q_j] x_i \mid x_j \;\big|\; i \in I, j \in J\}$ over the state terms $x_i \mid x_j$.



We will also show that under a suitable notion of \emph{provability},  $\Sigma$-distribution rules correctly specifies the semantics of closed distribution terms, meaning that they allow us to infer expressions of kind $\sigma(\Theta) \trans[q] t$ for a closed substitution $\sigma$ if and only if the distribution $\sigma(\Theta)$ assigns weight $q$ to the closed term $t$.

We remark that our approach can be extended to decompose formulae of any logic offering modalities for the specification of the probabilistic properties of processes.
Moreover, it can be easily generalized to cover the case of sub-distributions, which are usually considered alongside a weak semantics for processes \cite{LdV15,LdV16}.


\subsection{\texorpdfstring{$\Sigma$}{Sigma}-distribution rules}

A  \emph{distribution literal} is an expression of the form $\Theta \trans[q] t$, with $\Theta \in \openDTerms$, $q \in (0,1]$ and $t \in \openT$.
Given a set of (distribution) literals $L$ we denote by $\lhs(L)$ the set of the left-hand sides of the (distribution) literals in $L$ and by $\rhs(L)$ the set of right-hand sides of the (distribution) literals in $L$. 

A set of distribution literals $\{\Theta \trans[q_i] t_i \mid i \in I\}$ is a \emph{distribution over terms} if $\sum_{i \in I}q_i = 1$ and all terms $t_i$ are pairwise distinct.
This expresses that the possibly open distribution term $\Theta \in \openDTerms$ is the distribution over possibly open terms in $\openT$ giving weight $q_i$ to $t_i$. 
Given an open distribution term $\Theta \in \openDTerms$ and a distribution over terms $L = \{\Theta \trans[q_i] t_i \mid i \in I\}$ we denote the set of terms in $\rhs(L)$ by $\support(\Theta) = \{t_i \mid i \in I\} \subseteq \openT$.

Our target is to derive distributions over terms $\{\pi \trans[q_i] t_i \mid i \in I\}$ for a distribution $\pi  \in \Delta(\closedTerms)$ (which coincides with a closed distribution term) and closed terms $t_i \in \closedTerms$ such that:
\begin{enumerate*}[label=(\emph{\roman*})]
\item
$\{\pi \trans[q_i] t_i \mid i \in I\}$ if and only if $\pi(t_i) = q_i$ for all $i \in I$, and
\item
$\{\pi \trans[q_i] t_i \mid i \in I\}$ is obtained inductively wrt.\ the structure of $\pi$.
\end{enumerate*}
To this aim, we introduce the \emph{$\Sigma$-distribution rules} and the \emph{$\Sigma$-distribution specification}.

Let $\delta_{\SVar} := \{\delta_x \mid x \in \SVar\}$ denote the set of all instantiable Dirac distributions with a variable as term, and $\vartheta, \vartheta_i , \ldots$ denote distribution terms in $\openDTerms$ ranging over $\DVar \cup \delta_{\SVar}$.
Then, for arbitrary sets $S_1,\ldots,S_n$, we denote by $\bigtimes_{i = 1}^n S_i$ the set of tuples $k = [s_1,\ldots, s_n]$ with $s_i \in S_i$. The $i$-th element of $k$ is denoted by $k(i)$.

\begin{defi}
[\dgsos rules]
\label{def:dgsos_rule}
Assume a signature $\Sigma$. The set $R_{\Sigma}$ of the \emph{\dgsos rules} consists of the least set containing the following inference rules:
\begin{enumerate}
\item \label{def:dgsos_rule_delta}
\[
\SOSrule{}{\{\delta_x \trans[1] x\}}
\]
 for any state variable $x \in \SVar$;
\item \label{def:dgsos_rule_f}
\[
\SOSrule{\bigcup_{i = 1,\ldots,\n}\left\{ \vartheta_i \trans[q_{i,j}] x_{i,j} \mid j \in J_i \right\}}
{\left\{f(\vartheta_1, \ldots, \vartheta_{\n}) \trans[q_k] f(x_{1,k(1)}, \ldots, x_{\n, k(\n)}) \; \Big| \;
k \in \bigtimes_{i=1,\ldots,\n}J_i  \text{ and } q_k = \prod_{i =1,\ldots,\n} q_{i,k(i)}\right\}}
\]
where: 
\begin{enumerate}
\item $f \in \Sigma$, 
\item the distribution terms $\vartheta_1, \ldots, \vartheta_{\n}$ are in $\DVar \cup \delta_{\SVar}$ and are all distinct, 
\item \label{def:dgsos_rule_premises_are_distributions} for each $i=1,\ldots,\n$ the state variables $x_{i,j}$'s with $j \in J_i$ are all distinct,
\item \label{def:dgsos_rule_premises_are_distributions_bis} for each $i=1,\ldots,\n$ we have $\sum_{j \in J_i} q_{i,j} = 1$;
\end{enumerate}
\item \label{def:dgsos_rule_convex}
\[
\SOSrule{\bigcup_{i \in I}\left\{\vartheta_i \trans[q_{i,j}] x_{i,j} \mid j \in J_i \right\}}
{\left\{ \sum_{i \in I} p_i \vartheta_i \trans[q_x] x \; \Big| \;
x \in \{x_{i,j} \mid i \in I \wedge j \in J_i\} \text{ and }
q_x = \sum_{i \in I, j \in J_i \text{ s.t. } x_{i,j} = x} p_i \cdot q_{i,j}  \right\}}
\]
where: 
\begin{enumerate}
\item 
$I$ is an at most countable set of indexes, 
\item
the distribution terms $\vartheta_i$ with $i \in I$ are in $\DVar \cup \delta_{\SVar}$ and are all distinct,
\item \label{def:dgsos_rule_premises_are_distributions_sum}
for each $i \in I$ the state variables $x_{i,j}$'s with $j \in J_i$ are all distinct,
\item \label{def:dgsos_rule_premises_are_distributions_sum_bis}
for each $i\in I$ we have $\sum_{j \in J_i} q_{i,j} = 1$.
\end{enumerate}
\end{enumerate}
Then, the \emph{$\Sigma$-distribution specification} (\emph{$\Sigma$-DS}) is defined as the pair $D_{\Sigma} = (\Sigma, R_{\Sigma})$.
\end{defi}

For each $\Sigma$-distribution rule $r_{\D}$, all sets above the line are called \emph{premises}, notation $\prem{r_{\D}}$, and the set below the line is called \emph{conclusion}, notation $\conc{r_{\D}}$.
Then, we name the distribution term on the left side of all distribution literals in the conclusion of $r_{\D}$ as \emph{source} of $r_{\D}$, notation $\src{r_{\D}}$, and the set of the terms in the right side of all distribution literals in the conclusion as target, notation $\trg{r_{\D}}$. 

All premises in a $\Sigma$-distribution rule are distributions over terms. 
This is immediate for rules as in Definition~\ref{def:dgsos_rule}.\ref{def:dgsos_rule_delta}, follows by constraints \ref{def:dgsos_rule_premises_are_distributions} and \ref{def:dgsos_rule_premises_are_distributions_bis} for rules as in Definition~\ref{def:dgsos_rule}.\ref{def:dgsos_rule_f} and follows by constraints
 \ref{def:dgsos_rule_premises_are_distributions_sum} and  \ref{def:dgsos_rule_premises_are_distributions_sum_bis} for rules as in Definition~\ref{def:dgsos_rule}.\ref{def:dgsos_rule_convex}.
We can show that also the conclusion is a distribution over terms (the detailed proof can be found in Appendix).

\begin{prop}
\label{lem:sum_to_1}
The conclusion of any \dgsos rule is a distribution over terms.
\end{prop}

We conclude this section with an example of $\Sigma$-distribution rule for the distribution term $\mu \mid \nu$, where $\mid$ is the synchronous parallel composition operator introduced in Example~\ref{tab:ex_PGSOS}.

\begin{exa}
\label{ex:distribution_rule}
An example of a $\Sigma$-distribution rule with source $\mu \mid \nu$ is the following:
\[
\SOSrule{\{\mu \trans[1/4] x_1,  \quad \mu \trans[3/4] x_2\} \qquad \{\nu \trans[1/3] y_1, \quad \nu \trans[2/3] y_2\}}
{\{\mu \mid \nu \trans[1/12] x_1 \mid y_1, \quad \mu \mid \nu \trans[1/6] x_1 | y_2, \quad \mu \mid\nu \trans[1/4] x_2 \mid y_1, \quad \mu \mid \nu \trans[1/2] x_2 \mid y_2\}} \; \text{.}
\]
\end{exa}


\subsection{Reductions}

The following notion of reduction wrt.\ a substitution allows us to extend the notion of substitution to distributions over terms and, then, to $\Sigma$-distribution rules.

\begin{defi}
[Reduction wrt.\ a substitution]
\label{def:sigma_reduction}
Assume a substitution $\sigma$ and a set of distribution literals $L = \{ \Theta \trans[q_i] t_i \mid i \in I\}$.
We say that $\sigma$ \emph{reduces} $L$ to the set of distribution literals $L' = \{\sigma(\Theta) \trans[q_j] t_j \mid j \in J\}$, or that $L'$ is the \emph{reduction wrt.\ $\sigma$} of $L$, denoted by $\sigma(L) = L'$, if:
\begin{itemize}
\item $\{t_j \mid j \in J\} = \{\sigma(t_i) \mid i \in I\}$;
\item the terms $\{t_j \mid j \in J\}$ are pairwise distinct;
\item for each index $j \in J$, we have $q_j = \sum_{\{i \in I \mid \sigma(t_i) = t_j\}} q_i$.
\end{itemize}
\end{defi}

A reduction wrt. $\sigma$ of a distribution over terms is, in turn, a distribution over terms (the detailed proof can be found in Appendix).

\begin{prop}
\label{prop:sigma_reduction}
For a substitution $\sigma$ and a distribution over terms $L$, the set of distribution literals $\sigma(L)$ is a distribution over terms.
\end{prop}

\begin{defi}
[Reduced instance of a \dgsos rule]
\label{def:reduced_instance}
The \emph{reduced instance} of a $\Sigma$-distribution rule $r_{\D}$ wrt.\ a substitution $\sigma$ is 
the inference rule $\sigma(r_{\D})$ defined as follows:
\begin{enumerate}
\item \label{def:reduced_instance_delta}
If $r_{\D}$ is as in Definition~\ref{def:dgsos_rule}.\ref{def:dgsos_rule_delta} then $\sigma(r_{\D})$ is the $\Sigma$-distribution rule
\[
 \SOSrule{}{\{\delta_{\sigma(x)} \trans[1] \sigma(x)\}}.\\[1.0 ex]
\]
\item \label{def:reduced_instance_f}
If $r_{\D}$ is as in Definition~\ref{def:dgsos_rule}.\ref{def:dgsos_rule_f} then $\sigma(r_{\D})$ is the $\Sigma$-distribution rule
\[
\SOSrule{\bigcup_{i = 1,\ldots,\n}\{ \sigma(\vartheta_i) \trans[q_{i,h}] t_{i,h} \mid h \in H_i \}}
{\left\{f(\sigma(\vartheta_1), \ldots, \sigma(\vartheta_{\n})) \trans[q_{\kappa}] f(t_{1,\kappa(1)}, \ldots, t_{\n, \kappa(\n)}) \; \Big| \; 
\kappa \in \bigtimes_{i=1,\ldots\n} H_i  \text{ and } q_{\kappa} = \prod_{i =1,\ldots,\n} q_{i,\kappa(i)}   \right\}}
\]
where 
$\{ \sigma(\vartheta_i) \trans[q_{i,h}] t_{i,h} \mid h \in H_i\} = \sigma(\{\vartheta_i \trans[q_{i,j}] x_{i,j} \mid j \in J_i\})$.
\item \label{def:reduced_instance_Sum}
If $r_{\D}$ is as in Definition~\ref{def:dgsos_rule}.\ref{def:dgsos_rule_convex} then $\sigma(r_{\D})$ is the $\Sigma$-distribution rule
\[
\SOSrule{\bigcup_{i \in I}\{ \sigma(\vartheta_i) \trans[q_{i,h}] t_{i,h} \mid h \in H_i\}}
{\left\{ \sum_{i \in I} p_i \sigma(\vartheta_i) \trans[q_t] t \; \Big| \;
t \in \{t_{i,h} \mid  i \in I \wedge h \in H_i\}
\text{ and } 
q_t = \sum_{i \in I \wedge h \in H_i \text{ s.t. } t_{i,h} = t} p_i \cdot q_{i,h} \right\}}
\]
where $\{\sigma(\vartheta_i) \trans[q_{i,h}] t_{i,h} \mid h \in H_i\} = \sigma(\{\vartheta_i \trans[q_{i,j}] x_{i,j} \mid j \in J_i\})$.
\end{enumerate}
\end{defi}

\begin{exa}
\label{ex:reduction}
Consider the $\Sigma$-distribution rule $r_{\D}$ for the distribution term $\mu \mid \nu$ given in Example~\ref{ex:distribution_rule} and consider the substitution $\sigma$ with
\[
\sigma(x_1) = x \qquad \qquad \sigma(x_2) = x \qquad \qquad \sigma(y_1) = y \qquad \qquad \sigma(y_2) = \mathrm{nil} 
\]
where $\mathrm{nil}$ denotes the process that cannot perform any action.
Then we have that the reduced instance of $r_{\D}$ wrt.\ $\sigma$ is given by
\[
\sigma(r_{\D}) = \SOSrule{\{\sigma(\mu) \trans[1] x\} \qquad \{\sigma(\nu) \trans[1/3] y, \quad \sigma(\nu) \trans[2/3] \mathrm{nil}\}}
{\{\sigma(\mu) \mid \sigma(\nu) \trans[1/3] x \mid y, \qquad \sigma(\mu) \mid \sigma(\nu) \trans[2/3] x \mid \mathrm{nil}\}} \; \text{.}
\]
\end{exa}

Notice that Proposition~\ref{prop:sigma_reduction} ensures that the premises of $\sigma(r_{\D})$ are distributions over terms. 
Proposition~\ref{lem:sum_to_1} and Proposition~\ref{prop:sigma_reduction} ensure that also the conclusion of  $\sigma(r_{\D})$ is a distribution over terms.


\begin{prop}
\label{lem:sum_to_1_bis}
Let $D_{\Sigma}$ be the $\Sigma$-DS.
The conclusion of a reduced instance of a \dgsos rule in $D_{\Sigma}$ is a distribution over terms.
\end{prop}

\subsection{Semantics of distribution terms}

We conclude this section by showing that the $\Sigma$-distribution specification correctly defines the semantics of closed distribution terms as probability distributions over closed terms as in Section~\ref{sec:ptss}.

\begin{defi}
[Proof from the $\Sigma$-DS]
\label{def:proof_DS}
A \emph{proof} from the $\Sigma$-DS $D_{\Sigma}$ of a closed distribution over terms $L$ is a well-founded, upwardly branching tree, whose nodes are labeled by closed distributions over terms, such that the root is labeled $L$, and, if $\beta$ is the label of a node $\mathfrak{q}$ and $\K$ is the set of labels of the nodes directly above $\mathfrak{q}$, then $\K/\beta$ is a closed reduced instance of a $\Sigma$-distribution rule in $R_{\Sigma}$.

A closed distribution over terms $L$ is \emph{provable} from $D_{\Sigma}$, notation $D_{\Sigma} \vdash L$, if there exists a proof from $D_{\Sigma}$ for $L$.
\end{defi}

\begin{exa}
\label{ex:proof}
Consider any signature $\Sigma$ containing the operator of synchronous parallel composition $\mid$ and let $D_{\Sigma}$ be the $\Sigma$-DS built on it.
We want to show that 
the following distribution over terms is provable from the $\Sigma$-DS:
\begin{align*}
L = \bigg\{ \frac{2}{5} \Big( \frac{1}{4} \delta_{t_1} + \frac{3}{4} \delta_{t_2} \Big) + \frac{3}{5} \Big( (\frac{1}{3} \delta_{t_3} + \frac{2}{3} \delta_{t_4}) \mid \delta_{t_5} \Big) 
& \trans[\frac{1}{10}] t_1, \\
\frac{2}{5} \Big( \frac{1}{4} \delta_{t_1} + \frac{3}{4} \delta_{t_2} \Big) + \frac{3}{5} \Big( (\frac{1}{3} \delta_{t_3} + \frac{2}{3} \delta_{t_4}) \mid \delta_{t_5} \Big)
& \trans[\frac{3}{10}] t_2, \\
\frac{2}{5} \Big( \frac{1}{4} \delta_{t_1} + \frac{3}{4} \delta_{t_2} \Big) + \frac{3}{5} \Big( (\frac{1}{3} \delta_{t_3} + \frac{2}{3} \delta_{t_4}) \mid \delta_{t_5} \Big)
& \trans[\frac{1}{5}] t_3 \mid t_5, \\
\frac{2}{5} \Big( \frac{1}{4} \delta_{t_1} + \frac{3}{4} \delta_{t_2} \Big) + \frac{3}{5} \Big( (\frac{1}{3} \delta_{t_3} + \frac{2}{3} \delta_{t_4}) \mid \delta_{t_5} \Big)
& \trans[\frac{2}{5}] t_4 \mid t_5 \bigg\}
\end{align*}
To this aim, we first consider
the following upwardly tree-structure, whose nodes are $\Sigma$-distribution rules.
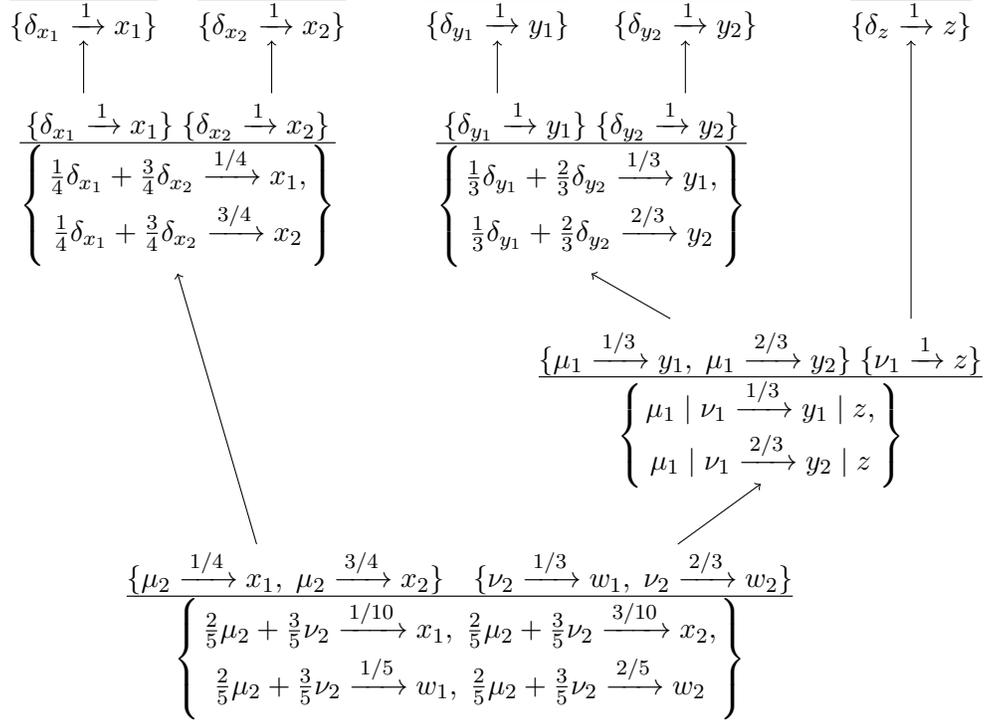
\begin{center}
\begin{tikzpicture}
\node at (0,8.2) {${\SOSrule{}{\{ \delta_{x_1} \trans[1] x_1\}}}$};
\node at (2.5,8.2) {${\SOSrule{}{\{ \delta_{x_2} \trans[1] x_2\}}}$};
\node at (5.5,8.2) {${\SOSrule{}{\{ \delta_{y_1} \trans[1] y_1\}}}$};
\node at (8,8.2) {${\SOSrule{}{\{ \delta_{y_2} \trans[1] y_2\}}}$};
\node at (11,8.2) {${\SOSrule{}{\{ \delta_{z} \trans[1] z\}}}$};
\draw[<-](0,7.9)--(0,7.2);
\draw[<-](2.5,7.9)--(2.5,7.2);
\draw[<-](5.5,7.9)--(5.5,7.2);
\draw[<-](8,7.9)--(8,7.2);
\draw[<-](11,7.9)--(11,4.2);
\node at (1.25,6){${\SOSrule{\{\delta_{x_1} \trans[1] x_1\} \; \{\delta_{x_2} \trans[1] x_2\}}{\left\{{\frac{1}{4}\delta_{x_1} + \frac{3}{4} \delta_{x_2} \trans[1/4] x_1,}\atop{\frac{1}{4}\delta_{x_1} + \frac{3}{4} \delta_{x_2} \trans[3/4] x_2}\right\}}}$};
\node at (6.75,6){${\SOSrule{\{\delta_{y_1} \trans[1] y_1\} \; \{\delta_{y_2} \trans[1] y_2\}}{\left\{{\frac{1}{3}\delta_{y_1} + \frac{2}{3} \delta_{y_2} \trans[1/3] y_1,}\atop{\frac{1}{3}\delta_{y_1} + \frac{2}{3} \delta_{y_2} \trans[2/3] y_2}\right\}}}$};
\draw[<-](1.25,4.8)--(2.3,1.2);
\draw[<-](6.75,4.8)--(7.8,4.2);
\node at (9,3){${\SOSrule{\{\mu_1 \trans[1/3] y_1,\; \mu_1 \trans[2/3] y_2\} \; \{\nu_1 \trans[1] z\}}{\left\{{\mu_1 \mid \nu_1 \trans[1/3] y_1 \mid z,}\atop{\mu_1 \mid \nu_1 \trans[2/3] y_2 \mid z}\right\}}}$};
\draw[<-](9,2)--(7.9,1.2);
\node at (5,0){${\SOSrule{\{\mu_2 \trans[1/4] x_1,\; \mu_2 \trans[3/4] x_2\} \quad \{\nu_2 \trans[1/3] w_1,\; \nu_2 \trans[2/3] w_2\}}{\left\{\frac{2}{5}\mu_2  + \frac{3}{5} \nu_2 \trans[1/10] x_1,\; \frac{2}{5}\mu_2  + \frac{3}{5} \nu_2 \trans[3/10] x_2,\atop{\frac{2}{5}\mu_2  + \frac{3}{5} \nu_2 \trans[1/5] w_1,\; \frac{2}{5}\mu_2  + \frac{3}{5} \nu_2 \trans[2/5] w_2}\right\}}}$};
\end{tikzpicture}
\end{center}
Then, the proof for $L$ confirming that $D_{\Sigma} \vdash L$ is the proof tree that is obtained by firstly replacing each of these $\Sigma$-distribution rules with its conclusion and, then, by applying to the obtained set of distribution literals the following closed substitution $\sigma$:
\[
\sigma(x_1) = t_1, \;
\sigma(x_2) = t_2, \;
 \sigma(y_1) = t_3,  \;
\sigma(y_2) = t_4, \;
\sigma(z) = t_5, \;
\sigma(w_1) = t_3 \mid t_5, \;
 \sigma(w_2) = t_4 \mid t_5
\]
\[
\sigma(\mu_1) = \frac{1}{3} \delta_{t_3} + \frac{2}{3} \delta_{t_4}, \; \;
 \sigma(\nu_1) = \delta_{t_5} , \;\;
\sigma(\mu_2) = \frac{1}{4} \delta_{t_1} + \frac{3}{4} \delta_{t_2}, \;\;
\sigma(\nu_2) = \left(\frac{1}{3} \delta_{t_3} + \frac{2}{3} \delta_{t_4}\right) \mid \delta_{t_5}.
\]
Notice that we decided to use as nodes the $\Sigma$-distribution rules instead of using the  $\sigma$-closed substitution instances of their conclusions to improve readability.
\end{exa}

Since \dgsos rules have only positive premises, the set of the distribution over terms provable from the $\Sigma$-DS is unique. 
The following result confirms that all probability distributions over $\closedTerms$ can be inferred through the $\Sigma$-DS.
This result is necessary for the decomposition of distribution formulae.

\begin{thm}
\label{prop:proof_sigma_q}
Assume a signature $\Sigma$.
Let $\pi \in \closedDTerms$ be a closed distribution term and 
$\{t_m\}_{m \in M} \subseteq \closedTerms$ a set of pairwise distinct closed terms.
Then 
\[
D_{\Sigma} \vdash \{\pi \trans[q_m] t_m \mid m \in M \} \Leftrightarrow \text{for all } m \in M \text{ it holds }\pi(t_m) = q_m \text{ and } \sum_{m \in M} q_m = 1.
\] 
\end{thm}


\begin{proof}
The first implication is proved by induction over the length of the closed proof over $D_{\Sigma}$ giving $\{\pi \trans[q_m] t_m \mid m \in M \}$. 
The second implication is by structural induction over $\pi$.

\begin{description}
  \item[($\Rightarrow$)] We aim to prove that 
\begin{equation}
\label{eq:proof_sigma_q_1}
D_{\Sigma} \vdash \{\pi \trans[q_m] t_m \mid m \in M \} \text{ implies }  \pi(t_m) = q_m
\text{ for all } m \in M \text{ and } \sum_{m \in M} q_m = 1.
\end{equation}
We proceed by induction over the length of a closed proof $\gamma$ of $\{\pi \trans[q_m] t_m \mid m \in M \}$ from $D_{\Sigma}$. 
\begin{itemize}
\item Base case $|\gamma| = 1$.
Since the only distributions over terms derivable in one step are the closed reduced substitution instances of distribution axioms, we have one of the following two cases:
\begin{enumerate}
\item $\pi = \delta_{t}$ for some $t \in \closedTerms$.
The only \dgsos rule defining the instantiable Dirac function $\delta_t$ is the distribution axiom $r_{\D} = \SOSrule{}{\{\delta_x \trans[1] x\}}$ (Definition~\ref{def:dgsos_rule}.\ref{def:dgsos_rule_delta}), which should be reduced by a closed substitution $\sigma$ such that $\sigma(x) = t$, thus giving $\sigma(r_{\D}) = \SOSrule{}{\{\delta_t \trans[1] t\}}$ by Definition~\ref{def:reduced_instance}.\ref{def:reduced_instance_delta}.
Consequently the hypothesis $D_{\Sigma} \vdash \{\pi \trans[q_m] t_m \mid m \in M \}$ instantiates to
$D_{\Sigma} \vdash \{\delta_{t} \trans[1] t\}$ for which the proof obligation Equation~\eqref{eq:proof_sigma_q_1} is straightforward.
\item $\pi = c$ for some constant operator $c \in \Sigma$.
From Definition~\ref{def:dgsos_rule}.\ref{def:dgsos_rule_f} and considering that by convention $\prod_{\emptyset} = 1$, it is not hard to see that the only \dgsos rule defining the behavior of constant operator $c$ is the distribution axiom $r_{\D} = \SOSrule{}{\{c \trans[1] c\}}$, which independently on the substitution $\sigma$ is reduced to $\sigma(r_{\D}) = \SOSrule{}{\{c \trans[1] c\}}$ by Definition~\ref{def:reduced_instance}.\ref{def:reduced_instance_f}.
Therefore, we can conclude that the hypothesis $D_{\Sigma} \vdash \{\pi \trans[q_m] t_m \mid m \in M \}$ instantiates to
$D_{\Sigma} \vdash \{c \trans[1] c\}$ for which the proof obligation Equation~\eqref{eq:proof_sigma_q_1} is straightforward.
\end{enumerate}
\item Inductive step $|\gamma| > 1$.
We can distinguish two cases, based on the structure of the closed distribution term $\pi$. 
\begin{enumerate}
\item $\pi = f(\pi_1,\ldots, \pi_{\n})$, for some $f \in \Sigma$ and $\pi_i \in \closedDTerms$ for $i = 1,\ldots, \n$.
Then, the bottom of the closed proof $\gamma$ is constituted by the closed reduced instance of a \dgsos rule $r_{\D} \in R_{\Sigma}$ of the form
\[
\SOSrule{\bigcup_{i = 1}^{\n}\{ \vartheta_i \trans[q_{i,j}] x_{i,j} \mid j \in J_i \}}
{\left\{f(\vartheta_1, \ldots, \vartheta_{\n}) \trans[q_k] f(x_{1,k(1)}, \ldots, x_{\n, k(\n)}) \; \Big| \;
 k \in \bigtimes_{i=1}^{\n}J_i  \text{ and }  q_k = \prod_{i =1}^{\n} q_{i,k(i)}  \right\}}
\]
(see Definition~\ref{def:dgsos_rule}.\ref{def:dgsos_rule_f}) with respect to a closed substitution $\sigma$ with $\sigma(\vartheta_i) = \pi_i$. 
By Definition~\ref{def:reduced_instance}.\ref{def:reduced_instance_f} we get that $\sigma(r_{\D})$ has the form 
\[
\SOSrule{\bigcup_{i = 1}^{\n}\{ \pi_i \trans[q_{i,h}] t_{i,h} \mid h \in H_i \}}
{\left\{f(\pi_1, \ldots, \pi_{\n}) \trans[q_{\kappa}] f(t_{1,\kappa(1)}, \ldots, t_{\n, \kappa(\n)}) \; \Big| \;
 \kappa \in \bigtimes_{i=1}^{\n} H_i  \text{ and } q_{\kappa} = \prod_{i =1}^{\n} q_{i,\kappa(i)}  \right\}}
\]
where
\begin{itemize}
\item 
$t_{i,h}$ is a closed term in $\in \closedTerms$ for all $i \in I$ and $h \in H_i$, since $\sigma$ is closed;
\item 
for each $i = 1,\ldots,\n$, the closed terms $t_{i,h}$ are pairwise distinct for $h \in H_i$, since
$\{ \pi_i \trans[q_{i,h}] t_{i,h} \mid h \in H_i\}$ is obtained as $\sigma(\{\vartheta_i \trans[q_{i,j}] x_{i,j} \mid j \in J_i\})$ and we apply Proposition~\ref{lem:sum_to_1_bis}.
\item 
there is a bijection $\f \colon \bigtimes_{i=1}^{\n} H_i \to M$ with $f(t_{1,\kappa(1)}, \ldots,t_{\n,\kappa(\n)}) = t_{\f(\kappa)}$  and $q_{\kappa} = q_{\f(\kappa)}$ for each $\kappa \in \bigtimes_{i=1}^{\n} H_i$.
\end{itemize}
For each $i = 1,\ldots,\n$ there is a proof shorter than $\gamma$ for $\{ \pi_i \trans[q_{i,h}] t_{i,h} \mid h \in H_i \}$ from $D_{\Sigma}$.
By the inductive hypothesis, this implies that 
\[
q_{i,h} = \pi_i(t_{i,h}) \text{ for all } h \in H_i \text{ and } \sum_{h \in H_i} q_{i,h} = 1.
\]
In particular, we have that for each $\kappa \in \bigtimes_{i=1}^{\n} H_i$
\begin{equation}
\label{eq:distribution_theta_f}
q_{i,\kappa(i)} = \pi_i(t_{i,\kappa(i)}) 
\end{equation}
from which we draw
\begin{align*}
& q_{\f(\kappa)} \\
={} & q_{\kappa} & \text{(by definition of $\f$)}\\
={} & \prod_{i = 1}^{\n} q_{i,\kappa(i)} & \text{(by definition of $q_{\kappa}$)}\\
={} & \prod_{i=1}^{\n} \pi_i(t_{i,\kappa(i)}) & \text{(by Equation~\eqref{eq:distribution_theta_f})}\\
={} & \pi(f(t_{1,\kappa(1)}, \ldots, t_{\n,\kappa(\n)})) & \text{(} \pi = f(\pi_1,\ldots,\pi_{\n})\text{)} \\
={} & \pi(t_{\f(\kappa)}) & \text{(by definition of $\f$)}.
\end{align*}
Summarizing, we have obtained that $q_m = \pi(t_m)$ for each $m \in M$.
Moreover, we have that 
\begin{align*}
\sum_{m \in M} q_m ={} & \sum_{m \in M} q_{\f^{-1}(m)}\\
={} & \sum_{\kappa \in \bigtimes_{i=1}^{\n} H_i} q_{\kappa}\\
={} & 1 & \text{(by Proposition~\ref{lem:sum_to_1_bis})}
\end{align*}
thus giving Equation~\eqref{eq:proof_sigma_q_1}.

\item $\pi = \sum_{i \in I} p_i \pi_i$ for some $\pi_i \in \closedDTerms$, $p_i \in (0,1]$ for each $i \in I$ and $\sum_{i \in I} p_i = 1$.
Then, the bottom of the closed proof $\gamma$ is constituted by the closed reduced instance of a \dgsos rule $r_{\D} \in R_{\Sigma}$ of the form
\[
\SOSrule{\bigcup_{i \in I}\{ \vartheta_i \trans[q_{i,j}] x_{i,j} \mid j \in J_i \}}
{\left\{ \sum_{i \in I} p_i \vartheta_i \trans[q_x] x \; \Big| \;
x \in \{x_{i,j} \mid i \in I \wedge j \in J_i\} \text{ and }
q_x = \sum_{i \in I, j \in J_i \text{ s.t. } x_{i,j} = x} p_i q_{i,j}  \right\}}
\]
(see Definition~\ref{def:dgsos_rule}.\ref{def:dgsos_rule_convex}) with respect to a closed substitution $\sigma$ with $\sigma(\vartheta_i) = \pi_i$. 
By Definition~\ref{def:reduced_instance}.\ref{def:reduced_instance_Sum} we get that $\sigma(r_{\D})$ is of the form
\[
\SOSrule{\bigcup_{i \in I}\{ \pi_i \trans[q_{i,h}] t_{i,h} \mid h \in H_i \}}
{\left\{ \sum_{i \in I} p_i \pi_i \trans[q_u] u \; \Big| \;
u \in \{t_{i,h} \mid i \in I \wedge h \in H_i\} \text{ and } q_u = \sum_{i \in I, h \in H_i \text{ s.t. } t_{i,h} = u} p_i q_{i,h}\right\}}
\]
where
\begin{itemize}
\item 
$t_{i,h}$ is a closed term in $\closedTerms$ for all $h \in H_i$, since $\sigma$ is closed;
\item 
for each $i \in I$ the closed terms $t_{i,h}$ are pairwise distinct for $h \in H_i$, since
$\{ \pi_i \trans[q_{i,h}] t_{i,h} \mid h \in H_i\}$ is obtained as $\sigma(\{\vartheta_i \trans[q_{i,j}] x_{i,j} \mid j \in J_i\})$ and we apply Proposition~\ref{lem:sum_to_1_bis};
\item 
there is a bijection $\f \colon \{t_{i,h} \mid i \in I \wedge h \in H_i\} \to M$ with $u = t_{\f(u)}$  and $q_{u} = q_{\f(u)}$ for each $u \in \{t_{i,h} \mid i \in I \wedge h \in H_i\}$.
\end{itemize}
For each $i \in I$ there is a proof shorter than $\gamma$ for $\{\pi_i \trans[q_{i,h}] t_{i,h} \mid h \in H_i\}$ from $D_{\Sigma}$.
By the inductive hypothesis, this implies that 
\begin{equation}
\label{eq:distribution_theta_sum}
q_{i,h} = \pi_i(t_{i,h}) \text{ for all } h \in H_i \text{ and } \sum_{h \in H_i} q_{i,h} = 1.
\end{equation}
Then, we have
\begin{align*}
q_{\f(u)} ={} & q_{u}  & \text{(by definition of $\f$)}\\
={} & \sum_{i \in I, h \in H_i, \text{ s.t. } t_{i,h} = u} p_i\, q_{i,h}\\
={} & \sum_{i \in I, h \in H_i, \text{ s.t. } t_{i,h} = u} p_i\, \pi_i(t_{i,h}) & \text{(by Equation~\eqref{eq:distribution_theta_sum})}\\
={} & \sum_{i \in I, h \in H_i, \text{ s.t. } t_{i,h} = u} p_i\, \pi_i(u)\\
={} & \sum_{i \in I} p_i\, \pi_i(u) \\
={} & \pi(u)\\
={} & \pi(t_{\f(u)}) & \text{(by definition of $\f$)}.
\end{align*}
Thus, we have obtained that $q_m = \pi(t_m)$ for each $m \in M$.
Moreover, we have
\begin{align*}
\sum_{m \in M} q_m ={} & \sum_{m \in M} q_{\f^{-1}(m)}\\
={} & \sum_{u \in \{t_{i,h} \mid h \in H_i, i \in I\}} q_{u}\\
={} & 1 & \text{(by Proposition~\ref{lem:sum_to_1_bis})}
\end{align*}
thus giving Equation~\eqref{eq:proof_sigma_q_1}.
\end{enumerate}
\end{itemize}

\item[($\Leftarrow$)]
We aim to prove that
\begin{equation}
\label{eq:proof_sigma_q_2}
\pi(t_m) = q_m  \text{ for all } m \in M \text{ and } \sum_{m \in M} q_m = 1 \text{ imply } D_{\Sigma} \vdash \{\pi \trans[q_m] t_m \mid m \in M \}.
\end{equation}
We proceed by structural induction over $\pi \in \closedDTerms$. 
\begin{itemize}
\item 
Base case $\pi = \delta_{t}$ for some $t \in \closedTerms$.
Consider the \dgsos rule $r_{\D}$
$\SOSrule{}{\{\delta_x \trans[1] x\}}$ (Definition~\ref{def:dgsos_rule}.\ref{def:dgsos_rule_delta}) and a closed substitution $\sigma$ such that $\sigma(x) = t$.
By Definition~\ref{def:reduced_instance}.\ref{def:reduced_instance_delta} we get that $\sigma(r_{\D})$ is of the form
$\SOSrule{}{\{\delta_t \trans[1] t\}}$, from which we can directly conclude that $D_{\Sigma} \vdash \{\delta_{t} \trans[1] t\}$, thus giving Equation~\eqref{eq:proof_sigma_q_2}.

\item 
Inductive step $\pi = f(\pi_1, \ldots, \pi_{\n})$ for some $\pi_i \in \closedDTerms$ for each $i = 1,\ldots, \n$ and $f \in \Sigma$.
For each $i = 1,\ldots,\n$ there is a set of indexes $M_i$ such that:
\begin{enumerate}
\item \label{item:sum_q_1_f}  $\pi_i(t_{i,m}) = q_{i,m}$ for all $m \in M_i$,
\item $\sum_{m \in M_i} q_{i,m}= 1$ and 
\item \label{item:distinct_f} the terms $t_{i,m} \in \closedTerms$ are pairwise distinct for each $m \in M_i$.
\end{enumerate}
Let $M = \bigtimes_{i=1}^{\n} M_i$.
We have $\support(\pi) = \{f(t_{1,\kappa(1)}, \ldots, t_{\n,\kappa(\n)}) \mid \kappa \in M\}$ and 
\[
q_{\kappa} := \pi \big( f(t_{1,\kappa(1)}, \ldots, t_{\n,\kappa(\n)}) \big) = \prod_{i = 1}^{\n} \pi_i(t_{i,\kappa(i)}) = \prod_{i = 1}^{\n} q_{i,\kappa(i)}
\]
for each $\kappa \in M$.
Hence, to prove Equation~\eqref{eq:proof_sigma_q_2} we need to exhibit a proof of
\begin{displaymath}
\{f(\pi_1,\ldots,\pi_{\n}) \trans[q_{\kappa}] f(t_{1,\kappa(1)}, \ldots, t_{\n,\kappa(\n)}) \mid \kappa \in M\}
\end{displaymath}
from $D_{\Sigma}$.

By the inductive hypothesis, for each $i = 1,\ldots, \n$ 
items~\eqref{item:sum_q_1_f}--\eqref{item:distinct_f} above give 
\begin{equation}
\label{eq:premises_provable_f}
D_{\Sigma} \vdash \{\pi_i \trans[q_{i,m}] t_{i,m} \mid m \in M_i\}.
\end{equation}

Consider the \dgsos rule $r_{\D}$
\[
\SOSrule{\bigcup_{i = 1}^{\n}\{ \vartheta_i \trans[q_{i,m}] x_{i,m} \mid m \in M_i \}}
{\left\{f(\vartheta_1, \ldots, \vartheta_{\n}) \trans[q_{\kappa}] f(x_{1,\kappa(1)}, \ldots, x_{\n, \kappa(\n)}) \; \Big| \;
\kappa \in M \text{ and } q_{\kappa} = \prod_{i =1}^{\n} q_{i,\kappa(i)}\right\}}
\]
as in Definition~\ref{def:dgsos_rule}.\ref{def:dgsos_rule_f} and a closed substitution $\sigma$ with $\sigma(\vartheta_i) = \pi_i$ and $\sigma(x_{i,m}) = t_{i,m}$ for each $m \!\in\! M_i$, $i \!\in\!\{1,..., \n\}$, s.t.\ the closed reduced instance $\sigma(r_{\D})$ is of the form:
\[
\SOSrule{\bigcup_{i = 1}^{\n}\{ \pi_i \trans[q_{i,m}] t_{i,m} \mid m \in M_i \}}
{\left\{f( \pi_1, \ldots, \pi_{\n}) \trans[q_{\kappa}] f(t_{1,\kappa(1)}, \ldots, t_{\n, \kappa(\n)}) \; \Big| \;
\kappa \in M \text{ and } q_{\kappa} = \prod_{i =1}^{\n} q_{i,\kappa(i)}  \right\}}.
\]
We observe that $\trg{\sigma(r_{\D})} = \support(\pi)$ and since the premises of $\sigma(r_{\D})$ are provable from $D_{\Sigma}$ (Equation~\eqref{eq:premises_provable_f}) we can conclude that 
\[
D_{\Sigma} \vdash \{ f(\pi_1,\ldots,\pi_{\n}) \trans[q_{\kappa}] f(t_{1,\kappa(1)}, \ldots, t_{\n,\kappa(\n)}) \mid \kappa \in M \}
\]
thus proving Equation~\eqref{eq:proof_sigma_q_2}.

\item Inductive step $\pi = \sum_{i \in I} p_i \pi_i$ for some $\pi_i \in \closedDTerms$, $p_i \in (0,1]$ and $\sum_{i \in I} p_i = 1$.
For each $i \in I$ there is a set of indexes $M_i$ such that for each $m \in M_i$ such that 
\begin{enumerate}
\item \label{item:sum_q_1_convex} $\pi_i(t_{i,m}) = q_{i,m}$,
\item $\sum_{m \in M_i} q_{i,m}= 1$ and 
\item \label{item:distinct_convex} the terms $t_{i,m} \in \closedTerms$ are pairwise distinct for each $m \in M_i$.
\end{enumerate}
Let $T = \{t_{i,m} \mid i \in I\ \text{ and }m \in M_i\}$.
We have $\support(\pi) = T$ and, for each $u \in T$,
\[
q_{u} := \pi(u) = \sum_{i \in I} p_i \pi_i(u) = \sum_{i \in I, m \in M_i,  t_{i,m} = u} p_i q_{i,m}.
\]
To prove Equation~\eqref{eq:proof_sigma_q_2} we need to exhibit a proof of $\{\pi \trans[q_{u}] u \mid u \in T\}$ from $D_{\Sigma}$.

By the inductive hypothesis, for all $i \in I$ by items~\eqref{item:sum_q_1_convex}--\eqref{item:distinct_convex} above we get 
\begin{equation}
\label{eq:premises_provable_convex}
D_{\Sigma} \vdash \{\pi \trans[q_{i,m}] t_{i,m} \mid m \in M_i\}.
\end{equation}

Consider the \dgsos rule $r_{\D}$
\[
\SOSrule{\bigcup_{i \in I}\{ \vartheta_i \trans[q_{i,m}] x_{i,m} \mid m \in M_i \}}
{\left\{ \sum_{i \in I} p_i \vartheta_i \trans[q_{x}] x \; \Big| \;
x \in \{x_{i,m} \mid i \in I \wedge m \in M_i\}  \text{ and } 
q_{x} = \sum_{i \in I, m \in M_i \text{ s.t. } x_{i,m} = x} p_i  q_{i,m}\right\}}
\]
as in Definition~\ref{def:dgsos_rule}.\ref{def:dgsos_rule_convex} and a closed substitution $\sigma$  with $\sigma(\vartheta_i) = \pi_i$ and $\sigma(x_{i,m}) = t_{i,m}$ for each $i \in I$ and $m \in M_i$ s.t.\ the closed reduced instance $\sigma(r_{\D})$ is of the form:
\[
\SOSrule{\bigcup_{i \in I}\{ \pi_i \trans[q_{i,m}] t_{i,m} \mid m \in M_i \}}
{\left\{ \sum_{i \in I} p_i \pi_i \trans[q_{u}] u \; \Big| \;
u \in T  \text{ and }
q_{u} = \sum_{i \in I, m \in M_i \text{ s.t. } t_{i,m} = u} p_i  q_{i,m} \right\}}
\]
We observe that $\trg{\sigma(r_{\D})} = \support(\pi)$ and since the premises of $\sigma(r_{\D})$ are provable from $D_{\Sigma}$ (Equation~\eqref{eq:premises_provable_convex}) we can conclude that  
\[
D_{\Sigma} \vdash \{ \sum_{i \in I} p_i \pi_i \trans[q_{u}] u \mid u \in T \text{ and } q_{u} = \sum_{i\in I, m \in M_i \text{ s.t. } t_{i,m} = u} p_i q_{i,m}\}
\]
thus proving Equation~\eqref{eq:proof_sigma_q_2}. \qedhere
\end{itemize}
\end{description}
\end{proof}


\section{Ruloids and distribution ruloids.}
\label{sec:ruloids}

In this section we introduce the concept of \emph{ruloid} \cite{BFvG04,BIM95}, namely a derived inference rule 
with an arbitrary term as source
allowing us to deduce the behavior of that source term directly from the behavior of the variables occurring in it.
This feature makes ruloids fundamental for the decomposition method.
The characterization theorems (Theorem~\ref{thm:adequacy_DD11} and Theorem~\ref{thm:ready_sim_adequate}) assert that each formula satisfied by a process captures a different aspect of its behavior.
Hence, the aim of a decomposition method, which we recall is to reduce the satisfaction problem of a formula for a process to the satisfaction problem of derived formulae for its subprocesses, can be restated by saying that we need to find a method to relate the behavior of a process to the behavior of its subprocesses.
This is where ruloids play their r\^ole: they give us the constraints, expressed as premises of an inference rule, that the closed substitution instances of the variables occurring in the source term of the ruloid must satisfy in order to guarantee that the closed substitution instance of the source term behaves accordingly to the considered formula.

Formally, in Section~\ref{sec:decompose_ruloids} we introduce $P$-\emph{ruloids}, namely the class of ruloids built from a PGSOS-PTSS $P$ and in Section~\ref{sec:decompose_distribution_ruloids} we introduce $\Sigma$-\emph{distribution ruloids}, namely derived $\Sigma$-distribution rules allowing us to infer the behavior of any distribution term directly from the behavior of the variables occurring in it.
We prove that both classes of ruloids are \emph{sound and specifically witnessing} \cite{BIM95}, i.e.\ a closed literal $\alpha$ (resp.\ a distribution over terms $L$) is provable from a PGSOS-PTSS $P$ (resp.\ the $\Sigma$-DS) if{f} $\alpha$ (resp.\ $L$) is a closed substitution instance of the conclusion of a $P$-ruloid (resp.\ $\Sigma$-distribution ruloid) (Theorem~\ref{thm:term_ruloid_theorem} and Theorem~\ref{thm:distribution_ruloid_theorem}).


\subsection{Ruloids}
\label{sec:decompose_ruloids}

Ruloids are a generalization of PGSOS rules that allow us to infer the behavior of all open terms directly from the behavior of their variables.
Informally, given an arbitrary open term, for instance $x +_p (y \mid z)$ for $x,y,z \in \SVar$, we aim to construct an inference rule, the ruloid, allowing us to derive the behavior of $x +_p (y \mid z)$ from the behavior of $x$, $y$ and $z$ (as done in Example~\ref{ex:state_ruloid} below).
Notice that this purpose cannot be met by using only the SOS rules, since in the source of rules only one operator is admitted, and therefore there is no rule with source $x +_p (y \mid z)$.
A ruloid has the chosen open term as source, and positive and negative premises for the variables occurring in that term.
Ruloids are defined by an inductive composition of PGSOS rules.
In detail, from a rule $r$ and a substitution $\sigma$, a ruloid $\rho$ with conclusion $\sigma(\conc{r})$ is built as follows: 
\begin{enumerate*}
\item for each positive premise $\alpha$ in $\sigma(r)$, either we put $\alpha$ among the premises of $\rho$, if the left side of $\alpha$ is a variable, or, otherwise,
we take any ruloid having $\alpha$ as conclusion and we put its premises among the premises of $\rho$; 
\item for each negative premise $\alpha$ in $\sigma(r)$, either we put $\alpha$ among the premises of $\rho$, if the left side of $\alpha$ is a variable, or, otherwise, for each ruloid $\rho'$ having any literal denying $\alpha$ as conclusion, we select any premise $\beta$ of $\rho'$, we take any literal $\beta'$ denying $\beta$, and we put $\beta'$ among the premises of $\rho$. 
\end{enumerate*}

For a PGSOS-PTSS $P=(\Sigma,\Act,R)$, let $\lit{P}$ denote the set of literals that can be built with terms in $\openTerms$ and actions in $\Act$.

\begin{defi}
[Ruloids]
\label{def:term_ruloids}
Let $P=(\Sigma,\Act,R)$ be a PGSOS-PTSS.
The set of \emph{$P$-ruloids} $\ReT$ is the smallest set such that:
\begin{itemize}
\item 
$\ddedrule{x\trans[a]\mu}{x\trans[a]\mu}$ is a $P$-ruloid for all $x \in \SVar$, $a \in \Act$ and $\mu \in \DVar$;
\item
For a 
PGSOS rule $r \in R$ of the form
\[
\SOSrule{\{x_i\trans[a_{i,m}]\mu_{i,m} \mid i\in I, m\in M_i\} \qquad \{x_i\ntrans[a_{i,n}] \mid i\in I, n\in N_i\}}{f(x_1, \ldots, x_{\n})\trans[a] \Theta'}
\]
and a substitution $\sigma$ with $\sigma(x_i) = t_i$ for $i=1,\ldots,\n$ and $\sigma(\Theta') = \Theta$, the inference rule
\[
\SOSrule{\bigcup_{i \in I, m\in M_i} \HH_{i,m} \cup \bigcup_{i \in I, n\in N_i} \HH_{i,n}}{f(t_1, \ldots, t_{\n})\trans[a] \Theta}
\]
is a $P$-ruloid if the following constraints are satisfied:
  \begin{itemize}
  \item for every positive premise $x_i\trans[a_{i,m}]\mu_{i,m}$ of $r$
    \begin{itemize}
    \item either $\sigma(x_i)$ is a variable and $\HH_{i,m}=\{\sigma(x_i)\trans[a_{i,m}]\sigma(\mu_{i,m})\}$, 
    \item or there is a $P$-ruloid $\rho_{i,m}= \SOSrule{\HH_{i,m}}{\sigma(x_i)\trans[a_{i,m}]\sigma(\mu_{i,m})}$;
    \end{itemize}
  \item for every negative premise $x_i\ntrans[a_{i,n}]$ of $r$ 
    \begin{itemize}
    \item either $\sigma(x_i)$ is a variable and $\HH_{i,n}=\{\sigma(x_i)\ntrans[a_{i,n}]\}$, 
    \item or $\HH_{i,n}=\opp(\pick(\ReT_{\sigma(x_i), a_{i,n}}))$, where:
    \begin{enumerate}[label=\emph{\roman*}.]
 \item $\ReT_{\sigma(x_i), a_{i,n}}\in \powset{\powset{\lit{P}}}$ is the set containing the sets of premises of all $P$-ruloids with conclusion $\sigma(x_i)\trans[a_{i,n}] \theta$ for any distribution term $\theta \in\openDTerms$, formally
\[
\ReT_{\sigma(x_i), a_{i,n}} = \{\prem{\rho} \mid \rho\in\ReT \text{ and } \conc{\rho}= \sigma(x_i) \trans[a_{i,n}] \theta \text{ for some } \theta \in\openDTerms\} \text{,}
\]
     \item $\pick \colon \powset{\powset{\lit{P}}} \to \powset{\lit{P}}$ is any mapping such that, given any sets of literals $L_k$ with $k \in K$, $\pick(\{L_k \mid k \in K\}) = \{l_k \mid k \in K \wedge l_k \in L_k\}$,  namely $\pick$ selects exactly one literal from each set $L_k$,
     \item $\opp \colon \powset{\lit{P}} \to \powset{\lit{P}}$ is any mapping satisfying 
        $\opp(L) = \{\opp(l) \mid l \in L\}$ for all sets of literals $L$, where 
        $\opp(t' \trans[a] \theta) = t' \ntrans[a]$, and 
        $\opp(t' \ntrans[a]) = t' \trans[a] \theta$ for some fresh distribution term $\theta$, namely $\opp$ applied to any literal returns a denying literal;
      \end{enumerate}
\end{itemize}
\item the sets of the right hand side variables in $\HH_{i,m}$ and $\HH_{i,n}$ are all pairwise disjoint, formally $\rhs(\HH_{i,h}) \cap \rhs(\HH_{j,k}) \neq \emptyset$ for any $h \in M_i \cup N_i$ and $k \in M_{j} \cup N_j$ implies $h=k$ and $i=j$.
\end{itemize}
\end{itemize}
\end{defi}

\begin{exa}
\label{ex:state_ruloid}
From the rules in Example~\ref{tab:ex_PGSOS}, we derive the following ruloids for term $x +_p (y \mid z)$:
\[
\begin{array}{cc}
\SOSrule{x \trans[a] \mu \quad y \ntrans[a]}{x +_p (y \mid z) \trans[a] \mu}
\qquad & \qquad
\SOSrule{x \trans[a] \mu \quad z \ntrans[a]}{x +_p (y \mid z) \trans[a] \mu} \\[2em]
\SOSrule{x \ntrans[a] \quad y \trans[a] \nu \quad z \trans[a] \upsilon}{x +_p (y\mid z) \trans[a] \nu \mid \upsilon}
\qquad & \qquad
\SOSrule{x \trans[a] \mu \quad y \trans[a] \nu \quad z \trans[a] \upsilon}{x +_p (y \mid z) \trans[a] p\mu + (1-p)(\nu \mid \upsilon)}\; .
\end{array}
\]
We describe the construction of the first ruloid: 
\[
\SOSrule{\textcolor{white}{\SOSrule{}{\textcolor{black}{x \trans[a] \mu}}} \quad\quad \SOSrule{y \ntrans[a]\;}{y|z \ntrans[a]}\;}{x +_p (y|z) \trans[a] \mu} \text{.}
\]
Accordingly to the second PGSOS rule in Example~\ref{tab:ex_PGSOS}, whenever the term $x$ makes an $a$-move to the distribution variable $\mu$ and the term $y | z$ cannot execute action $a$, then the term $x +_p (y|z)$ makes an $a$-move to $\mu$.
As the left-hand side of the positive premise $x \trans[a]\mu$ is already a variable, then there is nothing more to do.
Conversely, the left-hand side of the negative premise $y|z \ntrans[a]$ is a term.
By Definition~\ref{def:term_ruloids} we need to consider all the PGSOS rules having a literal $y | z \trans[a] \Theta$, for some $\Theta$ in $\openDTerms$, as conclusion, namely any proper instance of the first rule in Example~\ref{tab:ex_PGSOS}.
Then we need to choose one premise for each of those rules, for instance the one having $y$ as left-hand side, and deny the ones we have selected.
In our example, from this construction we obtain the single negative premise $y \ntrans[a]$ whose left-hand side is a variable and thus concludes the construction of the first $P$-ruloid for the term $x +_p (y|z)$.  
\end{exa}

We can show that if the PGSOS-PTSS $P$ is positive then also the derived $P$-ruloids are positive.
This ensures that if $P$ is positive then all formulae obtained by the decomposition method will not contain any negation.

\begin{lem}
\label{lem:positive_ruloid}
Let $P$ be a positive PGSOS-PTSS. 
Then all the $P$-ruloids in $\ReT$ are positive.
\end{lem}


\begin{proof}
The proof follows immediately from Definition~\ref{def:term_ruloids} by noticing that since no rule in $P$ contains negative premises, then the function $\opp$ is never applied.
Therefore positive literals are never transformed into negative.
\end{proof}


The following result states that ruloids define completely the behavior of all open terms.
More precisely, Theorem~\ref{thm:term_ruloid_theorem} shows that ruloids allows us to infer the behavior of the closed substitution instances of any open term $t$ from the behaivor of the closed substitution instances of its variables.
This is crucial to support the decomposition method, which will decompose state formulae for $t$ into state formulae for its variables by exploiting the ruloids having $t$ as source.

\begin{thm}
[Ruloid theorem]
\label{thm:term_ruloid_theorem}
Assume a PGSOS-PTSS $P$, a closed substitution $\sigma$, a term $t \in \openT$ and a closed distribution term $\Theta' \in \closedDTerms$.
Then $P \vdash \sigma(t) \trans[a] \Theta'$ if and only if there are a $P$-ruloid $\ddedrule{\HH}{t \trans[a] \Theta}$ and a closed substitution $\sigma'$ with $\sigma'(t) = \sigma(t)$, $\sigma'(\Theta)=\Theta'$ and $P \vdash \sigma'(\HH)$. 
\end{thm}


\begin{proof}
We proceed by structural induction on the term $t \in \openT$.
\begin{description}
\item[Base case] \; $t = x \in \SVar$. 
  \begin{description}
    \item[($\Rightarrow$)]
The thesis follows immediately for the $P$-ruloid $\ddedrule{x \trans[a] \mu}{x \trans[a] \mu}$ and any closed substitution $\sigma'$ with $\sigma'(x) = \sigma(x)$ and $\sigma'(\mu) = \Theta'$.

\item[($\Leftarrow$)]
Accordingly to Definition~\ref{def:term_ruloids}, a $P$-ruloid having $x$ as source is of the form $\ddedrule{x \trans[a] \mu}{x \trans[a] \mu}$.
Thus, from $\sigma'(x) = \sigma(x)$, $\sigma'(\mu) = \Theta'$ and $P \vdash \sigma'(x) \trans[a] \sigma'(\mu)$ we can immediately infer that $P \vdash \sigma(x) \trans[a] \Theta'$.
\end{description}

\item[Inductive step] $t = f(t_1,\dots,t_{\n}) \in \openT$ for some $\n$-ary operator $f$.
\begin{description}
\item[($\Rightarrow$)]
We proceed by structural induction over a closed proof $\gamma$ of $\sigma(t)\trans[a] \Theta'$ from $P$.

The bottom of the closed proof $\gamma$ is constituted by a PGSOS rule $r \in R$ of the form
\[
\SOSrule{\{x_i \trans[a_{i,m}] \mu_{i,m} \mid i \in I , m \in M_i\} \cup \{x_i \ntrans[a_{i,n}] \mid i \in I , n \in N_i\}}
{f(x_1, \ldots, x_{\n}) \trans[a] \upsilon}
\] 
together with a closed substitution $\varsigma$ such that:
\begin{enumerate}
\item $\varsigma(x_i) = \sigma(t_i)$ for each $i \in I$;
\item $\varsigma(\upsilon) = \Theta'$;
\item \label{thm:rt_ip3} for all $i \in I$ and $m \in M_i$ there is a proof shorter than $\gamma$ of $\varsigma(x_i) \trans[a_{i,m}] \varsigma( \mu_{i,m})$ from $P$;
\item \label{thm:rt_ip4} for all $i \in I$ and $n \in N_i$ there is a proof shorter than $\gamma$ of $\varsigma(x_i) \ntrans[a_{i,n}]$ from $P$.
\end{enumerate}

Let $\varsigma_0$ be any substitution with $\varsigma_0(x_i) = t_i$ for each $i \in I$.
Considering that $\varsigma(x_i) = \sigma(t_i) = \sigma( \varsigma_0(x_i))$, from items~\eqref{thm:rt_ip3} and~\eqref{thm:rt_ip4} above we get that $P \vdash \sigma(\varsigma_0(x_i)) \trans[a_{i,m}] \varsigma(\mu_{i,m})$, for $i \in I$ and $m \in M_i$, and $P \vdash \sigma(\varsigma_0(x_i)) \ntrans[a_{i,n}]$, for $i \in I$ and $n \in N_i$.

Consider any $\sigma(\varsigma_0(x_i)) \trans[a_{i,m}] \varsigma(\mu_{i,m})$.
By the inductive hypothesis, there are a $P$-ruloid 
\[
\SOSrule{\HH_{i,m}}{\varsigma_0(x_i) \trans[a_{i,m}] \Theta_{i,m}}
\]
 and a closed substitution $\sigma'_{i,m}$ with 
\begin{itemize}
\item $\sigma'_{i,m}(\varsigma_0(x_i)) = \sigma(\varsigma_0(x_i))$,
\item $\sigma'_{i,m}(\Theta_{i,m}) = \varsigma(\mu_{i,m})$, and
\item $P \vdash \sigma'_{i,m}(\HH_{i,m})$.
\end{itemize}

Let us consider now any $\varsigma_0(x_i)\ntrans[a_{i,n}]$.
By definition, $P \vdash \sigma(\varsigma_0(x_i)) \ntrans[a_{i,n}]$ only if $P \not\vdash \sigma(\varsigma_0(x_i)) \trans[a_{i,n}] \pi$ for any $\pi \in \closedDTerms$.
By structural induction on $\varsigma_0(x_i) = t_i$, this implies that for all $P$-ruloids of the form 
\[
\SOSrule{\HH_{\Theta_{i,n}}}{\varsigma_0(x_i) \trans[a_{i,n}] \Theta_{i,n}}
\]
and for all closed substitutions $\sigma''$ with $\sigma''(\varsigma_0(x_i)) = \sigma(\varsigma_0(x_i))$, it holds that $P \not\vdash \sigma''(\HH_{\Theta_{i,n}})$.
We can distinguish two cases.
\begin{enumerate}[label=\emph{\alph*})]
\item There is a negative literal $\alpha_{\Theta_{i,n}}$ in $\HH_{\Theta_{i,n}}$ such that $P \not\vdash \sigma''(\alpha_{\Theta_{i,n}})$ for any closed substitution $\sigma''$ with $\sigma''(\varsigma_0(x_i)) = \sigma(\varsigma_0(x_i))$.
Then the completeness of $P$ ensures that there are at least one positive literal $\beta_{\Theta_{i,n}}$ denying $\alpha_{\Theta_{i,n}}$ and one closed substitution $\sigma'_{i,n}$ with $\sigma'_{i,n}(\varsigma_0(x_i)) = \sigma(\varsigma_0(x_i))$ s.t. $P \vdash \sigma'_{i,n}(\beta_{\Theta_{i,n}})$.
\item The closed substitution instances of negative literals possibly occurring in $\HH_{\Theta_{i,n}}$, wrt.\ all closed substitutions $\sigma''$ with $\sigma''(\varsigma_0(x_i)) = \sigma(\varsigma_0(x_i))$, are provable from $P$.
In this case, since the condition $P \not \vdash \sigma''(\HH_{\Theta_{i,n}})$ holds for all closed substitutions $\sigma''$ as above, we can infer that there is at least one positive literal in $\HH_{\Theta_{i,n}}$, say $\alpha_{\Theta_{i,n}}$, s.t. $P \not \vdash \sigma''(\alpha_{\Theta_{i,n}})$ for all such closed substitutions $\sigma''$.
In detail, if we assume wlog.\ that $\alpha_{\Theta_{i,n}}$ is of the form $y \trans[a] \nu$ for some $y \in \var{\varsigma_0(x_i)}$ and $\nu \in \DVar$, then we have obtained that given any closed substitution $\sigma''$, with $\sigma''(\varsigma_0(x_i)) = \sigma(\varsigma_0(x_i))$, we have $P \not \vdash \sigma''(y) \trans[a] \pi$ for any $\pi \in \closedDTerms$.
By completeness of $P$, this implies that $P \vdash \sigma''(y) \ntrans[a]$.
In general, given a literal $\beta_{\Theta_{i,n}}$ denying $\alpha_{\Theta_{i,n}}$ and any closed substitution $\sigma'_{i,n}$ with $\sigma'_{i,n}(\varsigma_0(x_i)) = \sigma(\varsigma_0(x_i))$, we obtain that $P \vdash \sigma'_{i,n}(\beta_{\Theta_{i,n}})$.
\end{enumerate}

Therefore, if we consider $\HH_{i,n} = \bigcup_{\Theta_{i,n}}\beta_{\Theta_{i,n}}$ and we take a closed substitution $\sigma'_{i,n}$ as described in the two cases above, then we obtain
\[
P \vdash \sigma'_{i,n}(\HH_{i,n}).
\]

We remark that since we are working with a countable set of variables, we can always assume that the variables in $\rhs(\HH_{i,m})$ for $i \in I$ and $m \in M_i$ and the variables in $\rhs(\HH_{i,n})$ for $i \in I$ and $n \in N_i$ are pairwise disjoint.
Moreover, all those variables are disjoint from $\var{t}$.
Therefore, we can define a closed substitution $\sigma'$ as follows:
\begin{enumerate}
\item $\sigma'(y) = \sigma(y)$ for all $y \in \var{t}$;
\item $\sigma'(\mu) = \sigma'_{i,m}(\mu)$ for all $\mu \in \rhs(\HH_{i,m})$, with $i \in I$ and $m \in M_i$;
\item $\sigma'(\mu) = \sigma'_{i,n}(\mu)$ for all $\mu \in \rhs(\HH_{i,n})$, with $i \in I$ and $n \in N_i$.
\end{enumerate}
Then define
\[
\HH =  \bigcup_{i \in I, m \in M_i} \HH_{i,m} \cup \bigcup_{i \in I,n \in N_i} \HH_{i,n} .
\]
Moreover, let $\varsigma_1$ be a substitution with $\varsigma_1(x_i) = t_i$ and $\varsigma_1(\mu_{i,m}) = \Theta_{i,m}$ for all $i \in I$ and $m \in M_i$.
We can show that the $P$-ruloid
\[
\SOSrule{\HH}{f(t_1, \ldots, t_{\n}) \trans[a] \varsigma_1(\upsilon)}
\]
together with the substitution $\sigma'$ satisfies the required properties:
\begin{enumerate}
\item First we prove that $\sigma'(f(t_1, \ldots, t_{\n}) ) = \sigma(f(t_1, \ldots, t_{\n}) )$. 
This immediately follows from $\sigma'(y) = \sigma(y)$ for all $y \in \var{f(t_1, \ldots, t_{\n}) }$.
\item Then we prove that $P \vdash \sigma'(\HH)$, which is derived from the following considerations:
\begin{enumerate}
\item \label{subst_agree_1}
Substitutions $\sigma'$ and $\sigma'_{i,m}$ agree on all variables occurring in $\ddedrule{\HH_{i,m}}{\varsigma_0(x_i) \trans[a_{i,m}] \Theta_{i,m}}$ for all $i \in I$ and $m \in M_i$.
Indeed, assume any $i \in I$ and $m \in M_i$.
Since $\var{f(t_1, \ldots, t_{\n})} = \bigcup_{i = 1}^{\n} \var{t_i} = \bigcup_{i=1}^{\n}\var{\varsigma_0(x_i)}$, and, moreover, $\sigma$ and $\sigma'$ agree on $\var{f(t_1, \ldots, t_{\n})}$ we obtain that $\sigma' (\varsigma_0(x_i)) = \sigma(\varsigma_0(x_i))$ for each $i \in I$.
Moreover, by construction we have that $\sigma'_{i,m}(\varsigma_0(x_i)) = \sigma(\varsigma_0(x_i))$, thus giving $\sigma'(\varsigma_0(x_i)) = \sigma'_{i,m}(\varsigma_0(x_i))$, namely $\sigma'$ and $\sigma'_{i,m}$ agree on $\var{\varsigma_0(x_i)}$.
Then, by definition $\sigma'$ and $\sigma'_{i,m}$ agree on all variables in $\HH_{i,m}$.
Finally, as $\var{\Theta_{i,m}} \subseteq \var{\varsigma_0(x_i)} \cup \rhs(\HH_{i,m})$ we can infer that $\sigma'$ and $\sigma'_{i,m}$ agree also on $\var{\Theta_{i,m}}$. 
\item \label{subst_agree_2}
With a similar argument we obtain that $\sigma'$ and $\sigma'_{i,n}$ agree on all variables occurring in $\ddedrule{\HH_{i,n}}{\varsigma_0(x_i) \ntrans[a_{i,n}]}$ for all $i \in I$ and $n \in N_i$.
\item By item~\ref{subst_agree_1} above, for all $i \in I$ and $m \in M_i$ $\sigma'$ agrees with $\sigma'_{i,m}$ on all variables in $\HH_{i,m}$, hence $P \vdash \sigma'_{i,m}(\HH_{i,m})$ implies $P \vdash \sigma'(\HH_{i,m})$.
Analogously, by item~\ref{subst_agree_2} above,
for all $i \in I$ and $n \in N_i$ $\sigma'$ agrees with $\sigma'_{i,n}$ on all variables in $\HH_{i,n}$, hence $P \vdash \sigma'_{i,n}(\HH_{i,n})$ implies $P \vdash \sigma'(\HH_{i,n})$.
Then, since $\HH = \bigcup_{i \in I, m \in M_i} \HH_{i,m} \cup \bigcup_{i \in I n \in N_i} \HH_{i,n}$ we can conclude that $P \vdash \sigma'(\HH)$.
\end{enumerate}
\item Finally, we prove that $\sigma'(\varsigma_1(\upsilon)) = \Theta'$.
Notice that the substitutions $\varsigma_0$ and $\varsigma_1$ agree on $\var{f(t_1,\ldots,t_{\n})}$ thus giving $\sigma(\varsigma_0(x_i)) = \sigma(\varsigma_1(x_i))$ for all $i \in I$.
Then we have that $\sigma'(\varsigma_1(x_j)) = \sigma'(t_j) = \sigma(t_j) = \varsigma(x_j)$ for $j = 1, \ldots, \n$.
Moreover, since $\sigma'$ and $\sigma'_{i,m}$ agree on $\var{\Theta_{i,m}}$, we can infer that $\sigma'(\varsigma_1(\mu_{i,m})) = \sigma'(\Theta_{i,m}) = \sigma'_{i,m}(\Theta_{i,m}) = \varsigma(\mu_{i,m})$ for all $i \in I$ and $m \in M_i$.
As $\var{\upsilon} \subseteq \{x_1,\ldots, x_{\n}\} \cup \{\mu_{i,m} \mid m \in M_i, i \in I\}$, it follows that $\sigma'(\varsigma_1(\upsilon)) = \varsigma(\upsilon)= \Theta'$.
\end{enumerate}

\item[($\Leftarrow$)]
Assume that there a $P$-ruloid $\rho = \ddedrule{\HH}{t \trans[a] \Theta}$ and a closed substitution $\sigma'$ with $P \vdash \sigma'(\HH)$, $\sigma'(t) = \sigma(t)$ and $\sigma'(\Theta) = \Theta'$.
We note that the thesis $P \vdash \sigma(t) \trans[a] \Theta'$ is equivalent to $P \vdash \sigma'(t) \trans[a] \sigma'(\Theta)$.

Accordingly to Definition~\ref{def:term_ruloids}, let $r$ and $\sigma_0$ be resp.\ the PGSOS rule and the substitution from which $\rho$ is built, namely let $r$ be of the form
\[
r = \SOSrule{\{x_i \trans[a_{i,m}] \mu_{i,m} \mid i \in I, m \in M_i\} \qquad \{x_i \ntrans[a_{i,n}] \mid i \in I, n \in N_i\}}{f(x_1,\dots,x_{\n}) \trans[a] \Theta''}
\]
for $I = \{1,\dots,\n\}$, and $\sigma_0$ be such that $\sigma_0(x_i) = t_i$ and $\sigma_0(\Theta'') = \Theta$. 
Then $\rho$ is of the form
\[
\rho = \SOSrule{\bigcup_{i \in I, m\in M_i} \HH_{i,m} \cup \bigcup_{i \in I, n\in N_i} \HH_{i,n}}{f(t_1, \ldots, t_{\n})\trans[a] \Theta}
\]
where:
\begin{itemize}
\item For every positive premise $x_i \trans[a_{i,m}] \mu_{i,m}$ of $r$:
\begin{itemize}
\item Either $\sigma_0(x_i)$ is a variable and $\HH_{i,m} = \{\sigma_0(x_i) \trans[a_{i,m}] \sigma_0(\mu_{i,m})\} = \{t_i \trans[a_{i,m}] \sigma_0(\mu_{i,m})\}$.
Hence from $P \vdash \sigma'(\HH)$ we can directly infer that $P \vdash \sigma'(t_i) \trans[a_{i,m}] \sigma'(\sigma_0(\mu_{i,m}))$.
\item Or there is a $P$-ruloid $\rho_{i,m} = \ddedrule{\HH_{i,m}}{\sigma_0(x_i) \trans[a_{i,m}] \sigma_0(\mu_{i,m})} = \ddedrule{\HH_{i,m}}{t_i \trans[a_{,m}] \sigma_0(\mu_{i,m})}$.
Since $P \vdash \sigma'(\HH)$ implies $P \vdash \sigma'(\HH_{i,m})$, by structural induction on $t_i$ we can infer that $P \vdash \sigma'(t_i) \trans[a_{i,m}] \sigma'(\sigma_0(\mu_{i,m}))$.
\end{itemize}
We can therefore conclude that the closed substitution instances wrt.\ $\sigma' \circ \sigma_0$ of the positive premises of $r$ are provable from $P$.

\item For every negative premise $x_i \ntrans[a_{i,n}]$ of $r$:
\begin{itemize}
\item Either $\sigma_0(x_i)$ is a variable and $\HH_{i,n} = \{\sigma_0(x_i)\ntrans[a_{i,n}]\} = \{t_i \ntrans[a_{i,n}]\}$.
Hence from $P \vdash \sigma'(\HH)$ we can immediately infer that $P \vdash \sigma'(t_i) \ntrans[a_{i,n}]$.
\item Or $\HH_{i,n} = \opp(\pick(\ReT_{\sigma_0(x_i), a_{i,n}}))$, namely (see Definition~\ref{def:term_ruloids}) for each $P$-ruloid $\rho'$ such that $\conc{\rho'} = \sigma_0(x_i) \trans[a_{i,n}] \theta$, for any $\theta \in \openDTerms$, we have that $\HH_{i,n}$ contains at least one literal denying a literal in $\prem{\rho'}$.
Hence, since $P \vdash \sigma'(\HH)$ implies $P \vdash \sigma'(\HH_{i,n})$, we can infer that $P \not \vdash \sigma'(\prem{\rho'})$.
Hence, the structural induction on $\sigma_0(x_i) = t_i$ (case ($\Rightarrow$)) gives that $P \not\vdash \sigma'(t_i) \trans[a_{i,n}] \sigma'(\sigma_0(\theta))$, for any $\theta \in \openDTerms$, thus implying $P \vdash \sigma'(t_i) \ntrans[a_{i,n}]$.
\end{itemize}
We can therefore conclude that the closed substitution instances wrt.\ $\sigma' \circ \sigma_0$ of the negative premises of $r$ are provable from $P$.
\end{itemize}
We have obtained that all the closed substitution instances wrt.\ $\sigma' \circ \sigma_0$ of the premises of $r$ are provable from $P$ and therefore we can infer that there is a proof from $P$ of $\sigma'(t) \trans[a] \sigma'(\Theta)$, which concludes the proof.\qedhere
\end{description}
\end{description}
\end{proof}


If the PGSOS-PTSS $P$ is positive, then the ruloid theorem can be reformulated by stressing that the involved $P$-ruloids are positive.
This will be necessary in the decomposition of formulae in the sublogic $\logicplus$.
Indeed, the absence of negative premises in the $P$-ruloids will ensure that by decomposing formulae in $\logicplus$ one gets formulae that have no negation and, therefore, are still in $\logicplus$.

\begin{cor}
\label{cor:positive_ruloid}
Let $P$ be a positive PGSOS-PTSS. 
Then $P \vdash \sigma(t) \trans[a] \Theta'$ for $t \in \openT$, $\Theta' \in \closedDTerms$ and $\sigma$ a closed substitution, if{f} there are a positive $P$-ruloid $\HH / t \trans[a] \Theta$ and a closed substitution $\sigma'$ with $P \vdash \sigma'(\HH)$, $\sigma'(t) = \sigma(t)$ and $\sigma'(\Theta)=\Theta'$.
\end{cor}


\begin{proof}
The proof follows immediately from Lemma~\ref{lem:positive_ruloid} and Theorem~\ref{thm:term_ruloid_theorem}.
\end{proof}


\subsection{Distribution ruloids}
\label{sec:decompose_distribution_ruloids}

$\Sigma$-distribution ruloids are a generalization of $\Sigma$-distribution rules and define the behavior of arbitrary open distribution terms.
More precisely, they allow us to infer the behavior of a distribution term as a probability distribution over terms from the distribution over terms that characterize the behavior of the variables occurring in it.
For instance, distribution ruloids allow us to infer the behavior of a distribution term of the form $\frac{2}{5} \mu + \frac{3}{5}(\nu \mid v)$ from the behavior of the variables $\mu$, $\nu$ and $v$.
Notice that distribution rules are not enough to meet this purpose, since in the source of distribution rules only one operator over distributions is admitted, and therefore there is no $\Sigma$-distribution rule with source $\frac{2}{5} \mu + \frac{3}{5}(\nu \mid v)$.
Similarly to $P$-ruloids, a $\Sigma$-distribution ruloid is defined by an inductive composition of $\Sigma$-distribution rules and the left-hand sides of its premises are the variables occurring in the source, which is an arbitrary open distribution term.
As the $\Sigma$-DS is positive, the definition of $\Sigma$-distribution ruloids results technically simpler than that of $P$-ruloids.

\begin{defi}
[$\Sigma$-distribution ruloids]
\label{def:distribution_ruloids}
Let $D_{\Sigma} = (\Sigma,R_{\Sigma})$ be the $\Sigma$-DS.
The set of \emph{$\Sigma$-distribution ruloids} $\ReD$ is the smallest set such that:
\begin{itemize}
\item
The inference rule
\[
\SOSrule{\{\delta_x \trans[1] x\}}{\{\delta_x \trans[1] x\}}
\]
is a $\Sigma$-distribution ruloid for any $x \in \SVar$;
\item
The inference rule
\[
\SOSrule{\{ \mu \trans[q_i] x_i  \mid i \in I\}}{\{ \mu \trans[q_i] x_i \mid i \in I\}}
\] 
is a $\Sigma$-distribution ruloid for any $\mu \in \DVar$, provided that $\sum_{i \in I}q_i = 1$ and all variables $x_i$ with $i \in I$ are distinct;
\item
For a \dgsos rule $r_{\D} \in R_{\Sigma}$ of the form
\[
\SOSrule{\bigcup_{i = 1,\ldots,\n}\{ \vartheta_i \trans[q_{i,j}] x_{i,j} \mid j \in J_i \}}
{\Big\{f(\vartheta_1, \ldots, \vartheta_{\n}) \trans[q_k] f(x_{1,k(1)}, \ldots, x_{\n, k(\n)}) \; \Big| \; 
 k \in \bigtimes_{i=1,\ldots,\n}J_i   \text{ and } q_k = \prod_{i =1,\ldots,\n} q_{i,k(i)} \Big\}}
\]
as in Definition~\ref{def:dgsos_rule}.\ref{def:dgsos_rule_f} and a substitution $\sigma$ with $\sigma(r_{\D})$ of the form
\[
\SOSrule{\bigcup_{i = 1,\ldots,\n}\{ \Theta_i \trans[q_{i,h}] t_{i,h} \mid h \in H_i \}}
{\Big\{f(\Theta_1,\ldots, \Theta_{\n}) \trans[q_{\kappa}] f(t_{1,\kappa(1)},\ldots, t_{\n, \kappa(\n)}) \; \Big| \;
\kappa \in \bigtimes_{i=1,\ldots,\n} H_i  \text{ and } q_{\kappa} = \prod_{i =1,\ldots,\n} q_{i,\kappa(i)}\Big\}}
\]
(see Definition~\ref{def:reduced_instance}.\ref{def:reduced_instance_f}), the inference rule 
\[
\SOSrule{ \bigcup_{i =1,\ldots,\n} \HH_i }
{\Big\{f(\Theta_1,\ldots, \Theta_{\n}) \trans[q_{\kappa}] f(t_{1,\kappa(1)},\ldots, t_{\n, \kappa(\n)}) \; \Big| \;
 \kappa \in \bigtimes_{i=1,\ldots,\n} H_i \text{ and } q_{\kappa} = \prod_{i =1,\ldots,\n} q_{i,\kappa(i)}\Big\}}
\]
is a $\Sigma$-distribution ruloid if for each $i=1,\ldots,\n$ we have that:
\begin{itemize}
\item either $\Theta_i$ is a variable or a Dirac distribution and $\HH_i =  \{ \Theta_i \trans[q_{i,h}] t_{i,h} \mid h \in H_i\}$,
\item or there is a $\Sigma$-distribution ruloid $\rho^{\D}_i = \SOSrule{\HH_i}{\{\Theta_i \trans[q_{i,h}] t_{i,h} \mid h \in H_i\}}$;
\end{itemize}
\item
For a \dgsos rule $r_{\D} \in R_{\Sigma}$ of the form
\[
\SOSrule{\bigcup_{i \in I}\{ \vartheta_i \trans[q_{i,j}] x_{i,j} \mid j \in J_i \}}
{\Big\{ \sum_{i \in I} p_i \vartheta_i \trans[q_x] x \; \Big| \;
x \in \{x_{i,j} \mid i \in I \wedge  j \in J_i\} \text{ and } q_x = \sum_{i \in I, j \in J_i \text{ s.t. } x_{i,j} = x} p_i \cdot q_{i,j}\Big\}}
\]
as in Definition~\ref{def:dgsos_rule}.\ref{def:dgsos_rule_convex} and a substitution $\sigma$ with $\sigma(r_{\D})$ of the form
\[
\SOSrule{\bigcup_{i \in I}\{ \Theta_i \trans[q_{i,h}] t_{i,h} \mid h \in H_i \}}
{\Big\{ \sum_{i \in I} p_i \Theta_i \trans[q_u] u \; \Big| \;
u \in \{t_{i,h} \mid i \in I \wedge h \in H_i\} \text{ and } q_u = \sum_{i \in I, h \in H_i \text{ s.t. } t_{i,h} = u} p_i \cdot q_{i,h}\Big\}}
\]
(see Definition~\ref{def:reduced_instance}.\ref{def:reduced_instance_Sum}), the inference rule
\[
\SOSrule{ \bigcup_{i \in I} \HH_i }{\Big\{ \sum_{i \in I} p_i \Theta_i \trans[q_u] u \; \Big| \;
u \in \{t_{i,h} \mid i \in I \wedge h \in H_i\} \text{ and }
q_u = \sum_{i \in I, h \in H_i \text{ s.t. } t_{i,h} = u} p_i \cdot q_{i,h}\Big\}}
\]
is a $\Sigma$-distribution ruloid if for every $i \in I$ we have that:
\begin{itemize}
\item either $\Theta_i$ is a variable or a Dirac distribution and $\HH_i =  \{ \Theta_i \trans[q_{i,h}] t_{i,h} \mid h \in H_i\}$,
\item or there is a $\Sigma$-distribution ruloid $\rho^{\D}_i = \SOSrule{\HH_i}{\{\Theta_i \trans[q_{i,h}] t_{i,h} \mid h \in H_i\}}$.
\end{itemize}
\end{itemize}
\end{defi}


\begin{exa}
\label{ex:distribution_ruloid}
Consider the distribution term $\frac{2}{5} \mu + \frac{3}{5}(\nu | \upsilon)$ (which is an instance of the target of the fourth $P$-ruloid in Example~\ref{ex:state_ruloid}).
Then, we can build the following $\Sigma$-distribution ruloid:
\[
\SOSrule{\textcolor{white}{\SOSrule{}{\textcolor{black}{\{\mu \trans[1/4] x_1 \;\;\; \mu \trans[3/4] x_2\}}}} \qquad\quad \SOSrule{\{\nu \trans[1/3] y_1, \;\;\; \nu \trans[2/3] y_2\} \;\;\; \{\upsilon \trans[1] z\}}{\{\nu | \upsilon \trans[1/3] y_1|z \;\;\; \nu| \upsilon \trans[2/3] y_2 | z\}}}
{\Big\{\frac{2}{5} \mu + \frac{3}{5}(\nu|\upsilon) \trans[\frac{1}{10}] x_1,
\frac{2}{5} \mu + \frac{3}{5}(\nu|\upsilon) \trans[\frac{3}{10}] x_2, 
\frac{2}{5} \mu + \frac{3}{5}(\nu|\upsilon) \trans[\frac{1}{5}] y_1|z, 
\frac{2}{5} \mu + \frac{3}{5}(\nu|\upsilon) \trans[\frac{2}{5}] y_2|z \Big\}}.
\]
\end{exa}

\begin{prop}
\label{lem:sum_to_1_tris}
The conclusion of a \dgsos ruloid is a distribution over terms.
\end{prop}

The following structural property of \dgsos ruloids will be exploited to prove that these ruloids define completely the behavior of all open distribution terms (Theorem~\ref{thm:distribution_ruloid_theorem}).

\begin{lem}
\label{lem:ruloid_variables}
Any \dgsos ruloid $\ddedrule{\HH}{\{\Theta \trans[q_m] t_m \mid m \in M\}}$ is such that:
\begin{enumerate}
\item \label{lem:ruloid_variables_lhs_mu}
for all $\mu \in \DVar$, $\mu \in \var{\Theta}$ if{f} $\mu$ is the left-hand side of a 
premise in $\HH$;
\item \label{lem:ruloid_variables_lhs_delta}
for all $x \in \SVar$, $x \in \var{\Theta}$ if{f} $\delta_{x}$ is the left-hand side of a 
premise in $\HH$;
\item \label{lem:ruloid_variables_rhs}
 $\bigcup_{m \in M} \var{t_m} = \rhs(\HH)$.
\end{enumerate}
\end{lem}


\begin{proof}
The proof follows by structural induction over the source term $\Theta \in \openDTerms$.
\end{proof}


We are now ready to show that \dgsos ruloids define completely the behavior of all open distribution terms.
More precisely, Theorem~\ref{thm:distribution_ruloid_theorem} shows that distribution ruloids allows us to infer the behavior of the closed substitution instances of any open distribution term $\Theta$ from the behaviour of the closed substitution instances of its variables.
This is crucial to support the decomposition method, which will decompose distribution formulae for $\Theta$ into state and distribution formulae for its variables by exploiting the distribution ruloids having $\Theta$ as source. 

\begin{thm}
[Distribution ruloid theorem]
\label{thm:distribution_ruloid_theorem}
Assume the $\Sigma$-DS $D_{\Sigma}$, a closed substitution $\sigma$, 
a distribution term $\Theta \in \openDTerms$ and closed terms $t_m \in \closedTerms$ with $m \in M$ pairwise distinct. 
Then $D_{\Sigma} \vdash \{\sigma(\Theta) \trans[q_m] t_m \mid m \in M \}$  if and only if there are a $\Sigma$-distribution ruloid $\ddedrule{\HH}{\{ \Theta \trans[q_m] u_m \mid m\in M\}}$ and a closed substitution $\sigma'$ with $\sigma'(\Theta)= \sigma(\Theta)$, $\sigma'(u_m) = t_m$ for each $m \in M$ and $D_{\Sigma} \vdash \sigma'(\HH)$.
\end{thm}


\begin{proof}
We proceed by structural induction over $\Theta \in \openDTerms$.

\begin{enumerate}
\item
Base case: $\Theta$ is a Dirac distribution $\Theta = \delta_x$ for some $x \in \SVar$.
\begin{description}
  \item[($\Rightarrow$)]
The thesis follows immediately for the \dgsos ruloid 
\[
\SOSrule{\{\delta_x \trans[1] x\}}{\{\delta_x \trans[1] x\}}
\]
and the closed substitution $\sigma'=\sigma$.

\item[($\Leftarrow$)]
By Definition~\ref{def:distribution_ruloids} the only possible \dgsos ruloid for $\Theta$ has the form 
\[
\SOSrule{\{\delta_x \trans[1] x\}}{\{\delta_x \trans[1] x\}}.
\]
Thus the thesis follows immediately from $D_{\Sigma} \vdash \sigma'(\{\delta_x \trans[1] x\})$ and the choice of $\sigma'$.
\end{description}

\item
Base case: $\Theta$ is a variable $\mu \in \DVar$.
\begin{description}
  \item[($\Rightarrow$)]
The thesis immediately follows for the \dgsos ruloid 
\[
\SOSrule{\{ \mu \trans[q_m] x_m \mid m \in M \}}{\{ \mu \trans[q_m] x_m \mid m \in M\}}
\]
and the closed substitution $\sigma'$ with $\sigma'(\mu) = \sigma(\mu)$ and $\sigma'(x_m) = t_m$ for each $m \in M$.

\item[($\Leftarrow$)]
By Definition~\ref{def:distribution_ruloids} the considered \dgsos ruloid for $\Theta$ has the form 
\[
\SOSrule{\{\mu \trans[q_m] x_m \mid m \in M\}}{\{\mu \trans[q_m] x_m \mid m \in M\}}.
\]
Thus the thesis follows immediately from $D_{\Sigma} \vdash \sigma'(\{\mu \trans[q_m] x_m \mid m \in M\})$ and the choice of $\sigma'$.
\end{description}

\item
Inductive step $\Theta = f(\Theta_1,\ldots, \Theta_{\n})$ for some $f \in \Sigma$ and $\Theta_i \in \openDTerms$ for $i =1,\ldots, \n$.
\begin{description}
  \item[($\Rightarrow$)]
First of all, we recall that by Theorem~\ref{prop:proof_sigma_q} we have $D_{\Sigma} \vdash \{\sigma(\Theta) \trans[q_m] t_m \mid m \in M \}$ if{f} $\sigma(\Theta)(t_{m}) = q_m$ for each $m \in M$ and $\sum_{m \in M} q_m = 1$.
Thus, for the particular choice of $\sigma(\Theta)$, we have that the closed terms $t_m$ are of the form $t_m = f(t_{1,m}, \ldots, t_{\n,m})$ for some $\{t_{i,m} \mid i=1,\ldots,\n\} \subseteq \closedTerms$, for $m \in M$, so that $t_{i,m} \in \support(\sigma(\Theta_i))$ for each $m \in M$.
Next, let us consider a closed proof $\gamma$ of $\{\sigma(\Theta)\trans[q_m] t_m \mid m \in M \}$ from $D_{\Sigma}$.
The bottom of $\gamma$ is constituted by the closed reduced instance of a \dgsos rule $r_{\D} \in R_{\Sigma}$ of the form
\[
\SOSrule{\bigcup_{i = 1}^{\n}\{ \vartheta_i \trans[q_{i,j}] x_{i,j} \mid j \in J_i \}}
{\Big\{f(\vartheta_1, \ldots, \vartheta_{\n}) \trans[q_k] f(x_{1,k(1)}, \ldots, x_{\n, k(\n)}) \; \Big| \;
k \in \bigtimes_{i=1}^{\n}J_i  \text{ and } q_k = \prod_{i =1}^{\n} q_{i,k(i)}\Big\}}
\] 
wrt.\ a closed substitution $\varsigma$ with $\varsigma(\vartheta_i) = \sigma(\Theta_i)$ for $i = 1, \ldots, \n$.
More precisely, let $\varsigma(r_{\D})$  be the inference rule of the form
\[
\SOSrule{\bigcup_{i = 1}^{\n}\{ \sigma(\Theta_i) \trans[q_{i,h}] t_{i,h} \mid h \in H_i \}}
{\Big\{f(\sigma(\Theta_1), \ldots, \sigma(\Theta_{\n})) \trans[q_{\kappa}] f(t_{1,\kappa(1)}, \ldots, t_{\n, \kappa(\n)}) \; \Big| \;
 \kappa \in \bigtimes_{i=1}^{\n} H_i  \text{ and } q_{\kappa} = \prod_{i =1}^{\n} q_{i,\kappa(i)}\Big\}}
\] 
where 
\begin{itemize}
\item each set $\{ \sigma(\Theta_i) \trans[q_{i,h}] t_{i,h} \mid h \in H_i\}$ is the reduction wrt.\ $\sigma$ of the corresponding set $\{\vartheta_i \trans[q_{i,j}] x_{i,j} \mid j \in J_i\}$,
\item there is bijection $\f \colon \bigtimes_{i=1}^{\n} H_i \to M$ with $t_{i,\kappa(i)} = t_{i,\f(\kappa)}$ for each $i = 1,\ldots, \n$,
\item for all $i = 1, \ldots, \n$ there is a proof shorter than $\gamma$ of $\{\sigma(\Theta_i) \trans[q_{i,h}] t_{i,h} \mid h \in H_i\}$ from $D_{\Sigma}$.
\end{itemize}
Let $\varsigma_0$ be a substitution with $\varsigma_0(\vartheta_i) = \Theta_i$ for $i = 1, \ldots, \n$.
Considering that $\varsigma(\vartheta_i) = \sigma(\Theta_i) = \sigma(\varsigma_0(\vartheta_i))$, we have $\varsigma(\vartheta_i) = \sigma(\varsigma_0(\vartheta_i))$ for $i = 1,\ldots, \n$.
As a consequence, $\{\sigma(\varsigma_0(\vartheta_i)) \trans[q_{i,h}] t_{i,h} \mid h \in H_i\}$ for $i = 1,\ldots, \n$, is provable from $D_{\Sigma}$ with a proof shorter than $\gamma$. 
Hence, by structural induction over each $\Theta_i = \varsigma_0(\vartheta_i)$, for each $i = 1,\ldots, \n$ there are a \dgsos ruloid 
\[
\SOSrule{\HH_{i}}{\{\varsigma_0(\vartheta_i) \trans[q_{i,h}] u_{i,h}\mid h \in H_i\}}
\]
 and a closed substitution $\sigma_i$ with 
\begin{enumerate}
\item $\sigma_{i}(\varsigma_0(\vartheta_i)) = \sigma(\varsigma_0(\zeta_i))$,
\item $\sigma_{i}(u_{i,h}) = t_{i,h}$, and
\item $D_{\Sigma} \vdash \sigma_{i}(\HH_{i})$.
\end{enumerate}
Consider a closed substitution $\sigma'$ with 
\begin{itemize}
\item $\sigma'(\zeta) = \sigma(\zeta)$ for all $\zeta \in \var{\Theta}$,
\item $\sigma'(\rhs(\HH_i)) = \sigma_{i}(\rhs(\HH_i))$ for all $i=1,\ldots, \n$
\end{itemize}
and let $\HH = \bigcup_{i = 1}^{\n} \HH_{i}$.
Moreover, let $\varsigma_1$ be a substitution with $\varsigma_1(\vartheta_i) = \Theta_i$ and $\varsigma_1(x_{i,j}) = u_{i,h}$ for some $h \in H_i$ accordingly to the reduced instance $\varsigma(r_{\D})$, for all $i=1,\ldots, \n$ and $j \in J_i$.
We recall that $\sigma_i(u_{i,\kappa(i)}) = t_{i,\kappa(i)} = t_{i,\f(\kappa)}$ for each $i = 1,\ldots, \n$ and we show that the \dgsos ruloid
\[
\SOSrule{\HH}{\{f(\Theta_1,\ldots, \Theta_{\n}) \trans[q_\kappa] f(u_{1,\kappa(1)},\ldots, u_{\n,\kappa(\n)}) \mid \kappa \in \bigtimes_{i=1}^{\n} H_i\}}
\]
together with the substitution $\sigma'$ satisfies the required properties:
\begin{enumerate}
\item First we prove that $\sigma'(\Theta) = \sigma(\Theta)$. 
This immediately follows from $\sigma'(\zeta) = \sigma(\zeta)$ for all $\zeta \in \var{\Theta}$.
\item Then we show that $D_{\Sigma} \vdash \sigma'(\HH)$, which is derived from the following considerations:
\begin{enumerate}
\item Notice that $\var{\Theta} = \bigcup_{i = 1}^{\n} \var{\Theta_i} = \bigcup_{i=1}^{\n} \var{\varsigma_0(\vartheta_i)}$.
Thus, since $\sigma$ and $\sigma'$ agree on $\var{\Theta}$ we obtain that $\sigma' (\varsigma_0(\vartheta_i)) = \sigma(\varsigma_0(\vartheta_i))$ for each $i = 1,\ldots,\n$.
Moreover, by construction we have that $\sigma_{i}(\varsigma_0(\vartheta_i)) = \sigma(\varsigma_0(\vartheta_i))$ for each $i =1,\ldots,\n$, thus giving $\sigma'(\varsigma_0(\vartheta_i)) = \sigma_{i}(\varsigma_0(\vartheta_i))$ for each $i = 1,\ldots,\n$.
Further, by definition $\sigma'$ and $\sigma_{i}$ agree on all variables in $\rhs(\HH_{i})$.
As by Lemma~\ref{lem:ruloid_variables}.\ref{lem:ruloid_variables_rhs}, $\rhs(\HH_i) = \bigcup_{h \in H_i} \var{u_{i,h}}$, we can conclude that $\sigma'$ and $\sigma_{i}$ agree on all variables occurring in $\ddedrule{\HH_{i}}{\{\varsigma_0(\vartheta_i) \trans[q_{i,h}] u_{i,h}\mid h \in H_i\}}$ for each $i = 1,\ldots, \n$.
\item As by the previous item we know that $\sigma'$ agrees with $\sigma_{i}$ on all variables in $\HH_{i}$ and $D_{\Sigma} \vdash \sigma_{i}(\HH_{i})$, we infer that $D_{\Sigma} \vdash \sigma'(\HH_i)$, for $i=1,\ldots, \n$.
Then, from $\HH = \bigcup_{i = 1}^{\n} \HH_i$, we can immediately conclude that $D_{\Sigma} \vdash \sigma'(\HH)$. 
\end{enumerate}
\item Finally, we prove that $\sigma'(f(u_{1,\kappa(1)},\ldots, u_{\n,\kappa(\n)})) = t_{\f(\kappa)}$ for each $\kappa \in \bigtimes_{i=1}^{\n} H_i$.
By Lemma~\ref{lem:ruloid_variables}.\ref{lem:ruloid_variables_rhs} we have that $\var{f(u_{1,\kappa(1)},\ldots, u_{\n,\kappa(\n)})} \subseteq \rhs(\HH)$.
In addition, we have
\begin{itemize}
\item $\var{u_{i,\kappa(i)}} \subseteq \rhs(\HH_i)$;
\item $\sigma'$ agrees with $\sigma_{i}$ on all variables in $\rhs(\HH_i)$, for all $i = 1,\ldots, \n$;
\item $\rhs(\HH) = \bigcup_{i = 1}^{\n} \rhs(\HH_i)$.
\end{itemize}
Therefore, we have that $\sigma'(u_{i,\kappa(i)}) = \sigma_i(u_{i,\kappa(i)}) = t_{i,\kappa(i)} = t_{i,\f(\kappa)}$ for each $i = 1,\ldots, \n$ and for each $\kappa \in \bigtimes_{i=1}^{\n} H_i$.
Hence, we can conclude that for each $\kappa \in \bigtimes_{i=1}^{\n}H_i$ we have $\sigma' \big( f(u_{1,\kappa(1)},\ldots, u_{\n,\kappa(\n)}) \big) = t_{\f(\kappa)}$.
\end{enumerate}

\item[($\Leftarrow$)]
We aim to show that $D_{\Sigma} \vdash \{\sigma(\Theta) \trans[q_m] t_m \mid m \in M \}$.
To this aim it is enough to show that $D_{\Sigma} \vdash \{\sigma'(\Theta) \trans[q_m] \sigma'(u_m) \mid m \in M\}$ which, since the closed terms $t_m$ are pairwise distinct by the hypothesis, by the choice of $\sigma'$ is equivalent to $D_{\Sigma} \vdash \{\sigma(\Theta) \trans[q_m] t_m \mid m \in M\}$.

Notice that by the choice of $\Theta$ we have that the open terms $u_m$ are of the form $u_m = f(u_{1,m},\dots,u_{\n,m})$ for some $\{u_{i,m} \mid i = 1,\dots,\n\} \subseteq \closedTerms$ for $m \in M$, so that $u_{i,m} \in \support(\Theta_i)$ for each $m \in M$.

Accordingly to Definition~\ref{def:distribution_ruloids}, let $r_{\D}$ and $\sigma_0$ be resp.\ the \dgsos rule and the substitution from which $\rho^{\D}$ is built, namely let $r_{\D}$ be of the form
\[
\SOSrule{\bigcup_{i = 1,\ldots,\n}\{ \vartheta_i \trans[q_{i,j}] x_{i,j} \mid j \in J_i
\}}
{\Big\{f(\vartheta_1, \ldots, \vartheta_{\n}) \trans[q_k] f(x_{1,k(1)}, \ldots, x_{\n, k(\n)}) \; \Big| \; 
 k \in \bigtimes_{i=1,\ldots,\n}J_i   \text{ and } q_k = \prod_{i =1,\ldots,\n} q_{i,k(i)} \Big\}}
\]
as in Definition~\ref{def:dgsos_rule}.\ref{def:dgsos_rule_f} and $\sigma_0$ be such that $\sigma_0(r_{\D})$ is of the form
\[
\SOSrule{\bigcup_{i = 1,\ldots,\n}\{ \Theta_i \trans[q_{i,h}] u_{i,h} \mid h \in H_i\}}
{\Big\{f(\Theta_1,\ldots, \Theta_{\n}) \trans[q_{\kappa}] f(u_{1,\kappa(1)},\ldots, u_{\n, \kappa(\n)}) \; \Big| \;
\kappa \in \bigtimes_{i=1,\ldots,\n} H_i  \text{ and } q_{\kappa} = \prod_{i =1,\ldots,\n} q_{i,\kappa(i)}\Big\}}
\]
(see Definition~\ref{def:reduced_instance}.\ref{def:reduced_instance_f}) and there is a bijection $\f \colon \bigtimes_{i = 1,\dots,\n} H_i \to M$ so that $u_{i,\kappa(i)} = u_{i,\f(\kappa)}$ for each $i = 1,\dots,\n$, and $q_{\kappa} = q_{\f(\kappa)}$ for each $\kappa \in \bigtimes_{i=1,\dots,\n} H_i$.

Then $\rho^{\D}$ is of the form 
\[
\rho^{\D} = \SOSrule{\bigcup_{i = 1,\dots,\n} \HH_i}{\{f(\Theta_1,\dots,\Theta_{\n}) \trans[q_m] u_m \mid m \in M\}}
\]
where for each $i = 1,\dots,\n$ we have that:
\begin{itemize}
\item Either $\sigma_0(\vartheta_i) = \Theta_i$ is a variable or a Dirac distribution and $\HH_i = \{\Theta_i \trans[q_i,h] u_{i,h} \mid h \in H_i\}$.
Hence from $D_{\Sigma} \vdash \sigma'(\HH)$ we can immediately infer that $D_{\Sigma} \vdash \sigma'(\{\Theta_i \trans[q_i,h] u_{i,h} \mid h \in H_i\})$.
\item Or there is a \dgsos ruloid $\rho^{\D}_i = \ddedrule{\HH_i}{\{\sigma_0(\vartheta) \trans[q_{i,h}] u_{i,h} \mid h \in H_i\}}$.
Since $D_{\Sigma} \vdash \sigma'(\HH)$ implies $D_{\Sigma} \vdash \sigma'(\HH_i)$, by structural induction on $\Theta_i$ we can infer that $D_{\Sigma} \vdash \sigma'(\{\Theta_i \trans[q_{i,h}] u_{i,h} \mid h \in H_i\})$.
\end{itemize}
Hence, we have obtained that the closed substitution instances wrt.\ $\sigma' \circ \sigma_0$ of the premises of $r_{\D}$ are provable from $D_{\Sigma}$ and therefore we can infer that there is a proof from $D_{\Sigma}$ of $\{\sigma'(\Theta) \trans[q_m] \sigma'(u_m) \mid m \in M\}$.
By the choice of $\sigma'$, we can conclude that $D_{\Sigma} \vdash \{\sigma(\Theta) \trans[q_m] t_m \mid m \in M\}$.
\end{description}

\item Inductive step $\Theta = \sum_{i \in I} p_i \Theta_i$ for some $\Theta_i \in \openDTerms$, $p_i \in [0,1]$ for $i \in I$ and $\sum_{i \in I} p_i = 1$.

  \begin{description}
    \item[($\Rightarrow$)]
First of all, we recall that by Theorem~\ref{prop:proof_sigma_q} $D_{\Sigma} \vdash \{\sigma(\Theta)\trans[q_m] t_m \mid m \in M \}$ if{f} $\sigma(\Theta)(t_m) = q_m$ and $\sum_{m \in M} q_m = 1$.
Thus, for the particular choice of $\sigma(\Theta)$, we have that the closed terms $t_m$ are such that $\{t_{m} \mid m \in M\} = \bigcup_{i \in I}\support(\sigma(\Theta_i))$.
Next, let us consider a closed proof $\gamma$ of $\{\sigma(\Theta)\trans[q_m] t_m \mid m \in M \}$ from $D_{\Sigma}$.
The bottom of $\gamma$ is constituted by the closed reduced instance of a \dgsos rule $r_{\D} \in R_{\Sigma}$ of the form
\[
\SOSrule{\bigcup_{i \in I}\{ \vartheta_i \trans[q_{i,j}] x_{i,j} \mid j \in J_i\}}
{\Big\{ \sum_{i \in I} p_i \vartheta_i \trans[q_x] x \, \big| \;
 x \in \{x_{i,j} \mid i \in I \wedge j \in J_i\} \text{ and }
q_x = \sum_{i \in I, j \in J_i \text{ s.t. } x_{i,j} = x} p_i q_{i,j} \Big\}}
\] 
wrt.\ a closed substitution $\varsigma$ with $\varsigma(\vartheta_i) = \sigma(\Theta_i)$ for $i \in I$.
More precisely, let $\varsigma(r_{\D})$ be the inference rule of the form
\[
\SOSrule{\bigcup_{i \in I}\{ \sigma(\Theta_i) \trans[q_{i,h}] t_{i,h} \mid h \in H_i\}}
{\Big\{ \sum_{i \in I} p_i \sigma(\Theta_i) \trans[q_u] u \; \Big| \;
u \in \{t_{i,h} \mid i \in I \wedge h \in H_i\}  \text{ and }
q_u = \sum_{i \in I, h \in H_i \text{ s.t. } t_{i,h} = u} p_i q_{i,h}\Big\}}
\] 
where 
\begin{itemize}
\item each set $\{ \sigma(\Theta_i) \trans[q_{i,h}] t_{i,h} \mid h \in H_i\}$ is the reduction wrt.\ $\sigma$ of the corresponding set $\{ \varsigma(\vartheta_i) \trans[q_{i,j}] \varsigma(x_{i,j}) \mid j \in J_i\}$,
\item there is bijection $\f \colon \{t_{i,h} \mid h \in H_i, i \in I\} \to M$ so that $u = t_{\f(u)}$ for each $u \in \{t_{i,h} \mid h \in H_i, i \in I\}$ and 
\item for each $i \in I$ there is a proof shorter than $\gamma$ of $\{\sigma(\Theta_i) \trans[q_{i,h}] t_{i,h} \mid h \in H_i\}$ from $D_{\Sigma}$.
\end{itemize}
Let $\varsigma_0$ be a substitution with $\varsigma_0(\vartheta_i) = \Theta_i$ for each $i \in I$.
Considering that $\varsigma(\vartheta_i) = \sigma(\Theta_i) = \sigma(\varsigma_0(\vartheta_i))$, we have $\varsigma(\vartheta_i) = \sigma(\varsigma_0(\vartheta_i))$ for each $i \in I$.
As a consequence, $\{\sigma(\varsigma_0(\vartheta_i)) \trans[q_{i,h}] t_{i,h} \mid h \in H_i\}$ for each $i \in I$, is provable from $D_{\Sigma}$ with a proof shorter than $\gamma$. 
Hence, by structural induction over each $\Theta_i = \varsigma_0(\vartheta_i)$, for each $i \in I$ there are a \dgsos ruloid 
\[
\SOSrule{\HH_{i}}{\{\varsigma_0(\vartheta_i) \trans[q_{i,h}] u_{i,h}\mid h \in H_i\}}
\]
 and a closed substitution $\sigma_i$ with 
\begin{enumerate}
\item $\sigma_{i}(\varsigma_0(\vartheta_i)) = \sigma(\varsigma_0(\vartheta_i))$,
\item $\sigma_{i}(u_{i,h}) = t_{i,h}$, and
\item $D_{\Sigma} \vdash \sigma_{i}(\HH_{i})$.
\end{enumerate}
So let us consider a closed substitution $\sigma'$ with 
\begin{itemize}
\item $\sigma'(\zeta) = \sigma(\zeta)$ for all $\zeta \in \var{\Theta}$,
\item $\sigma'(\rhs(\HH_i)) = \sigma_{i}(\rhs(\HH_i))$ for all $i \in I$
\end{itemize}
and let $\HH = \bigcup_{i \in I} \HH_{i}$.
Moreover, let $\varsigma_1$ be a substitution with $\varsigma_1(\vartheta_i) = \Theta_i$ and $\varsigma_1(x_{i,j}) = u_{i,h}$ for some $h \in H_i$ accordingly to the reduced instance $\varsigma(r_{\D})$, for all $j \in J_i, i\in I$.
We recall that $\sigma_i(u_{i,h}) = t_{i,h}$ for each $h \in H_i, i\in I$.
and we prove that the \dgsos ruloid
\[
\SOSrule{\HH}{\Big\{\sum_{i \in I}p_i \Theta_i \trans[q_u] u \mid u \in \{t_{i,h} \mid h \in H_i, i \in I\} \Big\}}
\]
together with the substitution $\sigma'$ satisfies the required properties:
\begin{enumerate}
\item First we prove that $\sigma'(\Theta) = \sigma(\Theta)$. 
This immediately follows from $\sigma'(\zeta) = \sigma(\zeta)$ for all $\zeta \in \var{\Theta}$.
\item Then we prove that $D_{\Sigma} \vdash \sigma'(\HH)$, which is derived from the following considerations:
\begin{enumerate}
\item Notice that $\var{\Theta} = \bigcup_{i \in I} \var{\Theta_i} = \bigcup_{i \in I} \var{\varsigma_0(\vartheta_i)}$.
Thus, since $\sigma$ and $\sigma'$ agree on $\var{\Theta}$ we obtain that $\sigma' (\varsigma_0(\vartheta_i)) = \sigma(\varsigma_0(\vartheta_i))$ for each $i \in I$.
Moreover, by construction we have that $\sigma_{i}(\varsigma_0(\vartheta_i)) = \sigma(\varsigma_0(\vartheta_i))$ for each $i \in I$, thus giving $\sigma'(\varsigma_0(\vartheta_i)) = \sigma_{i}(\varsigma_0(\vartheta_i))$ for each $i \in I$.
Furthermore, by definition $\sigma'$ and $\sigma_{i}$ agree on all variables in $\rhs(\HH_{i})$.
As by Lemma~\ref{lem:ruloid_variables}.\ref{lem:ruloid_variables_rhs}, $\rhs(\HH_i) = \bigcup_{h \in H_i} \var{u_{i,h}}$, we can conclude that $\sigma'$ and $\sigma_{i}$ agree on all variables occurring in $\ddedrule{\HH_{i}}{\{\varsigma_0(\vartheta_i) \trans[q_{i,h}] u_{i,h}\mid h \in H_i\}}$ for each $i \in I$.
\item As by the previous item $\sigma'$ agrees with $\sigma_{i}$ on all variables in $\HH_{i}$ and $D_{\Sigma} \vdash \sigma_{i}(\HH_{i})$, we infer $D_{\Sigma} \vdash \sigma'(\HH_i)$, for each $i \in I$.
Then, from $\HH = \bigcup_{i \in I} \HH_i$, we can immediately conclude that $D_{\Sigma} \vdash \sigma'(\HH)$. 
\end{enumerate}
\item Finally, we prove that $\sigma'(u) = t_{\f(u)}$ for each $u \in \{t_{i,h} \mid h \in H_i, i \in I\}$.
 By Lemma~\ref{lem:ruloid_variables}.\ref{lem:ruloid_variables_rhs} we have that $\var{u} \subseteq \rhs(\HH)$.
Furthermore, we have that
\begin{itemize}
\item $\var{u_{i,h}} \subseteq \rhs(\HH_i)$;
\item $\sigma'$ agrees with $\sigma_{i}$ on all variables in $\rhs(\HH_i)$, for all $i \in I$;
\item $\rhs(\HH) = \bigcup_{i \in I} \rhs(\HH_i)$.
\end{itemize}
Therefore, we have that $\sigma'(u_{i,h}) = \sigma_i(u_{i,h}) = t_{i,h}$ for each $h \in H_i, i \in I$.
Hence, we can conclude that $\sigma' (u) = t_{\f(u)}$ for each $u \in \{t_{i,h} \mid h \in H_i, i \in I\}$.
\end{enumerate}

\item[($\Leftarrow$)]
We aim to show that $D_{\Sigma} \vdash \{\sigma(\Theta) \trans[q_m] t_m \mid m \in M \}$.
To this aim it is enough to show that $D_{\Sigma} \vdash \{\sigma'(\Theta) \trans[q_m] \sigma'(u_m) \mid m \in M\}$ which, since the closed terms $t_m$ are pairwise distinct by the hypothesis, by the choice of $\sigma'$ is equivalent to $D_{\Sigma} \vdash \{\sigma(\Theta) \trans[q_m] t_m \mid m \in M\}$.

Accordingly to Definition~\ref{def:distribution_ruloids}, let $r_{\D}$ and $\sigma_0$ be resp.\ the \dgsos rule and the substitution from which $\rho^{\D}$ is built, namely let $r_{\D}$ be of the form
\[
\SOSrule{\bigcup_{i \in}\{ \vartheta_i \trans[q_{i,j}] x_{i,j} \mid j \in J_i\}}
{\Big\{\sum_{i \in I} p_i \vartheta_i \trans[q_x] x \; \Big| \; 
 x \in \{x_{i,j} \mid i \in I \wedge j \in J_i\}   \text{ and } q_x = \sum_{i \in I, j \in J_i \text{ s.t. } x_{i,j} = x} p_i \cdot q_{i,j} \Big\}}
\]
as in Definition~\ref{def:dgsos_rule}.\ref{def:dgsos_rule_convex} and $\sigma_0$ be such that $\sigma_0(r_{\D})$ is of the form
\[
\SOSrule{\bigcup_{i \in}\{ \Theta_i \trans[q_{i,h}] u_{i,h} \mid h \in H_i\}}
{\Big\{\sum_{i \in I} p_i \Theta_i \trans[q_m] u_m \; \Big| \; 
 u_m \in \{u_{i,h} \mid i \in I \wedge h \in H_i\}   \text{ and } q_m = \sum_{i \in I, h \in H_i \text{ s.t. } u_{i,h} = u_m} p_i \cdot q_{i,h} \Big\}}
\]
(see Definition~\ref{def:reduced_instance}.\ref{def:reduced_instance_Sum}).
Then $\rho^{\D}$ is of the form 
\[
\rho^{\D} = \SOSrule{\bigcup_{i \in I} \HH_i}{\{ \sum_{i \in I} p_i \Theta_i \trans[q_m] u_m \mid m \in M\}}
\]
where for each $i \in I$ we have that:
\begin{itemize}
\item Either $\sigma_0(\vartheta_i) = \Theta_i$ is a variable or a Dirac distribution and $\HH_i = \{\Theta_i \trans[q_i,h] u_{i,h} \mid h \in H_i\}$.
Hence from $D_{\Sigma} \vdash \sigma'(\HH)$ we can immediately infer that $D_{\Sigma} \vdash \sigma'(\{\Theta_i \trans[q_i,h] u_{i,h} \mid h \in H_i\})$.
\item Or there is a \dgsos ruloid $\rho^{\D}_i = \ddedrule{\HH_i}{\{\sigma_0(\vartheta) \trans[q_{i,h}] u_{i,h} \mid h \in H_i\}}$.
Since $D_{\Sigma} \vdash \sigma'(\HH)$ implies $D_{\Sigma} \vdash \sigma'(\HH_i)$, by structural induction on $\Theta_i$ we can infer that $D_{\Sigma} \vdash \sigma'(\{\Theta_i \trans[q_{i,h}] u_{i,h} \mid h \in H_i\})$.
\end{itemize}
Hence, we have obtained that the closed substitution instances wrt.\ $\sigma' \circ \sigma_0$ of the premises of $r_{\D}$ are provable from $D_{\Sigma}$ and therefore we can infer that there is a proof from $D_{\Sigma}$ of $\{\sigma'(\Theta) \trans[q_m] \sigma'(u_m) \mid m \in M\}$.
By the choice of $\sigma'$, we can conclude that $D_{\Sigma} \vdash \{\sigma(\Theta) \trans[q_m] t_m \mid m \in M\}$. \qedhere
\end{description}
\end{enumerate}
\end{proof}


\begin{exa}
\label{ex:distribution_ruloid_theorem}
Consider the distribution term $\Theta = \frac{2}{5} \mu + \frac{3}{5} (\nu | \upsilon)$ and the closed substitution $\sigma$ with $
\sigma(\Theta) = \frac{2}{5}( \frac{1}{4} \delta_{t_1} + \frac{3}{4} \delta_{t_2} ) + \frac{3}{5} \big( (\frac{1}{3} \delta_{t_3} + \frac{2}{3} \delta_{t_4}) \mid \delta_{t_5} \big)$.
Notice that $\sigma(\Theta)$ is the source term of the distribution over terms $L$ in Example~\ref{ex:proof}.
Thus, we know that 
\[
D_{\Sigma} \vdash \{\sigma(\Theta) \trans[1/10] t_1,\; \sigma(\Theta) \trans[3/10] t_2,\; \sigma(\Theta) \trans[1/5] t_3 | t_5,\; \sigma(\Theta) \trans[2/5] t_4 | t_5\}.
\]
Consider the $\Sigma$-distribution ruloid $\rho^{\D}$ for $\Theta$ given in Example~\ref{ex:distribution_ruloid} 
\[
\SOSrule{\{\mu \trans[1/4] x_1 \quad \mu \trans[3/4] x_2\} \qquad \{\nu \trans[1/3] y_1, \quad \nu \trans[2/3] y_2\} \qquad \{\upsilon \trans[1] z\}}
{\Big\{\frac{2}{5} \mu + \frac{3}{5}(\nu|\upsilon) \trans[\frac{1}{10}] x_1,
\frac{2}{5} \mu + \frac{3}{5}(\nu|\upsilon) \trans[\frac{3}{10}] x_2, 
\frac{2}{5} \mu + \frac{3}{5}(\nu|\upsilon) \trans[\frac{1}{5}] y_1|z, 
\frac{2}{5} \mu + \frac{3}{5}(\nu|\upsilon) \trans[\frac{2}{5}] y_2|z \Big\}}.
\]
We want to exhibit a proper closed substitution $\sigma'$ such that $\rho^{\D}$ and $\sigma'$ satisfy Theorem~\ref{thm:distribution_ruloid_theorem} wrt.\ $\sigma(\Theta)$.
Let
\[
\begin{array}{c}
\sigma'(x_1) = t_1 \qquad \sigma'(x_2) = t_2 \qquad \sigma'(y_1) = t_3 \qquad \sigma'(y_2) = t_4 \qquad \sigma'(z) = t_5 \\[.7ex]
\sigma'(\mu) = \frac{1}{4} \delta_{t_1} + \frac{3}{4} \delta_{t_2} \qquad\qquad 
\sigma'(\nu) = \frac{1}{3} \delta_{t_3} + \frac{2}{3} \delta_{t_4} \qquad\qquad
\sigma'(\upsilon) = \delta_{t_5}.
\end{array}
\]
Then we have 
\[
\sigma'(\Theta) 
= \frac{2}{5} \sigma'(\mu) + \frac{3}{5} \sigma'(\nu | \upsilon) 
= \frac{2}{5} ( \frac{1}{4} \delta_{t_1} + \frac{3}{4} \delta_{t_2} ) + \frac{3}{5} \Big( (\frac{1}{3} \delta_{t_3} + \frac{2}{3} \delta_{t_4})\, |\, \delta_{t_5} \Big). 
\]
Moreover
\[
\sigma'(y_1 | z) = t_3 | t_5 \qquad \qquad \sigma'(y_2 | z) = t_4 | t_5
\]
thus giving that $\sigma'(\trg{\rho^{\D}}) = \rhs(L)$.
Finally, we remark that
\begin{itemize}
\item the proof presented for $\{\sigma(\mu_2) \trans[1/4] t_1,\, \sigma(\mu_2) \trans[3/4] t_2\}$ with $\sigma(\mu_2) = \frac{1}{4} \delta_{t_1} + \frac{3}{4} \delta_{t_2}$ in Example~\ref{ex:proof} gives us $D_{\Sigma} \vdash \{\sigma'(\mu) \trans[1/4] t_1,\, \sigma'(\mu) \trans[3/4] t_2\}$;
\item the proof presented for $\{\sigma(\mu_1) \trans[1/3] t_3,\, \sigma(\mu_1) \trans[2/3] t_4\}$ with $\sigma(\mu_1) = \frac{1}{3} \delta_{t_3} + \frac{2}{3} \delta_{t_4}$ in Example~\ref{ex:proof} gives us $D_{\Sigma} \vdash \{\sigma'(\nu) \trans[1/3] t_3,\, \sigma'(\nu) \trans[2/3] t_4\}$;
\item the proof presented for $\{\sigma(\nu_1) \trans[1] t_5\}$ with $\sigma(\nu_1) = \delta_{t_5}$ in Example~\ref{ex:proof} gives us $D_{\Sigma} \vdash \{\sigma'(\upsilon) \trans[1] t_5\}$.
\end{itemize}
We have therefore obtained that $D_{\Sigma} \vdash \sigma'(\prem{\rho^{\D}})$ and thus that $\rho^{\D}$ and $\sigma'$ satisfy Theorem~\ref{thm:distribution_ruloid_theorem} wrt.\ $\sigma(\Theta)$.
\end{exa}


\subsection{Related work}

The only paper dealing with ruloids for specifications of probabilistic process calculi is \cite{GF12}.
As previously outlined, \cite{GF12} deals with reactive transition systems, which are less expressive than PTSs as they
do not admit internal nondeterminism.
Transitions are of the form $t \trans[a,p] t'$, denoting that $t$ evolves by $a$ to $t'$ with probability $p$.
Informally, our $P$-ruloids generalize those in \cite{GF12} in the same way PTSSs generalize reactive systems.
In fact, to deal with the quadruple $t \trans[a,p] t'$, ruloids in \cite{GF12} are defined by keeping track of rules and ruloids used in their construction, in order to assign a proper probability weight to their conclusion.
In detail, to guarantee the property of semi-stochasticity, i.e.\ the sum of the probabilities of all transitions for an action from a term is either $0$ or $1$, a partitioning over ruloids is needed in \cite{GF12}: given a term $t$ the ruloids in the partition for $t$ related to action $a$ allow one to derive $a$-labeled transitions from $t$ whose total probability is $1$.
To do so, one also has to constantly keep track of the rules and ruloids used in the construction of the ruloids in a partition, because the exact probability weight of a transition depends on this construction.
An analogous technicality was already necessary in the SOS transition rules in \cite{LT09}, on which \cite{GF12} builds.

Here we do not need this technicality, since probabilities are directly managed by $\Sigma$-distribution ruloids and we can use $P$-ruloids to derive the transitions leading to probability distributions.
More precisely, we should say that given a term $t$, all ruloids in one partition for $t$ of \cite{GF12} are captured by one of our $P$-ruloids and one $\Sigma$-distribution ruloid.
The $P$-ruloid captures all the requirements that the subterms of $t$ must satisfy to derive the transition to the desired probability distribution over terms.
The proper probability weights are then automatically assigned to terms by the $\Sigma$-distribution ruloid, without the need of keeping track of all the rules and ruloids used in the construction.


\section{The decomposition method}
\label{sec:decompose_decomposition}
In this section we present our method for decomposing formulae in $\logic$, $\logicready$ and $\logicplus$.
To this purpose we exploit the two classes of ruloids introduced in Section~\ref{sec:ruloids}.
In fact, the idea behind the decomposition of state (resp.\ distribution) formulae is to establish which properties the closed substitution instances of the variables occurring in a (distribution) term must satisfy to guarantee that the closed substitution instance of that (distribution) term satisfies the chosen state (resp.\ distribution) formula.
Thus, since ($\Sigma$-distribution) ruloids derive the behavior of a (distribution) term directly from the behavior of the variables occurring in it, the decomposition method is firmly related to them.

Formally, starting from the class $\logic$, the decomposition of state formulae follows those in 
\cite{BFvG04,FvG16,FvGL17,FvGW06,FvGW12,GF12} and consists in assigning to each term $t \in \openT$ and formula $\varphi \in \logicstate$, a set of functions $\xi \colon \SVar \to \logicstate$, called \emph{decomposition mappings}, assigning to each variable $x$ in $t$ a proper formula in $\logicstate$ such that for any closed substitution $\sigma$ it holds that $\sigma(t) \models \varphi$ if{f} $\sigma(x) \models \xi(x)$ for each $x \in \var{t}$ (Theorem~\ref{thm:decomposition}).
Each mapping $\xi$ will be defined on a $P$-ruloid having $t$ as source, $P$ being the considered PGSOS-PTSS.
Similarly, the decomposition of distribution formulae consists in assigning to each distribution term $\Theta \in \openDTerms$ and distribution formula $\psi \in \logicdist$ a set of decomposition mappings $\eta \colon \Var \to \logicdist \cup \logicstate$ such that for any closed substitution $\sigma$ we get that $\sigma(\Theta) \models \psi$ if{f} $\sigma(\zeta) \models \eta(\zeta)$ for each $\zeta \in \var{\Theta}$ (Theorem~\ref{thm:decomposition}).
Each mapping $\eta$ will be defined on a $\Sigma$-distribution ruloid having $\Theta$ as source.

Then, as $\logicready$ and $\logicplus$ are subclasses of $\logic$, we will show how we can easily derive the decomposition method for them from the one proposed for $\logic$ (Theorem~\ref{cor:decomposition}).


\subsection{Decomposition of formulae in \texorpdfstring{$\logic$}{L}}
In this section we consider the logic $\logic$.
First we need to introduce the notion of matching for a distribution over terms and a distribution formula, seen as a probability distribution over state formulae \cite{CGT16a,DD11}.

\begin{defi}
[Matching]
Assume a distribution over terms $L=\{\Theta \trans[q_m] t_m \mid m \in M\}$ and a distribution formula $\psi = \bigoplus_{i \in I} r_i \varphi_i \in \logicdist$.
Then a \emph{matching} for $L$ and $\psi$ is a distribution over the product space $\w \in \ProbDist{\openT \times \logicstate}$ having $L$ and $\psi$ as left and right marginals respectively, that is $\sum_{i \in I} \w(t_m, \varphi_i) = q_m$ for all $m \in M$ and $\sum_{m \in M} \w(t_m, \varphi_i) = r_i$ for all $i \in I$.  
We denote by $\W(L,\psi)$ the set of all matchings for $L$ and $\psi$.
\end{defi}

\begin{defi}
[Decomposition of formulae in $\logic$]
\label{def:decomposition}
Let $P = (\Sigma, \Act, R)$ be a PGSOS-PTSS and let $D_{\Sigma}$ be the $\Sigma$-DS.
We define the mappings 
\begin{itemize}
\item
$\cdot^{-1}\colon \openT \to (\logicstate \to \powset{\SVar \to \logicstate})$, and 
\item
$\cdot^{-1} \colon \openDTerms \to (\logicdist \to \powset{\Var \to \logic})$ 
\end{itemize}
as follows.
For each term $t \in \openT$ and state formula $\varphi \in \logicstate$, $t^{-1}(\varphi) \in \powset{\SVar \to \logicstate}$ is the set of \emph{decomposition mappings} $\xi \colon \SVar \to \logicstate$ such that for any univariate term $t$ we have:
\begin{enumerate}
\item \label{def:decomposition_top}
$\xi \in t^{-1}(\top)$ if{f} $\xi(x) = \top$ for all $x \in \SVar$;
\item \label{def:decomposition_neg}
$\xi \in t^{-1}(\neg \varphi)$ if{f} there is a function $\f \colon t^{-1}(\varphi) \to \var{t}$ such that
\[
\xi(x) = 
\begin{cases}
\displaystyle \bigwedge_{\xi' \in \f^{-1}(x)} \neg \xi'(x) & \text{if } x \in \var{t}  \\
\top  & \text{otherwise;} 
\end{cases}
\]
\item \label{def:decomposition_wedge}
$\xi \in t^{-1}(\bigwedge_{j \in J} \varphi_j)$ if{f} there exist decomposition mappings $\xi_j \in t^{-1}(\varphi_j)$ for all $j \in J$ such that
\[
\xi(x) = \bigwedge_{j \in J} \xi_j(x) \text{ for all } x \in \SVar \text{;}
\]
\item \label{def:decomposition_diam}
$\xi \in t^{-1}(\diam{a}\psi)$ if{f} there exist a $P$-ruloid $\dedrule{\HH}{t \trans[a] \Theta}$ and a decomposition mapping $\eta \in \Theta^{-1}(\psi)$ such that:
\[
\xi(x) =
\begin{cases} \displaystyle \bigwedge_{x \trans[b] \mu \in \HH} \diam{b} \eta(\mu) \quad \wedge \quad  \bigwedge_{x \ntrans[c] \in \HH} \neg \diam{c} \top \quad \wedge \quad \eta(x)  & \text{ if } x \in \var{t} \\[1.5 ex]
 \top  & \text{ otherwise;} 
\end{cases}
\]
\item \label{def:decomposition_multi}
$\xi \in (\sigma(t))^{-1}(\varphi)$ for a non injective substitution $\sigma \colon \var{t} \to \SVar$ if{f} there is a decomposition mapping $\xi' \in t^{-1}(\varphi)$ such that
\[
\xi(x) = \begin{cases} \displaystyle 
\bigwedge_{y \in \sigma^{-1}(x)} \xi'(y) \quad & \text{ if } x \in \var{t}\\
\top & \text{ otherwise.}
\end{cases}
\]
\end{enumerate}
Then, for each distribution term $\Theta \in \openDTerms$ and distribution formula $\psi \in \logicdist$, $\Theta^{-1}(\psi) \in \powset{\Var \to \logic}$ is the set of \emph{decomposition mappings} $\eta \colon \Var \to \logic$ such that for any univariate distribution term $\Theta$ we have:
\begin{enumerate}
\setcounter{enumi}{5}
\item  \label{def:decomposition_oplus}
$\eta \in \Theta^{-1}(\bigoplus_{i \in I} r_i \varphi_i)$ if{f} there are a $\Sigma$-distribution ruloid $\ddedrule{\HH}{\{ \Theta \trans[q_{m}] t_{m} \mid m \in M \}}$ and a matching $\w  \in \W(\{ \Theta \trans[q_{m}] t_{m} \mid m \in M \}, \bigoplus_{i \in I} r_i \varphi_i)$ such that for all $m \in M$ and $i \in I$ there is a decomposition mapping $\xi_{m,i}$ with 
$
\begin{cases}
\xi_{m,i} \in t_{m}^{-1}(\varphi_i) & \text{if }\w(t_m, \varphi_i) > 0 \\ 
\xi_{m,i} \in t_{m}^{-1}(\top) & \text{otherwise}
\end{cases}
$
and we have
\begin{enumerate}
\item 
\label{def:decomposition_oplus_mu}
for all $\mu \in \DVar$, 
$\eta(\mu) = 
\begin{cases}
\displaystyle 
\bigoplus_{\mu \trans[q_j] x_j 
 \in \HH} q_j \bigwedge_{i \in I \atop m \in M} \xi_{m,i}(x_j) &  \text{ if } \mu \in \var{\Theta}\\[1.0 ex]
1 \top & \text{ otherwise}\\[2.0 ex]
\end{cases}
$
\item 
\label{def:decomposition_oplus_delta}
for all $x \in \SVar$,
$
\eta(x) = 
\begin{cases}
\displaystyle 
\bigwedge_{i \in I \atop m \in M} \xi_{m,i}(x) &  \text{ if } x \in \var{\Theta}\\[1.0 ex]
\top & \text{ otherwise.}\\
\end{cases}
$
\end{enumerate}
\item \label{def:decomposition_oplus_multi}
$\eta \in (\sigma(\Theta))^{-1}(\psi)$ for a non injective substitution $\sigma \colon \var{\Theta} \to \Var$ if{f} there is a decomposition mapping $\eta' \in \Theta^{-1}(\psi)$ such that
for all $\zeta \in \var{\sigma(\Theta)}$ it holds that
$\eta'(z) = \eta'(z')$ for all $z,z' \in \sigma^{-1}(\zeta)$ and
 \[
\eta(\zeta) = 
\begin{cases}
\eta'(\tilde{z}) & \text{if } \zeta \in \var{\sigma(\Theta)} \text{ and } \tilde{z} \in \sigma^{-1}(\zeta)\\  
\top & \text{if } \zeta \not\in \var{\sigma(\Theta)}.
\end{cases}
\]
\end{enumerate}
\end{defi}

We explain our decomposition method for the diamond modality for state formulae (Definition~\ref{def:decomposition}.\ref{def:decomposition_diam}) and for distribution formulae (Definition~\ref{def:decomposition}.\ref{def:decomposition_oplus}).
For the other modalities on state formulae, which do not directly involve the quantitative properties of processes, we refer  to \cite{FvGW06}.

We discuss first the decomposition of a state formula $\varphi = \diam{a}\psi \in \logicstate$.
Given any term $t \in \openT$ and closed substitution $\sigma$, we need to identify in $\xi \in t^{-1}(\varphi)$ which properties each $\sigma(x)$ with $x \in \var{t}$ has to satisfy in order to guarantee $\sigma(t) \models \varphi$.
By Definition~\ref{def:satisfiability} we have that $\sigma(t) \models \varphi$ if and only if $P \vdash \sigma(t) \trans[a] \pi$ for some probability distribution $\pi$ such that $\pi \models \psi$.
By Theorem~\ref{thm:term_ruloid_theorem} there is such a transition if and only if there are a $P$-ruloid $\HH / t \trans[a] \Theta$ and a closed substitution $\sigma'$ with $\sigma'(t) = \sigma(t)$ and
\begin{enumerate*}[label=(\emph{\roman*})]
\item \label{item:condition_i} $P \vdash \sigma'(\HH)$ and
\item \label{item:condition_ii} $\sigma'(\Theta) \models \psi$.
\end{enumerate*}
The validity of condition~\eqref{item:condition_i} follows if, for each $x \in \var{t}$, the literals in $\HH$ having $x$ as left hand side test only the provable behavior of $\sigma'(x)$. 
More precisely, we need that $\sigma'(x) \models \neg\diam{c}\top$ for each $x \ntrans[c] \in \HH$ and that $\sigma'(x) \models \diam{b} \eta(\mu)$ for each $x \trans[b] \mu \in \HH$, for a chosen decomposition mapping $\eta \in \Theta^{-1}(\psi)$ with $\sigma'(\mu) \models \eta(\mu)$ for each $\mu \in \var{\Theta}$.
The decomposed formula $\xi(x)$ is then defined as the conjunction of such formulae.
Moreover, we also add in $\xi(x)$ a conjunct $\eta(x)$ to capture the potential behavior of $x$ as a subterm of the target term $\Theta$.
Further, the choice of $\eta$ and its use in $\xi$ also guarantees that condition~\eqref{item:condition_ii} holds.

We discuss now the decomposition of a distribution formula $\psi = \bigoplus_{i \in I} r_i \varphi_i \in \logicdist$. 
Given any distribution term $\Theta \in \openDTerms$ and a closed substitution $\sigma$, we need to identify in $\eta \in \Theta^{-1}(\psi)$ which properties each $\sigma(\zeta)$ with $\zeta \in \var{\Theta}$ has to satisfy in order to guarantee $\sigma(\Theta) \models \psi$.
By Definition~\ref{def:satisfiability} we have that $\sigma(\Theta) \models \psi$ if and only if $\sigma(\Theta) = \sum_{i \in I} r_i \pi_i$ with $t \models \varphi_i$ for all $t \in \support(\pi_i)$.
Assume $\support(\sigma(\Theta)) = \{t_m \mid m \in M\}$ and $\sigma(\Theta)(t_m) = q_m$.
By Theorem~\ref{prop:proof_sigma_q}, this is equivalent to have $D_{\Sigma} \vdash \{\sigma(\Theta) \trans[q_m] t_m \mid m \in \M\}$ which, by Theorem~\ref{thm:distribution_ruloid_theorem}, is equivalent to say that there are a $\Sigma$-distribution ruloid $\HH/\{\Theta \trans[q_m] u_m \mid m \in M\}$ and a closed substitution $\sigma'$ with $\sigma'(\Theta) = \sigma(\Theta)$ and 
\begin{enumerate*}[label=(\emph{\roman*})]
\item \label{spiegaprob3}
$D_{\Sigma} \vdash \sigma'(\HH)$ and
\item \label{spiegaprob2}
$\sigma'(u_m) \models \varphi_i$ whenever $\sigma'(u_m) \in \support(\pi_i)$. 
\end{enumerate*}
Since the weights $q_m$ are univocally determined by the distributions over terms in $\HH$ and moreover they already represent the exact probability weights of $\sigma(\Theta)$, we define, for each $\mu \in \var{\Theta} \cap \DVar$, the decomposition mapping $\eta(\mu)$ using as weights the $q_j$ in the distributions over terms $\{\mu \trans[q_j] x_j \} \in \HH$. 
Then, to guarantee condition~\eqref{spiegaprob2}, we define $\w(u_m, \varphi_i)$ to be positive if $\sigma'(u_m) \in \support(\pi_i)$ so that we can assign the proper decomposed formula $\xi_{m,i}(x)$ to each $x \in \var{u_m}$ such that $\sigma'(x) \models \xi_{m,i}(x)$.
Moreover, since each $\sigma'(u_m)$ may occur in the support of more than one $\pi_i$, we impose that each $x \in \var{u_m}$ satisfies the conjunction of all the decomposed formulae $\xi_{m,i}(x)$.
Therefore, also condition~\eqref{spiegaprob3} follows.

\begin{exa}
We exemplify two decomposition mappings in the set $t^{-1}(\varphi)$ for term $t= x +_{2/5} (y|z)$, which is the term considered in Example~\ref{ex:state_ruloid} with $p = 2/5$, and 
the formula $\varphi = \diam{a} \psi$, with $\psi= \frac{1}{2} \diam{a} \top \oplus \frac{1}{2} \neg \diam{a} \top$.
As this example is aimed at providing a deeper insight on the mechanism of our decomposition method, we will choose arbitrarily the ruloids and the matchings for the considered terms and formulae in order to minimize the number of mappings involved in the decomposition and improve readability.
Let $\rho$ be the last ruloid for $t$ in Example~\ref{ex:state_ruloid}, $\Theta = \frac{2}{5}\mu + \frac{3}{5}(\nu | \upsilon)$ denote its target, and $\rho^{\D}$ be the $\Sigma$-distribution ruloid for $\Theta$ showed in Example~\ref{ex:distribution_ruloid}.
By Definition~\ref{def:decomposition}.\ref{def:decomposition_diam}, the decomposition mappings $\xi \in t^{-1}(\varphi)$ built over $\rho$ are such that:
\begin{equation}
\label{eq:ex_decomposition}
\xi(x) = \diam{a}\eta(\mu) \qquad\qquad \xi(y) = \diam{a}\eta(\nu) \qquad\qquad \xi(z)= \diam{a}\eta(\upsilon)
\end{equation}
where $\eta \in \Theta^{-1}(\psi)$.
Consider the matching $\w \in \W(\conc{\rho^{\D}},\psi)$ for $\conc{\rho^{\D}}$ and $\psi$ defined by 
\[
\w(x_1,\diam{a}\top) = \frac{1}{10}  \quad  
\w(x_2,\neg\diam{a}\top) = \frac{3}{10}  \quad  
\w(y_1|z, \neg\diam{a}\top) = \frac{1}{5}  \quad  
\w(y_2|z,\diam{a}\top) = \frac{2}{5} \text{.}
\]
For the terms and the formulae to which $\w$ gives a positive weight, we obtain the decomposition mappings in Table~\ref{tab:ex_decomposition}, where $\xi_3$ and $\xi_4$ derive from Definition~\ref{def:decomposition}.\ref{def:decomposition_neg}. 

\begin{table}[h!]
\begin{center}
{\renewcommand{\arraystretch}{2}
\begin{tabular}{| l | l |}
\hline
$x_1^{-1}(\diam{a}\top) = \{\xi_1\}$ & $\xi_1(x_1) = \diam{a}\top$, $\xi_1(x) = \top$ for all other $x \in \SVar$\\
\hline
$x_2^{-1}(\neg\diam{a}\top) = \{\xi_2\}$ & $\xi_2(x_2) = \neg\diam{a}\top$, $\xi_2(x) = \top$ for all other $x \in \SVar$\\
\hline
\multirow{2}{*}
{$(y_1|z)^{-1}(\neg\diam{a}\top) = \{\xi_3, \xi_4\}$} & 
$\xi_3(y_1) = \neg \diam{a} \top$, $\xi_3(z) = \top$, $\xi_3(x) = \top$ for all other $x \in \SVar$\\
& $\xi_4(y_1) = \top$, $\xi_4(z) = \neg \diam{a}\top$, $\xi_4(x) = \top$ for all other $x \in \SVar$\\
\hline
$(y_2|z)^{-1}(\diam{a}\top) = \{\xi_5\}$ & $\xi_5(y_2) =\! \diam{a}\top$, $\xi_5(z) =\! \diam{a}\top$, $\xi_5(x) =\! \top$ for all other $x \in \SVar$\\
\hline
\end{tabular}}
\caption{Derived decomposition mappings.}
\label{tab:ex_decomposition}
\end{center}
\end{table}
Next, we construct the decomposition mappings for the variable $\nu$ in $\Theta$ wrt.\ $\rho^{\D}$ and $\w$.
By Definition~\ref{def:decomposition}.\ref{def:decomposition_oplus_mu} we consider the weights of the premises of $\rho^{D}$ having $\nu$ as left-hand side, namely $\HH_{\nu} = \{ \nu \trans[1/3] y_1, \quad \nu \trans[2/3] y_2\}$, and use them as weights of the $\bigoplus$ operator.
Then for each of the variables $y_1,y_2$ in the right side of $\HH_{\nu}$, we consider the conjunction of the formulae assigned to it by one decomposition mapping from each set in the first column of Table~\ref{tab:ex_decomposition}.
In detail, by omitting multiple occurrences of the $\top$ formulae in conjunctions, for $y_1$ we consider $\xi_1(y_1) \wedge \xi_2(y_1) \wedge \xi_3(y_1) \wedge \xi_5(y_1) = \neg \diam{a} \top$ and $\xi_1(y_1) \wedge \xi_2(y_1) \wedge \xi_4(y_1) \wedge \xi_5(y_1) = \top$, and for $y_2$ we consider $\xi_1(y_2) \wedge \xi_2(y_2) \wedge \xi_3(y_2) \wedge \xi_5(y_2) =  \diam{a} \top$ and $\xi_1(y_1) \wedge \xi_2(y_1) \wedge \xi_4(y_1) \wedge \xi_5(y_1) = \diam{a} \top$.
Hence the choice between $\xi_3$ or $\xi_4$ generates two dif{f}erent decomposition mappings in $\Theta^{-1}(\psi)$: by $\xi_3$ we obtain the decomposition mapping $\eta_1 \in \Theta^{-1}(\psi)$ with $\eta_1(\nu)= \frac{1}{3} \neg \diam{a} \top \oplus \frac{2}{3} \diam{a}\top$
and by $\xi_4$ we obtain the decomposition mapping $\eta_2 \in \Theta^{-1}(\psi)$ with $\eta_2(\nu)= \frac{1}{3} \top \oplus \frac{2}{3} \diam{a}\top$.
By applying the same reasoning to $\mu$ and $\upsilon$ we obtain
\[
\eta_1(\mu) = \frac{1}{4} \diam{a}\top \oplus \frac{3}{4} \neg \diam{a}\top \quad 
\eta_1(\nu)= \frac{1}{3} \neg \diam{a} \top \oplus \frac{2}{3} \diam{a}\top \quad 
\eta_1(\upsilon) = 1 (\top \wedge \diam{a}\top)
\]
\[
\eta_2(\mu) = \frac{1}{4} \diam{a}\top \oplus \frac{3}{4} \neg \diam{a}\top \quad 
\eta_2(\nu)= \frac{1}{3} \top \oplus \frac{2}{3} \diam{a}\top \quad 
\eta_2(\upsilon) = 1 (\neg \diam{a}\top \wedge \diam{a}\top)
\]
where we have omitted multiple occurrences of the $\top$ formulae in conjunctions.
Finally, we obtain two decomposition mappings in $t^{-1}(\varphi)$ by substituting $\eta$ with either $\eta_1$ or $\eta_2$ in Equation~\eqref{eq:ex_decomposition}, obtaining respectively
\[
\xi^1(x) = \diam{a}\Big( \frac{1}{4} \diam{a}\top \oplus \frac{3}{4} \neg \diam{a}\top \Big) \quad
\xi^1(y) = \diam{a} \Big( \frac{1}{3}\neg \diam{a} \top \oplus \frac{2}{3} \diam{a}\top \Big) \quad
\xi^1(z) = \diam{a} \Big( 1 (\diam{a}\top \wedge \top) \Big)
\]   
\[
\xi^2(x) = \diam{a}\Big( \frac{1}{4} \diam{a}\top \oplus \frac{3}{4} \neg \diam{a}\top \Big) \quad
\xi^2(y) = \diam{a} \Big( \frac{1}{3} \top \oplus \frac{2}{3} \diam{a}\top \Big) \quad
\xi^2(z) = \diam{a}  \Big( 1 (\neg \diam{a}\top \wedge \diam{a}\top) \Big).
\] 
\end{exa}

We show now that by decomposing formulae in $\logic$ we get formulae in $\logic$.
This is essential to derive the congruence theorem for probabilistic bisimulation (Theorem~\ref{thm:congruence}.\ref{thm:congruence_bis}):
the congruence theorem exploits the characterization of probabilistic bisimulation with the logic $\logic$ recalled in Theorem~\ref{thm:adequacy_DD11}, and requires that the formulae for composed terms and those obtained for their subterms through the decomposition method are all formulae in the characterizing logic $\logic$.

\begin{lem}
\label{lem:decomposition_mapping_in_logic}
Assume the terms $t \in \openT$ and $\Theta \in \openDTerms$ and the formulae $\varphi \in \logicstate$ and $\psi \in \logicdist$. Then:
\begin{enumerate}
\item For all $x \in \SVar$ we have $\xi(x) \in \logicstate$ for each $\xi \in t^{-1}(\varphi)$.
\item For all $\zeta \in \DVar$ we have $\eta(\zeta) \in \logicdist$ for each $\eta \in \Theta^{-1}(\psi)$.
\item For all $\zeta \in \SVar$ we have $\eta(\zeta) \in \logicstate$ for each $\eta \in \Theta^{-1}(\psi)$.
\end{enumerate}
\end{lem}


\begin{proof}
The proof follows immediately from Definition~\ref{def:decomposition}.
\end{proof}


The following result confirms that our decomposition method is correct.

\begin{thm}
[Decomposition theorem]
\label{thm:decomposition}

Let $P = (\Sigma, \Act, R)$ be a PGSOS-PTSS and let $D_{\Sigma}$ be the $\Sigma$-DS.
For any term $t \in \openT$, closed substitution $\sigma$ and state formula $\varphi \in \logicstate$ we have
\[
\sigma(t) \models \varphi \Leftrightarrow \exists\, \xi \in t^{-1}(\varphi) \text{ such that for all } x \in \var{t}\text{ it holds } \sigma(x) \models \xi(x)
\]
and for any distribution term $\Theta \in \openDTerms$, closed substitution $\sigma$ and distribution formula $\psi \in\logicdist$ we have
\[
\sigma(\Theta) \models \psi \Leftrightarrow \exists\, \eta \in \Theta^{-1}(\psi) \text{ such that for all } \zeta \in \var{\Theta} \text{ it holds } \sigma(\zeta) \models \eta(\zeta).
\]
\end{thm}


\begin{proof}
We start with univariate terms.
We proceed by structural induction over $\phi \in \logic$ to prove that for any univariate $t \in \openT$, closed substitution $\sigma$ and $\phi = \varphi \in \logicstate$ we have
\begin{equation}
\label{eq:decomposition_state}
\sigma(t) \models \varphi \Leftrightarrow \exists \xi \in t^{-1}(\varphi) \text{ such that } \forall x \in \var{t}\text{ it holds } \sigma(x) \models \xi(x)
\end{equation}
and for any univariate $\Theta \in \openDTerms$, closed substitution $\sigma$ and $\phi = \psi \in \logicdist$ we have
\begin{equation}
\label{eq:decomposition_distribution}
\sigma(\Theta) \models \psi \Leftrightarrow \exists \eta \in \Theta^{-1}(\psi) \text{ such that } \forall \zeta \in \var{\Theta}\text{ it holds } \sigma(\zeta) \models \eta(\zeta).
\end{equation}
\begin{itemize}
\item Base case $\phi = \top$.
Then by Definition~\ref{def:decomposition}.\ref{def:decomposition_top} we have that $\xi \in t^{-1}(\top)$ if{f} $\xi(x) = \top$ for all $x \in \SVar$.
Then the proof obligation Equation~\eqref{eq:decomposition_state} directly follows from the definition of $\models$ (Definition~\ref{def:satisfiability}).

\item Inductive step $\phi = \neg \varphi$ for some $\varphi \in \logicstate$.
We have
\begin{align*}
& \sigma(t) \models \neg \varphi \\
\Leftrightarrow\; & \sigma(t) \not\models \varphi \\
\Leftrightarrow\; & \forall\, \xi \in t^{-1}(\varphi) \; \exists\, x \in \var{t} \text{ s.t. } \sigma(x) \not \models \xi(x)\\
\Leftrightarrow\; & \exists\, \f \colon t^{-1}(\varphi) \to \var{t} \text{ s.t.  } \forall\, \xi' \in t^{-1}(\varphi) \text{ it holds } \sigma(\f(\xi')) \not\models \xi'(\f(\xi'))\\
\Leftrightarrow\; & \exists\, \f \colon t^{-1}(\varphi) \to \var{t} \text{ s.t.  } \forall\, x \in \var{t} \text{ it holds } \sigma(x) \models \bigwedge_{ \xi' \in \f^{-1}(x)} \neg \xi'(x)\\
\Leftrightarrow\; & \exists\, \xi \in t^{-1}(\neg\varphi) \text{ s.t.  } \forall\, x \in \var{t} \text{ it holds } \sigma(x) \models \xi(x) 
\end{align*}
where the second relation follows by the inductive hypothesis and the last relation follows by construction of $t^{-1}(\neg \varphi)$ (Definition~\ref{def:decomposition}.\ref{def:decomposition_neg}).
Hence, the proof obligation Equation~\eqref{eq:decomposition_state} holds also in this case.

\item Inductive step $\phi = \bigwedge_{j \in J} \varphi_j$ for some $j \in J$ and $\varphi_j \in \logicstate$.
We have
\begin{align*}
& \sigma(t) \models \bigwedge_{j \in J} \varphi_j\\
\Leftrightarrow\; & \sigma(t) \models \varphi_j, \text{ for all } j \in J\\
\Leftrightarrow\; & \exists\, \xi_j \in t^{-1}(\varphi_j) \text{ s.t.  } \forall\, x \in \var{t} \text{ it holds } \sigma(x) \models \xi_j(x), \text{ for all } j \in J\\
\Leftrightarrow\; & \exists\, \xi_j \in t^{-1}(\varphi_j) \text{ for all } j \in J \text{ s.t.  } \forall\, x \in \var{t} \text{ it holds } \sigma(x) \models \bigwedge_{j \in J} \xi_j(x)\\
\Leftrightarrow\; & \exists\, \xi \in t^{-1}(\bigwedge_{j \in J} \varphi_j) \text{ s.t.  } \forall\, x \in \var{t} \text{ it holds } \sigma(x) \models \xi(x)
\end{align*}
where the second relation follows by the inductive hypothesis and the last relation follows by construction of $t^{-1}(\bigwedge_{j \in J} \varphi_j)$ (Definition~\ref{def:decomposition}.\ref{def:decomposition_wedge}).
Hence, the proof obligation Equation~\eqref{eq:decomposition_state} holds also in this case.

\item Inductive step $\phi = \bigoplus_{i \in I} r_i \varphi_i$ for some $\varphi_i \in \logicstate$, with $r_i \in (0,1]$ for $i \in I$ and $\sum_{i \in I}r_i = 1$.
Notice that in this case we have $\phi \in \logicdist$ and therefore we need to show Equation~\eqref{eq:decomposition_distribution}.
To this aim, we prove the two implications separately.

\begin{description}
  \item[($\Rightarrow$)] Assume first that $\sigma(\Theta) \models \bigoplus_{i \in I} r_i \varphi_i$.
Then, by definition of $\models$ (Definition~\ref{def:satisfiability}), this implies that there exists a family of probability distributions $\{\pi_i\}_{i \in I} \subseteq \ProbDist{\closedTerms}$ with $\sigma(\Theta) = \sum_{i \in I} r_i \pi_i$ and whenever $t \in \support(\pi_i)$ for some $t \in \closedTerms$, then $t \models \varphi_{i}$.
Notice that $\support(\sigma(\Theta)) = \bigcup_{i \in I} \support(\pi_i)$.
Let us order the elements of the support of the distribution $\sigma(\Theta)$ through indexes in a suitable set $M$, namely $\support(\sigma(\Theta)) = \{t_{m} \mid m \in M\}$, with $t_m, t_{m'}$ pairwise distinct for all $m, m' \in M$ with $m \neq m'$.
We have $\sigma(\Theta) = \sum_{m \in M} q_m \delta_{t_m}$, for some $q_m \in (0,1]$ such that $\sum_{m \in M} q_m = 1$.
In particular, this gives $q_m = \sigma(\Theta)(t_m)$, which, by Theorem~\ref{prop:proof_sigma_q}, implies that $D \vdash \{\sigma(\Theta) \trans[q_m] t_m \mid m \in M\}$. 
By Theorem~\ref{thm:distribution_ruloid_theorem}, $D_{\Sigma} \vdash \{\sigma(\Theta) \trans[q_m] t_{m} \mid m \in M\}$ implies that there are a \dgsos ruloid $\rho^{\D} = \ddedrule{\HH}{\{\Theta \trans[q_{m}] u_{m} \mid m \in M\}}$ and a closed substitution $\sigma'$ with $D_{\Sigma} \vdash \sigma'(\HH)$, $\sigma'(\Theta) = \sigma(\Theta)$ and $\sigma'(u_{m}) = t_{m}$ for each $m \in M$.
Let us show that the rewriting of $\sigma'(\Theta)$ as convex combination of the $\{\pi_i\}_{i \in I}$ gives rise to a matching for $\conc{\rho^{\D}}$ and $\bigoplus_{i \in I} r_i \varphi_i$.
Define $\w \in \W(\conc{\rho^{\D}}, \bigoplus_{i \in I} r_i \varphi_i)$ as $\w(u_m, \varphi_i) = r_i \pi_i(\sigma'(u_m))$, then $\w$ is a matching with left marginal $\conc{\rho^{\D}}$, and right marginal the distribution formula $\bigoplus_{i \in I} r_i \varphi_i$.
More precisely, we have
\begin{align*}
& q_m \\
={} & \sigma(\Theta)(t_m) & \text{(by construction of $\sigma(\Theta)$)}\\
={} & \sum_{i \in I} r_i \pi_i(t_m) & \text{(by def.\ of convex combination of distributions)}\\
={} & \sum_{i \in I} r_i \pi_i(\sigma'(u_m)) & \text{(by construction of $\sigma'$)}\\
={} & \sum_{i \in I} \w(u_m, \varphi_i) & \text{(by definition of $\w$)}
\end{align*}
and
\begin{align*}
& \sum_{m \in M} \w(u_m, \varphi_i) \\
={} & \sum_{m \in M} r_i \pi_i(\sigma'(u_m)) & \text{(by definition of $\w$)} \\
={} & \sum_{m \in M} r_i \pi_i(t_m) & \text{(by construction of $\sigma'$)}\\
={} & r_i \, \sum_{t \in \support(\sigma(\Theta))} \pi_i(t) & \text{(by the choice of $M$)} \\
={} & r_i \, \sum_{t \in\support(\pi_i)} \pi_i(t)\\
={} & r_i\text{.}
\end{align*}
We derive that:
\begin{enumerate}
\item 
\label{item:agree_on_var} 
from $\sigma'(\Theta) = \sigma(\Theta)$ we obtain that $\sigma'(\zeta) = \sigma(\zeta)$ for all variables $\zeta \in \var{\Theta}$;
\item 
\label{item:matching} 
whenever $\w(u_m, \varphi_i) > 0$ it holds that $\sigma'(u_m) \in \support(\pi_i)$ and, therefore, we infer $\sigma'(u_{m}) \models \varphi_i$.
By the inductive hypothesis we derive that there is a decomposition mapping $\xi_{m,i} \in u_{m}^{-1}(\varphi_i)$ such that $\sigma'(x) \models \xi_{m,i}(x)$ for all $x \in \var{u_{m}}$;
\item 
\label{item:premises}
 from $D_{\Sigma} \vdash \sigma'(\HH)$ we obtain that for all premises $\{\zeta \trans[q_j] x_j \mid j \in J \} \in \HH$ we have $D_{\Sigma} \vdash \{\sigma'(\zeta) \trans[q_h] t'_h \mid h \in H\}$, where
$\{\sigma'(\zeta) \trans[q_h] t'_h \mid h \in H\}$ is $\sigma'(\{\zeta \trans[q_j] x_j \mid j \in J\} )$,
for a suitable set of indexes $H$ and proper terms $t_h'$.
By Theorem~\ref{prop:proof_sigma_q},  $D_{\Sigma} \vdash \{\sigma'(\zeta) \trans[q_h] t'_h \mid h \in H\}$ iff
$\sigma'(\zeta)(t'_h) = q_h$ and $\sum_{h \in H} q_h = 1$.
Hence we have that 
\begin{align*}
& \sigma'(\zeta) \\ ={} & \sum_{h \in H} q_h \delta_{t'_h} \\
={} & \sum_{h \in H} \big( \sum_{j \in J, \sigma'(x_j) = t'_h} q_j \big) \delta_{t'_h} & \text{(by Definition~\ref{def:sigma_reduction})}\\
={} & \sum_{h \in H} \big( \sum_{j \in J, \sigma'(x_j) = t'_h} q_j \delta_{\sigma'(x_j)} \big)\\
={} & \sum_{j \in J} q_j \delta_{\sigma'(x_j)} & \text{(the $t'_h$ are pairwise distinct).}
\end{align*}
We remark that this reasoning holds since we assumed that $\Theta$ is univariate, and therefore there is only one set of distribution premises in $\HH$ with left-hand side $\zeta$, for each $\zeta \in \var{\Theta}$.
\end{enumerate}

Let $\eta \in \Theta^{-1}(\bigoplus_{i \in I} r_i \varphi_i)$ be the decomposition mapping defined as in Definition~\ref{def:decomposition}.\ref{def:decomposition_oplus} by means of the \dgsos ruloid $\ddedrule{\HH}{\{\Theta \trans[q_{m}] u_{m}\mid m \in M\}}$ and the decomposition mappings $\xi_{m,i}$ as in item~\eqref{item:matching} above for each $m \in M$ and $i \in I$ such that $\w(u_m, \varphi_i) > 0$, and $\xi_{m,i}$ defined by $\xi_{m,i}(x) = \top$ for all $x \in \SVar$ for those $m,i$ such that $\w(u_m, \varphi_i) = 0$.
We aim to show that for this $\eta$ it holds that $\sigma'(\zeta)\models \eta(\zeta)$ for each $\zeta \in \var{\Theta}$.
By construction, 
\[
\eta(\zeta) = \begin{cases}
\displaystyle \bigoplus_{\{\zeta \trans[q_j] x_j \mid j \in J\} \in \HH} q_j 
			\bigwedge_{m \in M \atop i \in I} \xi_{m,i}(x_j)
			& \text{ if } \zeta \in \DVar\\
			\displaystyle \bigwedge_{m \in M \atop i \in I} \xi_{m,i}(x)
			& \text{ if } \zeta = x \in \SVar.
		\end{cases}
\]
For each variable $y \in \{x_j \mid j \in J\} \cup \{x\}$ and for each $m \in M$ and $i \in I$, we can distinguish three cases:
\begin{enumerate}
\setcounter{enumi}{3}
\item $y \in \var{u_{m}}$ and $\w(u_m, \varphi_i) > 0$. 
Then, by item~\eqref{item:matching} above, we have $\sigma'(y) \models \xi_{m,i}(y)$.
\item $y \in \var{u_{m}}$ and $\w(u_m, \varphi_i) = 0$. 
Then by construction $\xi_{m,i}(y) = \top$, thus giving that $\sigma'(y) \models \xi_{m,i}(y)$ holds trivially also in this case.
\item $y \not \in \var{u_{m}}$. Then, whichever is the value of $\w(u_m, \varphi_i)$, we have $\xi_{m,i}(y) = \top$ (see Definition~\ref{def:decomposition}) and consequently $\sigma'(y) \models \xi_{m,i}(y)$ holds trivially also in this case.
\end{enumerate}
Since these considerations apply to each $m \in M$ and $i \in I$ we can conclude that if $\zeta \in \DVar$ then for all $\{\zeta \trans[q_j] x_j \mid j \in J\} \in \HH$ it holds that for each $x_j$ with $j \in J$ we have $\sigma'(x_j) \models \bigwedge_{m \in M, i \in I} \xi_{m,i}(x_j)$.
Furthermore, by item~\eqref{item:premises} above, if $\{\zeta \trans[q_j] x_j \mid j \in J\} \in \HH$ then $D_{\Sigma} \vdash \sigma'(\HH)$ gives $\sigma'(\zeta) = \sum_{j \in J} q_j \delta_{\sigma'(x_j)}$, from which we can conclude that 
\[
\sigma'(\zeta) \models \bigoplus_{j \in J} q_j \bigwedge_{i \in I, m \in M} \xi_{m,i}(x_j) \text{, namely } \sigma'(\zeta) \models \eta(\zeta).
\]
Similarly, if $\zeta = x \in \SVar$ then 
\[
\sigma'(x) \models \bigwedge_{m \in M, i \in I} \xi_{m,i}(x) \text{, namely } \sigma'(x) \models \eta(x).
\]
Thus, we can conclude that for each $\zeta \in \var{\Theta}$ it holds that $\sigma'(\zeta) \models \eta(\zeta)$.
Since moreover $\sigma(\zeta) = \sigma'(\zeta)$ (item~\eqref{item:agree_on_var} above), we can conclude that $\sigma(\zeta) \models \eta(\zeta)$ as required.

\item[($\Leftarrow$)] Assume now that there is a decomposition mapping $\eta \in \Theta^{-1}(\bigoplus_{i \in I} r_i \varphi_i)$ such that $\sigma(\zeta) \models \eta(\zeta)$ for all $\zeta \in \var{\Theta}$.
Following Definition~\ref{def:decomposition}.\ref{def:decomposition_oplus}, the existence of such a decomposition mapping $\eta$ entails the existence of a \dgsos ruloid $\rho^{\D} = \ddedrule{\HH}{\{\Theta \trans[q_{m}] t_{m}\mid m \in M\}}$ with $\sum_{m \in M} q_{m} = 1$ (Proposition~\ref{lem:sum_to_1_tris}) and of a matching $\w \in \W( \conc{\rho^{\D}},\bigoplus_{i \in I} r_i \varphi_i)$ from which we can build the following decomposition mappings:
\[
\begin{cases}
\xi_{m,i} \in t_{m}^{-1}(\varphi_i) & \text{ if } \w(t_m, \varphi_i) > 0 \\
\xi_{m,i} \in t_{m}^{-1}(\top) & \text{ otherwise.}
\end{cases}
\]
In particular, we have that for each $\mu \in \var{\Theta}$
\[
\eta(\mu) = \bigoplus_{\{\mu \trans[q_j] x_j \mid \sum_{j \in J} q_j = 1\} \in \HH} q_j \bigwedge_{i \in I, m \in M} \xi_{m,i}(x_j)
\]
and for each $x \in \var{\Theta}$
\[
\eta(x) = \displaystyle \bigwedge_{i \in I, m \in M} \xi_{m,i}(x).
\]
We define a closed substitution $\sigma'$ such that $\sigma'(\zeta) = \sigma(\zeta)$ for each $\zeta \in \var{\Theta}$ and $\sigma'(x) = \sigma(x)$ for each $x \in \rhs(\HH)$. 
Then, the following properties hold:
\begin{enumerate}[label=(\emph{\alph*})]
\item \label{item:one} 
From $\sigma'(\zeta) = \sigma(\zeta)$ and $\sigma(\zeta) \models \eta(\zeta)$ we derive $\sigma'(\zeta) \models \eta(\zeta)$.
In particular we obtain that $\sigma'(x) \models \bigwedge_{i \in I, m \in M} \xi_{m,i}(x)$ for each $x \in \var{\Theta}$.
\item \label{item:two} 
As $\sigma'(\mu) \models \eta(\mu)$ for each $\mu \in \var{\Theta}$, by previous item~(\ref{item:a}), we derive that there are probability distributions $\pi_j$ such that $\sigma'(\mu) = \sum_{j \in J} q_j \pi_j$ and whenever $t \in \support(\pi_j)$, for some $t \in \closedTerms$, then $t \models \bigwedge_{i \in I, m \in M} \xi_{m,i}(x_j)$.
By Definition~\ref{def:decomposition}.\ref{def:decomposition_oplus_mu}, the weights of the distribution formula $\eta(\mu)$ coincide with the weights of the distribution literals in $\{\mu \trans[q_j] x_j \mid \sum_{j \in J} q_j = 1\} \in \HH$.
Therefore, we have that $\sigma'(\mu) = \sum_{j \in J} q_j \delta_{\sigma'(x_j)}$ from which we gather $\sigma'(x_j) \models \bigwedge_{i \in I, m \in M} \xi_{m,i}(x_j)$, for each $j \in J$. 
\item \label{item:three}
From $\sigma(\zeta) = \sigma'(\zeta)$ for each $\zeta \in \var{\Theta}$ we infer that $\sigma'(\Theta) = \sigma(\Theta)$.
Moreover, by Lemma~\ref{lem:ruloid_variables}.\ref{lem:ruloid_variables_rhs} we have that $\rhs(\HH) = \bigcup_{m \in M} \var{t_m}$, so that $\sigma'(x) = \sigma(x)$ for each $x \in \rhs(\HH)$ implies $\sigma'(t_m) = \sigma(t_m)$ for each $m \in M$.
\end{enumerate}
From items~(\ref{item:one}), (\ref{item:two}) above and by structural induction we gather $\sigma'(t_{m}) \models \varphi_i$ for each $m \in M, i \in I$ with $\w(t_m, \varphi_i) > 0$.
Moreover, from $\sigma'(\zeta) \models \eta(\zeta)$ for each $\zeta \in \var{\Theta}$, item~(\ref{item:one}) above, we obtain that $D_{\Sigma} \vdash \sigma'(\HH)$, namely $D_{\Sigma}$ proves the reduced instance w.r.t,\ $\sigma'$ of each set of distribution premises $\{\zeta \trans[q_j] x_j \mid \sum_{j \in J} q_j = 1\} \in \HH$.
This fact taken together with item~(\ref{item:three}) above and Theorem~\ref{thm:distribution_ruloid_theorem} gives that $D_{\Sigma}$ proves the reduced instance of $\{\Theta \trans[q_{m}] t_{m} \mid m \in M\}$ wrt.\ $\sigma$, that is $D_{\Sigma} \vdash \{\sigma(\Theta) \trans[q_h] t'_h \mid h \in H\}$ for a suitable set of indexes $H$ and a proper set of closed terms $t'_h$ such that for each $h \in H$ there is at least one $m \in M$ such that $t'_h = \sigma'(t_m)$ and moreover $q_h = \sum_{\{m \in M \mid \sigma'(t_m) = t'_h\}} q_m$ (Definition~\ref{def:sigma_reduction}).
In addition, by Theorem~\ref{prop:proof_sigma_q} it follows that $q_h = \sigma(\Theta)(t'_h)$ for each $h \in H$ and $\sum_{h \in H} q_h = 1$.
Since moreover $q_h \in (0,1]$ for each $h \in H$, this is equivalent to say that $\sigma(\Theta) = \sum_{h \in H} q_h \delta_{\sigma'(t_{h})}$.
Finally, we notice that 
\begin{align*}
\sigma(\Theta) ={} & \sum_{h \in H} q_{h} \delta_{t'_{h}} \\
={} & \sum_{h \in H} \Big( \sum_{\{m \in M \mid \sigma'(t_m) = t'_h\}} q_{m} \Big) \delta_{t'_{h}} \\
={} & \sum_{m \in M} q_{m} \delta_{\sigma'(t_{m})} & \text{($t'_h$ pairwise distinct)}\\
={} & \sum_{m \in M} \Big( \sum_{i \in I} \w(t_m, \varphi_i) \Big) \delta_{\sigma'(t_{m})} & \text{(}\sum_{i \in I} \w(t_m, \varphi_i) = q_m\text{)}\\
={} & \sum_{i \in I} \Big( \sum_{m \in M} \w(t_m, \varphi_i) \delta_{\sigma'(t_{m})} \Big)\\
={} & \sum_{i \in I} \Big( \sum_{m \in M} r_i \frac{\w(t_m, \varphi_i)}{r_i} \delta_{\sigma'(t_{m})} \Big)\\
={} & \sum_{i \in I} r_i \Big( \sum_{m \in M} \frac{\w(t_m, \varphi_i)}{r_i} \delta_{\sigma'(t_{m})} \Big)\\
={} & \sum_{i \in I} r_i \pi_i 
\end{align*}
where for each $i \in I$, $\pi_i = \sum_{m \in M} \displaystyle\frac{\w(t_m, \varphi_i)}{r_i} \delta_{\sigma'(t_{m})}$ is a probability distribution as it is obtained as a convex combination of probability distributions ($\sum_{m \in M} \displaystyle\frac{\w(t_m, \varphi_i)}{r_i} = 1 $).
Moreover, the $\pi_i$ are such that whenever $\sigma'(t) \in \support(\pi_i)$ it holds that $\sigma'(t) \models \varphi_i$.
In fact, we have that whenever $\w(t_m,\varphi_i) > 0$, then the only closed term in the support of $\delta_{\sigma'(t_m)}$ is indeed $\sigma'(t_m)$.
Furthermore, whenever $\sigma'(t_m) \not\models \varphi_i$ we are granted that $\w(t_m, \varphi_i) =0$, thus giving that $\sigma'(t_m)$ is not in the support of $\pi_i$.
Therefore, we can conclude that $\sigma(\Theta) \models \bigoplus_{i \in I} r_i \varphi_i$ as requested.

Hence, the proof obligation Equation~\eqref{eq:decomposition_distribution} follows from the two implications.
\end{description}
\item Inductive step $\phi = \diam{a}\psi$ for some $\psi \in \logicdist$ and $a \in \Act$.
Notice that in this case we have $\phi \in \logicstate$ and therefore we need to show Equation~\eqref{eq:decomposition_state}.
To this aim, we prove the two implications separately.
\begin{description}
\item[($\Rightarrow$)] Assume first that $\sigma(t) \models \diam{a}\psi$.
Then, by definition of relation $\models$ (Definition~\ref{def:satisfiability}), there exists a probability distribution $\pi \in \ProbDist{\closedTerms}$ with $P \vdash \sigma(t) \trans[a] \pi$ and $\pi \models \psi$.
By Theorem~\ref{thm:term_ruloid_theorem}, $P \vdash \sigma(t) \trans[a] \pi$ implies that there are a $P$-ruloid $\ddedrule{\HH}{t \trans[a] \Theta}$ and a closed substitution $\sigma'$ with $P \vdash \sigma'(\HH)$, $\sigma'(t) = \sigma(t)$ and $\sigma'(\Theta) = \pi$.
We infer the following facts:
\begin{enumerate}
\item \label{item:agree_on_state_var} from $\sigma'(t) = \sigma(t)$ we obtain that $\sigma'(x) = \sigma(x)$ for all $x \in \var{t}$;
\item \label{item:inductive_theta} from $\sigma'(\Theta) = \pi$ and $\pi \models \psi$, we gather $\sigma'(\Theta) \models \psi$ and by the inductive hypothesis we obtain that there exists a $\eta \in \Theta^{-1}(\psi)$ such that $\sigma'(\zeta) \models \eta(\zeta)$ for all $\zeta \in \var{\Theta}$;
\item \label{item:positive} from $P \vdash \sigma'(\HH)$ we obtain that whenever $x \trans[b] \mu \in \HH$ we have $P \vdash \sigma'(x) \trans[b] \sigma'(\mu)$.
Then, if $\mu \in \var{\Theta}$, by previous item~\eqref{item:inductive_theta}, we get $\sigma'(\mu) \models \eta(\mu)$.
Otherwise, if $\mu \not \in \var{\Theta}$, we have $\eta(\mu) = \top$ thus giving $\sigma'(\mu) \models \eta(\mu)$ also in this case.
Hence, $\sigma'(\mu) \models \eta(\mu)$ and $\sigma'(x) \models \diam{b} \eta(\mu)$ in all cases.
\item \label{item:negative} from $P \vdash \sigma'(\HH)$ we obtain that whenever $x\ntrans[c] \in \HH$ we have $P \vdash \sigma'(x) \ntrans[c]$, namely $P \not\vdash \sigma'(x) \trans[c] \upsilon$ for any $\upsilon \in \closedDTerms$, giving $\sigma'(x) \models \neg \diam{c} \top$.  
\end{enumerate}
Let $\xi \in t^{-1}(\diam{a}\psi)$ be defined as in Definition~\ref{def:decomposition}.\ref{def:decomposition_diam} by means of the $P$-ruloid $\ddedrule{\HH}{t \trans[a] \Theta}$ and the decomposition mapping $\eta$ introduced in item~\eqref{item:inductive_theta} above.
We aim to show that for this $\xi$ it holds that $\sigma'(x)\models \xi(x)$ for each $x \in \var{t}$.
By construction, 
\[
\xi(x) = \bigwedge_{x \trans[b] \mu \in \HH} \diam{b} \eta(\mu) \wedge \bigwedge_{x \ntrans[c] \in \HH}\neg\diam{c}\top \wedge \eta(x).
\]
By item~\eqref{item:positive} above we have $\sigma'(x) \models \diam{b}\eta(\mu)$ for each $x \trans[b] \mu \in \HH$.
By item~\eqref{item:negative} above we have $\sigma'(x) \models \neg \diam{c} \top$ for each $x \ntrans[c] \in \HH$.
Finally, if $x \in \var{\Theta}$ by item~\eqref{item:inductive_theta} above we get $\sigma'(x) \models \eta(x)$.
If $x \not \in \var{\Theta}$ then we have $\eta(x) = \top$ (Definition~\ref{def:decomposition}.\ref{def:decomposition_oplus_delta}) thus giving $\sigma'(x) \models \eta(x)$ also in this case.
Hence, $\sigma'(x) \models \eta(x)$ in all cases.
Thus, we can conclude that $\sigma'(x) \models \xi(x)$.
Since, by item~\eqref{item:agree_on_state_var} above, $\sigma(x) = \sigma'(x)$ we can conclude that $\sigma(x) \models \xi(x)$ as required.

\item[($\Leftarrow$)] Assume now that there is a $\xi \in t^{-1}(\diam{a}\psi)$ such that $\sigma(x) \models \xi(x)$ for all $x \in \var{t}$.
Following Definition~\ref{def:decomposition}.\ref{def:decomposition_diam}, we construct $\xi$ in terms of some $P$-ruloid $\ddedrule{\HH}{t \trans[a] \Theta}$ and decomposition mapping $\eta \in \Theta^{-1}(\psi)$.
In particular, we have that for each $x \in \var{t}$
\[
\xi(x) = \bigwedge_{x \trans[b] \mu \in \HH} \diam{b} \eta(\mu) \wedge \bigwedge_{x \ntrans[c] \in \HH} \neg \diam{c} \top \wedge \eta(x). 
\]
We define a closed substitution $\sigma'$ such that the following properties hold:
\begin{enumerate}[label=(\emph{\alph*})]
\item \label{item:a} $\sigma'(x) = \sigma(x)$ for all $x \in \var{t}$. 
As a consequence, from $\sigma(x) \models \xi(x)$ we derive $\sigma'(x) \models \xi(x)$.
\item \label{item:b} As $\sigma'(x) \models \xi(x)$, by previous item~\eqref{item:a}, we derive that $\sigma'(x) \models \diam{b} \eta(\mu)$ for each $x \trans[b] \mu \in \HH$.
This implies that for each positive premise in $\HH$ there exists a probability distribution $\pi_{b,\mu}$ such that $P \vdash \sigma'(x) \trans[b] \pi_{b,\mu}$ and $\pi_{b,\mu} \models \eta(\mu)$.
We define $\sigma'(\mu) = \pi_{b,\mu}$ thus obtaining that for each $x \trans[b] \mu \in \HH$ we have $P \vdash \sigma'(x) \trans[b] \sigma'(\mu)$ and $\sigma'(\mu) \models \eta(\mu)$.
\item \label{item:c} As $\sigma'(x) \models \xi(x)$, by previous item~\eqref{item:a}, we derive that $\sigma'(x) \models \neg \diam{c} \top$ for each $x \ntrans[c] \in \HH$.
Therefore, we obtain that $P \vdash \sigma'(x) \ntrans[c]$ for each $x \ntrans[c] \in \HH$.
\item \label{item:d} Since $\var{\Theta} \subseteq \var{t} \cup \rhs(\HH)$, previous items~\eqref{item:b} and \eqref{item:c} we obtain that $\sigma'(\mu)\models \eta(\mu)$ for each $\mu \in \DVar$.
\item \label{item:e} $\sigma'(x) \models \eta(x)$ for each $x \in \var{\Theta}$.
\end{enumerate}
From items~\eqref{item:d}, \eqref{item:e} and structural induction, we gather $\sigma'(\Theta) \models \psi$.
Moreover, items~\eqref{item:b} and \eqref{item:c} give $P \vdash \sigma'(\HH)$.
Hence, by Theorem~\ref{thm:term_ruloid_theorem} we obtain $P \vdash \sigma'(t) \trans[a] \sigma'(\Theta)$.
From item~\eqref{item:a} we have that $\sigma'(t) = \sigma(t)$ and, therefore, we can conclude that $\sigma(t) \models \diam{a} \psi$.

Hence, the proof obligation Equation~\eqref{eq:decomposition_state} follows from the two implications.
\end{description}
\end{itemize}

Finally, let us deal with terms that are not univariate.

Assume first that $t$ is not univariate, namely $t = \varsigma(s)$ for some univariate $s$ and non-injective substitution $\varsigma \colon \var{s} \to \SVar$.
Then, $\sigma(\varsigma(s)) \models \varphi$ if{f} there exists a decomposition mapping $\xi' \in s^{-1}(\varphi)$ such that $\sigma(\varsigma(y)) \models \xi'(y)$, which by Definition~\ref{def:decomposition}.\ref{def:decomposition_multi} is equivalent to require that there exists a decomposition mapping $\xi' \in s^{-1}(\varphi)$ such that for each $x \in \var{t}$ we have $\sigma(x) \models \bigwedge_{y \in \varsigma^{-1}(x)} \xi'(y)$.
By defining the decomposition mapping $\xi \in t^{-1}(\varphi)$ as $\xi(x) = \bigwedge_{y \in \varsigma^{-1}(x)} \xi'(y)$, we obtain the thesis. 

Assume now that $\Theta$ is not univariate, namely $\Theta = \varsigma(\Theta_1)$ for some univariate $\Theta_1$ and non-injective substitution $\varsigma \colon \var{\Theta_1} \to \DVar \cup \delta_{\SVar}$.
Then, $\sigma(\varsigma(\Theta_1)) \models \psi$ if{f} there exists a decomposition function $\eta_1 \in \Theta_{1}^{-1}(\psi)$ such that $\sigma(\varsigma(z)) \models \eta_1(z)$, which by Definition~\ref{def:decomposition}.\ref{def:decomposition_oplus_multi} is equivalent to require that there exists a decomposition mapping $\eta' \in \Theta_{1}^{-1}(\psi)$ such that for each $\zeta \in \var{\Theta}$ we have $\eta'(z) = \eta'(z')$ for all $z,z' \in \varsigma^{-1}(\zeta)$ and, for a chosen $\tilde{z} \in \varsigma^{-1}(\zeta)$, $\sigma(\zeta) \models \eta'(\tilde{z})$.
By defining the decomposition mapping $\eta \in \Theta^{-1}(\psi)$ as $\eta(\zeta) = \eta'(\tilde{z})$, for $\tilde{z} \in \varsigma^{-1}(\zeta)$, we obtain the thesis. 
\end{proof}



\subsection{Decomposition of formulae in \texorpdfstring{$\logicready$}{L\_r} and \texorpdfstring{$\logicplus$}{L\_plus}}
In this section we consider the logics $\logicready$ and $\logicplus$, whose decomposition 
 can be derived from the one for $\logic$. 

\begin{defi}
[Decomposition of formulae in $\logicready$ and $\logicplus$]
\label{def:decomposition_ready}
Let $P = (\Sigma, \Act, R)$ be a PGSOS-PTSS and let $D_{\Sigma}$ be the $\Sigma$-DS.
The mappings $\cdot^{-1}\colon \openT \to (\logicstateready \to \powset{\SVar \to \logicstateready})$ and $\cdot^{-1} \colon \openDTerms \to (\logicdistready \to \powset{\Var \to \logicready})$ are 
obtained as in Definition~\ref{def:decomposition} by rewriting Definition~\ref{def:decomposition}.\ref{def:decomposition_neg} and Definition~\ref{def:decomposition}.\ref{def:decomposition_diam}, respectively,  by
\begin{enumerate}
\setcounter{enumi}{1}
\item \label{def:decomposition_ready_neg}
$\xi \in t^{-1}(\bar{a})$ if{f} there is a function $\f \colon t^{-1}(\diam{a} \top) \to \var{t}$ such that
\[ 
\xi(x) = 
\begin{cases}
\displaystyle \bigwedge_{\xi' \in \f^{-1}(x)} \neg \xi'(x) & \text{if } x \in \var{t} \\
\top & \text{otherwise;}
\end{cases}
\]
\setcounter{enumi}{3}
\item \label{def:decomposition_ready_diam}
$\xi \in t^{-1}(\diam{a}\psi)$ if{f} there are a ruloid $\dedrule{\HH}{t \trans[a] \Theta}$ and a decomposition mapping $\eta \in \Theta^{-1}(\psi)$ such that
\[
\xi(x) =
\begin{cases}
\displaystyle
\bigwedge_{x \trans[b] \mu \in \HH} \diam{b} \eta(\mu) \; \wedge \; \bigwedge_{x \ntrans[c] \in \HH} \bar{c} \; \wedge\; \eta(x) & \text{if } x \in \var{t} \\
\top & \text{otherwise.} 
\end{cases}
\]
\end{enumerate}
Moreover, if $P$ is positive, then the  mappings $\cdot^{-1}\colon \openT \to (\logicstateplus \to \powset{\SVar \to \logicstateplus})$ and $\cdot^{-1} \colon \openDTerms \to (\logicdistplus \to \powset{\Var \to \logicplus})$ are obtained as in Definition~\ref{def:decomposition} by removing 
Definition~\ref{def:decomposition}.\ref{def:decomposition_neg} and by rewriting Definition~\ref{def:decomposition}.\ref{def:decomposition_diam} by
\begin{enumerate}
\setcounter{enumi}{3}
\item $\xi \in t^{-1}(\diam{a}\psi)$ if{f} there are a positive $P$-ruloid $\dedrule{\HH}{t \trans[a] \Theta}$ and a decomposition mapping $\eta \in \Theta^{-1}(\psi)$ such that
\[
\xi(x) =
\begin{cases}
\displaystyle \bigwedge_{x \trans[b] \mu \in \HH} \diam{b} \eta(\mu) \;\wedge\; \eta(x) & \text{if } x \in \var{t} \\
\top&  \text{otherwise.}
\end{cases}
\]
\end{enumerate}
\end{defi}

We show now that by decomposing formulae in $\logicready$ (resp.\ $\logicplus$) we get formulae in $\logicready$ (resp.\ $\logicplus$).
Again, this is necessary for the precongruence theorem Theorem~\ref{thm:congruence}.\ref{thm:congruence_ready} (resp.\ Theorem~\ref{thm:congruence}.\ref{thm:congruence_sim}), which exploits the characterization of probabilistic ready simulation with the logic $\logicready$ (resp.\ the characterization of probabilistic simulation with the logic $\logicplus$) given in Theorem~\ref{thm:ready_sim_adequate} and requires that the formulae for composed terms and those obtained for their subterms through the decomposition method are all formulae in the characterizing logic $\logicready$ (resp.\ $\logicplus$).

\begin{lem}
\label{lem:decomposition_mapping_in_logicready}
Let $P$ be a PGSOS-PTSS and consider the term $t \in \openT$ and the formulae $\varphi \in \logicstateready$, $\psi \in \logicdistready$, $\varphi' \in \logicstateplus$ and $\psi' \in \logicdistplus$.
\begin{enumerate}
\item 
\begin{itemize}
\item For all $x \in \SVar$ we have $\xi(x) \in \logicstateready$ for each $\xi \in t^{-1}(\varphi)$.
\item For all $\zeta \in \DVar$ we have $\eta(\zeta) \in \logicdistready$ for each $\eta \in \Theta^{-1}(\psi)$.
\item For all $\zeta \in \SVar$ we have $\eta(\zeta) \in \logicstateready$ for each $\eta \in \Theta^{-1}(\psi)$.
\end{itemize}
\item If $P$ is positive, then 
\begin{itemize}
\item For all $x \in \SVar$ we have $\xi(x) \in \logicstateplus$ for each $\xi \in t^{-1}(\varphi')$.
\item For all $\zeta \in \DVar$ we have $\eta(\zeta) \in \logicdistplus$ for each $\eta \in \Theta^{-1}(\psi')$.
\item For all $\zeta \in \SVar$ we have $\eta(\zeta) \in \logicstateplus$ for each $\eta \in \Theta^{-1}(\psi')$.
\end{itemize}
\end{enumerate}
\end{lem}


\begin{proof}
The proofs of items (1) and (2) follow immediately from Definition~\ref{def:decomposition_ready}.
\end{proof}


Then, we can show that also the decomposition methods for $\logicready$ and $\logicplus$ are correct.

\begin{thm}
[Decomposition theorem II]
\label{cor:decomposition}
Let $P = (\Sigma, \Act, R)$ be a PGSOS-PTSS and $D_{\Sigma}$ be the $\Sigma$-DS. 
Assume the decomposition mappings as in Definition~\ref{def:decomposition_ready}. Then:
\begin{itemize}
\item The results in Theorem~\ref{thm:decomposition} hold for $\varphi \in \logicstateready$ and $\psi \in \logicdistready$.
\item Moreover, if $P$ is positive, then the results in Theorem~\ref{thm:decomposition} hold for $\varphi \in \logicstateplus$ and $\psi \in \logicdistplus$.
\end{itemize}
\end{thm}


\begin{proof}
The proof of both items can be obtained by following the one of Theorem~\ref{thm:decomposition} wrt.\ the decompositions of the two logics (Definition~\ref{def:decomposition_ready}).
In particular, we remark that in the proof for the diamond modality in $\logicplus$, we use Corollary~\ref{cor:positive_ruloid} in place of Theorem~\ref{thm:term_ruloid_theorem}.
\end{proof}



\subsection{(Pre)congruence theorems}
\label{sec:decompose_congruence}

To support the compositional reasoning, the congruence (resp.\ precongruence) property is required for any behavioral equivalence (resp.\ preorder) $\rel$.
It consists in verifying whether $f(t_1,\ldots,t_{\n}) \; \rel\; f(t_1',\ldots,t_{\n}')$ whenever $t_i \rel t_i'$ for $i = 1,\ldots,\n$.
In \cite{DGL14} it is proved that probabilistic bisimilarity is a congruence for all operators defined by a PGSOS-PTSS.
We can restate this result as a direct consequence of the characterization result of \cite{DD11} (Theorem~\ref{thm:adequacy_DD11}) combined with our first decomposition result in Theorem~\ref{thm:decomposition}.
Then, by our characterization results in Theorem~\ref{thm:ready_sim_adequate} and our decomposition results in Theorem~\ref{cor:decomposition} we can derive precongruence formats for both ready similarity  and similarity.

\begin{thm}[(Pre)congruence theorem]
\label{thm:congruence}
Let $P=(\Sigma, \Act, R)$ be a PGSOS-PTSS. Then:
\begin{enumerate}
\item 
\label{thm:congruence_bis}
Probabilistic bisimilarity is a congruence for all operators defined by $P$;
\item 
\label{thm:congruence_ready}
Probabilistic ready similarity is a precongruence for all operators defined by $P$;
\item 
\label{thm:congruence_sim}
If $P$ is positive, then probabilistic similarity is a precongruence for all operators defined by $P$.
\end{enumerate}
\end{thm} 


\begin{proof}
\begin{enumerate}
\item
Let $t \in \openT$ and let $\sigma, \sigma'$ be two closed substitutions.
We aim to show that
\begin{equation}
\label{eq:thm_congruence}
\text{whenever } \sigma(x) \sim \sigma'(x) \text{ for each } x \in \var{t} \text{ then it holds that } \sigma(t) \sim \sigma'(t).
\end{equation}
Considering the characterization result of $\logic$ for probabilistic bisimilarity (Theorem~\ref{thm:adequacy_DD11}), to prove the proof obligation Equation~\eqref{eq:thm_congruence} we simply have to show that $\sigma(t)$ and $\sigma'(t)$ satisfy the same formulae in $\logic$.
Assume that $\sigma(t) \models \varphi$, for some state formula $\varphi \in \logic$.
By Theorem~\ref{thm:decomposition}, there is a decomposition mapping $\xi \in t^{-1}(\varphi)$ such that $\sigma(x) \models \xi(x)$ for each $x \in \var{t}$.
From Lemma~\ref{lem:decomposition_mapping_in_logic} we gather that $\xi(x) \in \logicstate$ and moreover by Theorem~\ref{thm:adequacy_DD11} from $\sigma(x) \sim \sigma'(x)$ we obtain that $\sigma'(x) \models \xi(x)$ for each $x \in \var{t}$.
By applying Theorem~\ref{thm:decomposition} once again, we obtain that $\sigma'(t) \models \varphi$, thus proving Equation~\eqref{eq:thm_congruence}. 

\item The proof for probabilistic ready simulation is analogous to the one for item $1$ by exploiting Theorem~\ref{thm:ready_sim_adequate}.$1$ in place of Theorem~\ref{thm:adequacy_DD11}, Theorem~\ref{cor:decomposition}.$1$ in place of Theorem~\ref{thm:decomposition} and Lemma~\ref{lem:decomposition_mapping_in_logicready}.$1$ in place of Lemma~\ref{lem:decomposition_mapping_in_logic}.

\item Under the assumption of $P$ positive, the proof for probabilistic simulation is analogous to the one for item $1$ by exploiting Theorem~\ref{thm:ready_sim_adequate}.$2$ in place of Theorem~\ref{thm:adequacy_DD11}, Theorem~\ref{cor:decomposition}.$2$ in place of Theorem~\ref{thm:decomposition} and Lemma~\ref{lem:decomposition_mapping_in_logicready}.$2$ in place of Lemma~\ref{lem:decomposition_mapping_in_logic}. \qedhere
\end{enumerate}
\end{proof}


\section{Conclusions}
\label{sec:conclusions}

We developed a modal decomposition of formulae in $\logic$ and its subclasses $\logicready, \logicplus$, on nondeterministic probabilistic processes.
The modal logic $\logic$ was introduced in \cite{DD11} for the characterization of probabilistic bisimilarity and we have proved here that $\logicready$ and $\logicplus$ are powerful enough to characterize ready similarity and similarity, respectively.
Our decomposition method is novel with respect to the ones existing in the literature (see for instance \cite{BFvG04,FvG16,FvGL17,FvGW06,FvGW12,GF12}) as it is based on the structural operational semantics of nondeterministic probabilistic processes in the PTS model.

The dual nature of these processes, and of the classes of formulae characterizing them, enforced the introduction of a SOS framework tailored for the specification of distribution terms, namely the $\Sigma$-\emph{distribution specification} in which we have syntactically represented open distribution terms as probability distributions over open terms.
Moreover, the $\Sigma$-\emph{distribution ruloids}, built from this new specification, provide a general tool that can be used to support the decomposition of any modal logic with modalities specifying quantitative properties for the PTS model. 
Moreover, they can be easily adapted to models admitting subdistributions (see among others \cite{LdV15,LdV16}).

To prove the robustness of our decomposition method we have showed how the congruence theorems for probabilistic bisimilarity, ready similarity and similarity with respect to the PGSOS format can be restated as an application of our decomposition theorems.

As future work, we will investigate the application of our decomposition method to modal formulae characterizing dif{f}erent behavioral semantics for nondeterministic probabilistic processes, as trace \cite{BdNL14,CT17,S95tr}, testing \cite{BdNL14,DvGHM08} and weak semantics \cite{AW06,LdV15,LdV16}, and we will derive robust (pre)congruence formats for them from their modal characterizations, as done in \cite{BFvG04,FvG16,FvGL17,FvGW12} in the non probabilistic setting.

Moreover, in \cite{CGT16a} it is proved that by the modal logic $\logic$ we can provide a logical characterization of the bisimulation metric \cite{DCPP06,DGJP04,BW01b}.
Inspired by this result, we aim to start a new research line, that is deriving the compositional properties of a behavioral pseudometric from the modal decomposition of formulae characterizing them.
As the metric semantics provide notions of \emph{distance} over processes, the formats for them guarantee that a small variance in the behavior of the subprocesses leads to a bounded small variance in the behavior of the composed processes (\emph{uniform continuity} \cite{GLT15,GLT16}).
Then, we aim to use the decomposition method to re-obtain the formats for the bisimilarity metric proposed in \cite{GT14,GT15,GT18} and to automatically derive original formats for weak metric semantics~\cite{DJGP02} and metric variants of branching bisimulation equivalence~\cite{AW06}.

\bibliographystyle{alpha}
\bibliography{concur16_bib}

\newpage 

\appendix

\section{Proofs of auxiliary results}



\begin{apx-proof}{Proposition~\ref{lem:sum_to_1}}
We proceed by a case analysis over the form of \dgsos rules.
\begin{itemize}
\item For \dgsos rules $r_{\D} = \{\delta_x \trans[1] x\}$, for some $x \in \SVar$, and $r_{\D} = \{c \trans[1] c\}$, for some constant function $c \in \Sigma$, the thesis is immediate.
\item Consider a \dgsos rule $r_{\D}$ as in Definition~\ref{def:dgsos_rule}.\ref{def:dgsos_rule_f}. 
Then, to prove the thesis we need to show that $\sum_{k \in \bigtimes_{i = 1}^{\n} J_i} q_{k} = 1$.
We have
\begin{align*}
\sum_{k \in \bigtimes_{i = 1}^{\n} J_i} q_{k} ={} & \sum_{k \in \bigtimes_{i = 1}^{\n} J_i} \Big( \prod_{i=1}^{\n} q_{i,k(i)} \Big) \\
={} & \prod_{i = 1}^{\n} \Big( \sum_{j \in J_i} q_{i,j} \Big) \\
={} & \prod_{i = 1}^{\n} (1)\\
={} & 1
\end{align*}
where $\sum_{k \in \bigtimes_{i = 1}^{\n} J_i} \Big( \prod_{i=1}^{\n} q_{i,k(i)} \Big) =  \prod_{i = 1}^{\n} \Big( \sum_{j \in J_i} q_{i,j} \Big)$ follows by the distributive property of the summation wrt.\ the product and can be formally proved by induction over $\n$, with inductive step $\sum_{k \in \bigtimes_{i = 1}^{\n-1} J_i} \Big( \prod_{i=1}^{\n-1} q_{i,(i)} \Big) =  \prod_{i = 1}^{\n-1} \Big( \sum_{j \in J_i} q_{i,j} \Big)$, as follows:
\begin{align*}
& \sum_{k \in \bigtimes_{i = 1}^{\n} J_i} \Big( \prod_{i=1}^{\n} q_{i,k(i)} \Big) \\
={} & \sum_{j \in J_{\n}} q_{\n,j} \Bigg(\sum_{k \in \bigtimes_{i = 1}^{\n-1} J_i} \Big( \prod_{i=1}^{\n-1} q_{i,k(i)} \Big) \Bigg) \\
={} & \sum_{j \in J_{\n}} q_{\n,j} \Bigg( \prod_{i = 1}^{\n-1} \Big( \sum_{j \in J_i} q_{i,j} \Big) \Bigg) & \text{(inductive step)}\\
={} & \Big(\sum_{j \in J_{\n}} q_{\n,j}\Big) \cdot \Bigg( \prod_{i = 1}^{\n-1} \Big( \sum_{j \in J_i} q_{i,j} \Big) \Bigg) \\
={} & \prod_{i = 1}^{\n} \Big( \sum_{j \in J_i} q_{i,j} \Big).
\end{align*}
\item Finally, consider a \dgsos rule $r_{\D}$ as in Definition~\ref{def:dgsos_rule}.\ref{def:dgsos_rule_convex}. 
Then, to prove the thesis we need to show that $\sum_{x \in \{x_{i,j} \mid j \in J_i, i \in I\}} q_x = 1$.
We have
\begin{align*}
& \sum_{x \in \{x_{i,j} \mid j \in J_i, i \in I\}} q_x \\
={} & \sum_{x \in \{x_{i,j} \mid j \in J_i, i \in I\}} \Big( \sum_{i \in I, j \in J_i \atop \text{ s.t. } x_{i,j} = x} p_i q_{i,j} \Big) \\
={} & \sum_{i \in I} p_i \Big( \sum_{x \in \{x_{i,j} \mid j \in J_i, i \in I\} \atop j \in J_i \text{ s.t. } x_{i,j} = x} q_{i,j} \Big) \\
={} & \sum_{i \in I} p_i \Big( \sum_{j \in J_i} q_{i,j} \Big) & \text{(for each $i \in I$ the $x_{i,j}$ are distinct)}\\
={} & \sum_{i \in I} p_i \\
={} & 1.
\end{align*}
\end{itemize}
\end{apx-proof}




\begin{apx-proof}{Proposition~\ref{prop:sigma_reduction}}
The thesis follows directly by the definition of $\sigma(L)$.
In fact, if we let $\sigma(L) = \{\sigma(\Theta) \trans[q_j] t_j \mid j \in J\}$, then the targets $t_j$ are pairwise distinct by construction and moreover we have
\begin{align*}
& \sum_{j \in J} q_j\\
={} & \sum_{j \in J} \Big( \sum_{\{i \in I \mid \sigma(t_i) = t_j\}} q_i \Big) \\
={} & \sum_{i \in I} q_i \\
={} & 1 & \text{($L$ is a distribution over terms)}.
\end{align*}
\end{apx-proof}




\begin{apx-proof}{Proposition~\ref{lem:sum_to_1_tris}}
As the conclusion of a \dgsos ruloid coincides with the conclusion of a reduced instance of the \dgsos rule on which the \dgsos ruloid is built, the thesis follows immediately from Proposition~\ref{lem:sum_to_1_bis}. 
\end{apx-proof}


\end{document}